\documentclass[]{jfm_arXiv}
\usepackage{graphicx}
\usepackage{newtxtext}
\usepackage{newtxmath}
\usepackage{natbib}
\usepackage{subfig}
\usepackage{setspace}
\usepackage{hyperref}
\hypersetup{
    colorlinks = true,
    urlcolor   = blue,
    citecolor  = blue,
}

\newcommand{\RomanNumeralCaps}[1]
\linenumbers
\title{Turbulence Modeling of 3D High-speed Flows with Upstream-Informed Corrections}
\author{Chitrarth Prasad
  \corresp{\email{prasad.141@osu.edu}}
  \and Datta V. Gaitonde}

\affiliation{Department of Mechanical and Aerospace Engineering, The Ohio State University, OH 43210}

\usepackage{xargs}
\usepackage[colorinlistoftodos,prependcaption,textsize=tiny]{todonotes}
\newcommandx{\unsure}[2][1=]{\todo[linecolor=red,backgroundcolor=red!25,bordercolor=red,#1]{#2}}
\newcommandx{\needswork}[2][1=]{\todo[linecolor=blue,backgroundcolor=blue!25,bordercolor=blue,#1]{#2}}
\newcommandx{\info}[2][1=]{\todo[linecolor=OliveGreen,backgroundcolor=OliveGreen!25,bordercolor=OliveGreen,#1]{#2}}
\newcommandx{\improvement}[2][1=]{\todo[linecolor=Plum,backgroundcolor=Plum!25,bordercolor=Plum,#1]{#2}}
\newcommandx{\thiswillnotshow}[2][1=]{\todo[disable,#1]{#2}}
\usepackage{ulem}
\usepackage{soul}

\begin{document}
\maketitle

\begin{abstract} 
Turbulence modeling has the potential to revolutionize high-speed vehicle design by %providing accurate but cost-effective estimates of crucial aerothermal loading parameters. 
%High-speed vehicle design can be significantly advanced by accurate but cost-effective estimates of crucial aerothermal loading parameters.
serving as a co-equal partner to costly and challenging ground and flight testing. 
However, the fundamental assumptions that make turbulence modeling such an appealing alternative to its scale-resolved counterparts also degrade its accuracy for practical high-speed configurations, especially when fully 3D flows are considered.
The current investigation develops a methodology to improve the performance of turbulence modeling
%improve the surface predictions of 
%two-equation turbulence models 
for a complex Mach~8.3, 3D shock boundary layer interaction (SBLI) in a double fin geometry. %generated by $15^\circ$ fins on an adjacent flat plate. %
A representative two-equation model, with low-Reynolds number terms, is used as a test-bed.
Deficiencies in the baseline model %that inhibit accurate predictions 
are first elucidated using benchmark test cases involving a Mach~11.1 zero pressure gradient boundary layer and a Mach~6.17 flow over an axisymmetric compression corner. %, the deficiencies in the baseline model %that inhibit accurate predictions 
%are exposed.  %, the reasons responsible for degraded predictions are estabilished. 
%cause of flow over a series of increasingly complicated benchmark problems that collectively encompass many of the major phenomena expected in the hypersonic regime. 
%Emphasis is given to both axisymmetric and 3D configurations at Mach numbers in excess of 6.
%Using a flat plate and an axisymmetric 
%This is achieved by means of 
%Using the $k-\epsilon$ model as a baseline, 
From among different possibilities, two 
%user-defined 
coefficients are introduced to inhibit the non-physical over-amplification of (i) turbulence production and (ii) turbulence length-scale downstream of a shock wave.
%A key feature in this investigation is the use of 
The coefficients rely on terms already present in the original model, which simplifies implementation and maintains computational costs.
The values of the coefficients are predicated on the distribution of turbulence quantities upstream of the shock; this ensures that the modifications do not degrade the model predictions in simpler situations such as attached boundary layers, where they are unnecessary.  
%A crucial step is to select the values of these coefficients based on the distribution of turbulence quantities upstream of the shock; this ensures that these modifications do not degrade the model predictions in attached boundary layers where they are not necessary. 
% Another key feature of these modifications is the %do not introduce additional terms besides those 
% reliance on terms already present in the baseline model.
%making them straightforward to implement in any existing code.
%Thus, these corrections are straightforward to implement in any existing code and do not increase computational costs. 
%The variation of pre-existing terms in the baseline model upstream of the shock guides the values of correction coefficients; this ensures that these modifications are straightforward to implement in any existing code and do not degrade the performance in cases where they are not necessary. 
%The combined effect of these modifications %is observed to 
The effects of the modifications are shown to result in significant improvements in surface pressure and wall heat flux for the 3D SBLI test case, which contains numerous features not observed in 2D situations, such as 3D separation, skewed boundary layers and centerline vortices.
%while retaining the exact computational cost as the baseline. 
Considerations on the inflow values of turbulence variables and mesh resolution are provided

\noindent\textbf{Keywords:} Turbulence Modeling, Hypersonic
\end{abstract}

\begin{keywords}
%turbulence modeling, hypersonic
\end{keywords}

%\end{frontmatter}

%\linenumbers
%Lots of interest in RANS.
%Lots of codes exist.
%Performance leaves something to be desired.
%Lots of corrections have been developed. Conflict with baseline results.
%The one's that dont, for eg, Variable Prandtl number, Bowersox model, etc. These require drastic modification to the codes.
%We seek (i) simple corrections that can be implemented in existing codes, (ii) do not degrade performance in baseline situations.

%Three test cases, increasing order of complexity. We see that the 3D case exhibits features that cannot be simulated in 2D.
\section{Introduction} \label{sec:Intro}
%Why is RANS Important
With the renewed interest in hypersonic flight, %demands 
cost-effective estimates of mean surface loading parameters such as pressure, skin friction coefficient, and heat transfer rates %are highly desirable. 
can significantly accelerate design evolution.
%In particular, they may reduce the considerable cost-function associated with the parametric analysis and optimization of hypersonic vehicle components, for which ground and flight testing is difficult and costly. %and are extremely expensive.
%This motivates the continued vigorous development of Reynolds-Averaged Navier-Stokes (RANS) methods, since 
High-fidelity numerical approaches such as Direct Numerical Simulations (DNS) and Large-Eddy Simulations (LES) are prohibitively expensive  at high Reynolds numbers and remain well out of reach for practical systems for the foreseeable future; this motivates the continued vigorous development of Reynolds-Averaged Navier-Stokes (RANS) methods. %for high-speed configurations.
RANS approaches have the potential to greatly reduce the considerable cost-function associated with the parametric analyses and optimization of hypersonic vehicle components, for which ground and flight testing is difficult and costly.
%Although efforts such as Wall-Modeled LES (WMLES) and Detached Eddy Simulations (DES) show promise, their development is dependent on RANS models which are invoked in the near-wall region where scales are too small to be economically resolved at high Reynolds numbers~\cite{kawai2012wall}. 

%Popularity of RANS and Degradation of accuracy
In addition to determining quantities of design interest, suitably tailored RANS can also provide insights into the unsteadiness of complex flow-fields 
(i) through various perturbation techniques~\citep{ranjan2020robust,prasad2022time} % based on the recognition that larger unsteady scales can be recovered as stability modes of the mean flow-field \cite{crighton1976stability}, 
and (ii) as an essential near-wall component of Wall-Modeled LES (WMLES)~\citep{kawai2012wall} and Detached Eddy Simulations (DES)~\citep{spalart2009detached}.
These factors make RANS an attractive tool for the aerospace industry, as evident from its widespread adoption in commercial solvers. %to greatly reduce the considerable cost-function associated with the parametric analysis and optimization of hypersonic vehicle components.

%This great advantage of RANS, its cost-effective nature, is also its Achilles's heel.
%The comparatively lower expense of RANS stems from modeling, instead of resolving, smaller length and time scales.
Commonly used compressible RANS codes solve the Favre-averaged equations of motion. %this gives rise %
%in order to limit the number of terms that must be modelled. % have close analogies with their incompressible counterparts.
%Despite this simplification, 
Favre-averaging gives rise to additional terms in the governing equations %corresponding to 
representing Reynolds stresses, turbulent heat flux, molecular diffusion and turbulent transport; %, among others. 
%in the governing equations. % that are typically modeled based on 
%thus requiring the use of a turbulence model to close the governing equations. %assumptions  %the closure of terms involving Reynoldin the governing equations through the use of a turbulence model.
%This closure is achieved using assumptions based on 
%analogies between molecular and turbulent mixing~\cite{}, and 
these terms require the use of closure approximations through a turbulence model. %to close the governing equations. %are modeled based on observations from equilibrium turbulent flat-plate boundary layer flows.
The most widely used models involve closure approximations based on %analogies 
observations from equilibrium turbulent flat-plate boundary layer flows.
Popular prescriptions include the use of Boussinesq approximation to model the Reynolds stresses, the application of strong Reynolds analogy to model the turbulent heat flux and the adoption of gradient approximations to model molecular diffusion and turbulent transport. 
These simplifying assumptions form the foundation of RANS and significantly reduce its computational requirement %of RANS, making it an appealing alternative over 
when compared to 
its high-fidelity counterparts.
%These assumptions however, do not translate to practical high-speed flow configurations, which are characterized by high heat transfer rates and shock boundary layer interactions (SBLI)
However, in practical high-speed flow configurations, where high heat transfer rates and shock boundary layer interactions (SBLI) are ubiquitous, these defining assumptions of RANS severely compromise its ability to capture experimentally measured behavior. 
%Although these simplifying assumptions significantly reduce the computational requirement of RANS, making it an appealing alternative over its high-fidelity counterparts, they severely degrade the accuracy of RANS predictions for practical high-speed flow configurations where high heat transfer rates and shock boundary layer interactions (SBLI) are ubiquitous. 

%Existing analyses on the inaccuracy of RANS
%This lack of accuracy of RANS 
The degradation of RANS accuracy for practical high-speed configurations is well documented~\citep{roy2006review,georgiadis2014status,coratekin2004performance}, and several investigations have focused on improving predictions by examining the validity of these assumptions at high Mach numbers. %using resolved simulations. %and incorporating corrections. 
%%Boussinesq approximation
For instance, %Morgan \emph{et al.}~
\cite{morgan2013flow} studied the Reynolds stress transport budgets in SBLI flows at Mach~$2.28$ using a sequence of LES databases and argued that the inaccuracy of RANS stems from the inability of the Boussinesq approximation to represent the observed anisotropy in Reynolds stresses. 
They further indicated that although Reynolds stress transport (RST) models offered improved predictions than traditional eddy-viscosity-based turbulence models, their accuracy is still limited due to gradient approximations for turbulent transport.
%Zhang \emph{et al.}~
\cite{zhang2021formulation,zhang2022application} suggested a possible alternative to the RST-based-models by extending the application of a non-linear eddy-viscosity model~\citep{craft2000progress} to hypersonic flows. 
Although the inclusion of non-linearity in Boussinesq approximation improved predictions, they observed a $100\%$ over-estimation of wall heat flux at Mach~$7$.  
This over-prediction in the heat flux was reduced to $20\%$ by introducing an additional source term in the governing equations~\citep{zhang2022application}.

%Got to heat flux, so now focus on Reynolds analogy
%Xiao \emph{et al.}~
\cite{xiao2007role} took a fundamentally different approach to improve RANS heat transfer predictions by accounting for variable turbulent Prandtl number effects in RANS equations; the implication being that the strong Reynolds analogy is not valid across a shock wave~\citep{mahesh1997influence}.
This idea of variable turbulent Prandtl number was further expanded to flows without shocks by %Huang \emph{et al.}~
\cite{huang2020simulation}, who used DNS data of a Mach~$11$ flow over a cooled flat plate to observe that the algebraic energy flux model by %Bowersox~
\cite{bowersox2010algebraic} gave a significantly improved prediction of the turbulent heat flux compared to the commonly used constant turbulent Prandtl number assumption. %

%So many options exist, so what's the problem??
%These investigations demonstrate that %
In hindsight, it is rather unsurprising 
%These investigations demonstrate 
that RANS predictions can be improved by substituting one or more of its fundamental assumptions
%Replacing These assumptions are typically replaced 
with extra transport equations~\citep{xiao2007role,morgan2013flow,vemula2017reynolds,vyas2019reynolds} or phenomenological models~\citep{vuong1987modeling,sinha2005modeling,pathak2018phenomenological}. 
Although extra transport equations can be more physically meaningful, their implementation in existing codes is not straightforward and invariably increases computational costs; this restricts the computational advantage offered by RANS.
%Moreover, their implementation in existing RANS codes is not straightforward.
In contrast, phenomenological models are easier to implement and incur negligible extra costs, but their calibration in axisymmetric/2D problems usually does not extend accurately to 3D problems.
% are often calibrated for %a small subset of 
% axisymmetric/2D problems and thus require further %problem-dependent 
% adjustment for more complex 3D configurations. %development.
%require further development and calibration for problems with extra strain-rates.
The problem is further exacerbated because 
%Furthermore, 
some prescriptions may have the potential to conflict with others, and improvement in predictions for one problem may be accompanied by a degraded performance in others. 
This has led to a confusing array of possibilities and has inhibited the effectiveness of RANS in sophisticated production suites. 

%I see the problem, how do you address it
%The goal of this investigation is to improve the predictions of a generic baseline RANS model at high Mach numbers without challenging the key assumptions that may incur some extra computational cost. 
%This investigation aims to implement improve predictions from an existing RANS code at high Mach numbers. %models by 
This investigation aims to improve RANS predictions for 3D SBLI configurations at high Mach numbers without severely modifying the key assumptions responsible for the appeal of RANS techniques.
Specifically, we seek 
%The goal of this investigation is to seek 
corrections that (i) improve RANS predictions relative to existing approaches for flows where 3D features are dominant, (ii) do not degrade the performance in simpler cases where baseline RANS model is accurate, (iii) are straightforward to implement in existing RANS codes, and (iv) do not incur additional costs.
%For this, 
We select the two-equation Launder and Sharma $k-\epsilon$ model~\citep{gerolymos1990implicit} as a platform for further development; this selection is not restrictive based on its widespread availability in commercial solvers and its well known limitations in adverse pressure gradients~\citep{rodi1986scrutinizing,huang1994turbulence}.
A summary of the $k-\epsilon$ formulation is provided in $\S$~\ref{section:Turbmodel}.

The $k-\epsilon$ model is applied to a series of benchmark problems that collectively include the major phenomena expected in current and future hypersonic vehicles.
We focus our attention on test cases in excess of Mach~$6$.
%We focus on flow configurations above Mach 6.
Both two- and three-dimensional (2D, 3D) configurations are selected; a description of each of these test configurations and the underlying rationale for their selection is presented in $\S$~\ref{section:test}. %selection
Each configuration is well documented and has a range of published data, either from experiments or benchmark higher-fidelity simulations. %(wall-resolved LES or DNS). 
%Section~\ref{section:results} assesses 
The performance of the baseline $k-\epsilon$ model for each test configuration is examined in $\S$~\ref{section:results}. %and introduces corrections to improve results. 
Special care is taken to ensure that the solution is independent of factors that have no direct connection to the model itself, such as grid resolution and freestream turbulence. 
Based on the performance of the baseline model, corrections are introduced to improve predictions.
These corrections are carefully calibrated based on the upstream boundary layer to ensure they do not degrade the accuracy in other test cases where they are not necessary. 
%This strategy should contribute to efforts that seek 
Best practices for boundary conditions, mesh resolution and the use of corrections are documented to provide a degree of diagnostic analyses based on the phenomena of interest.
Concluding remarks are made in $\S$~\ref{section:conclude}.

\section{Turbulence Model Formulation} \label{section:Turbmodel}
%The present investigation considers primarily two-equation models. 
Following %Coakley and Huang 
\cite{coakley1992turbulence}, a general two-equation turbulence model can be written in the framework of two variables, $\chi_1$ and $\chi_2$ as follows %, %which determine the eddy viscosity, $\mu_t$.
%Popular models, including two-equation variants listed on the NASA Turbulence Modeling Resource~\cite{nasa} may be recovered by proper specification of terms.
%The governing equations for the turbulence variables are
\begin{equation}\label{eqn:two_eqn}
    \frac{\partial \left( \rho \chi_i \right)}{\partial t} + \frac{\partial}{\partial x_j} \left( \rho \chi_i u_j  -\left( \mu + \frac{\mu_t}{\sigma_i}\right) \frac{\partial \chi_i}{\partial x_j} \right) = \Psi_i, 
\end{equation}
where $\rho$ is the density, $\mu$ is the molecular viscosity and $\mu_t$ is the eddy viscosity. 
The subscript $i=1,2$ represents the equation for the corresponding turbulence variable $\chi_i$.
$\Psi_1$ and $\Psi_2$ are source terms whose general forms are given by
\begin{equation} \label{eqn: source}
    \Psi_i = \left[ C_{i1} C_\mu f_i \left( \frac{S}{\omega}\right)^2 -\alpha_i \frac{D}{\omega} -C_{i2}\right]\rho \omega \chi_i
\end{equation}
in terms of various constants $(C)$, functions $(f)$, and modeling parameters, $\alpha$. 
These source terms are a function of dilatation, $D=\partial u_k/\partial x_k$, specific rate of dissipation $(\omega)$ and the strain invariant $S^2$ given by
\begin{equation}
    S^2= \left( \frac{\partial u_i}{\partial x_j} + \frac{\partial u_j}{\partial x_i}\right)\frac{\partial u_i}{\partial x_j} - \frac{2}{3}\left( \frac{\partial u_k}{\partial x_k} \right)^2.
\end{equation}
The eddy viscosity, $\mu_t$, in its general form can be written as
\begin{equation}\label{eqn:eddy}
    \mu_t=C_\mu f_\mu \rho \hat{q} l,
\end{equation}
where, $\hat{q}=\sqrt{k}$ is the velocity scale and $l$ is a length scale that is determined by the variables $\chi_1$ and $\chi_2$. The values of $C_\mu$ and the damping function $f_\mu$ also depend on this choice. 

As stated previously, this investigation focuses on the Launder and Sharma $k-\epsilon$ model; this sets $(\chi_1,\chi_2)=(k,\epsilon)$ in eqn.~\ref{eqn:two_eqn}. 
The length scale in eqn.~\ref{eqn:eddy} is now given by $l=\frac{k^{3/2}}{\epsilon}$. 
The values of other model parameters are listed below in Table.~\ref{tab:parameters}.
\begin{table}
    \centering
    \caption{Model parameters for the baseline $k-\epsilon$ model.} \small
    \begin{tabular}{|l| l| l| l|} \hline \hline
        $\sigma_1=1.0$ & $C_{11}=1.0$ & $C_{\mu}=0.09$ & $f_1=f_\mu=\exp{\left(-3.4/(1.0+R^2_T/50)\right)}$\\
        
        $R_T=k^2/(\nu \epsilon)$ & $\alpha_1=\frac{2}{3}C_{11}$  
          & $\sigma_2=1.0$ & $C_{12}=1 + \frac{2 \nu}{\epsilon} \left[\left(\frac{\partial \hat{q}}{\partial x}\right)^2 + \left(\frac{\partial \hat{q}}{\partial y}\right)^2 + \left(\frac{\partial \hat{q}}{\partial z}\right)^2\right]$ \\
        
        $C_{21}=1.45$ & $\alpha_{2}=\frac{2}{3}C_{21}$ & $f_2=1-0.3\exp{\left(-R^2_T\right)}$& $C_{22}=1.92f_2 -(2\nu \nu_T/\epsilon^2) \left[\frac{\partial^2}{\partial x_k \partial x_k} \left((u_l u_l)^{1/2}\right)\right]^2$\\ \hline \hline
    \end{tabular}
    \label{tab:parameters}
\end{table}
%The wall damping terms are accommodated in equation~\ref{eqn: source} by adjusting $f_1$ and $f_2$. 
%Although the present investigation does not consider wall functions, wall damping terms can be easily replaced by wall functions, if required~\cite{grotjans1998wall}. 

%Currently, the focus is primarily on the Launder and Sharma $k-\epsilon$ model \cite{gerolymos1990implicit}.
All simulations are performed %in generalized curvilinear coordinates 
using an in-house finite difference solver %structured solver. %, 
that solves the full 3D Favre-averaged NS equations in generalized curvilinear coordinates. 
Steady state is achieved through time marching using an approximately factored~\citep{pulliam1981diagonal}, implicit second-order Beam Warming scheme~\citep{beam1978implicit}.
A third-order upwind biased Roe scheme~\citep{roe1981approximate} is used for the inviscid fluxes with the van Leer harmonic limiter~\citep{van1979towards}, whereas the viscous fluxes are computed using a second-order central differencing scheme. 
All calculations use the ideal gas relation with air as a working fluid $(Pr = 0.72)$. 
The molecular viscosity is calculated using Sutherland's law. 
The turbulent heat flux is modeled using the standard Reynolds analogy with a constant turbulent Prandtl number, $Pr_{T}=0.9$.
Details of the meshes and boundary conditions %used in this study are presented in %the next section. meshes and boundary conditions 
for each specific test case are presented in $\S$~\ref{section:results}.

\section{Test Configurations} \label{section:test}
\begin{figure}
    \centering
    \subfloat[ \label{fig:testcasesa}]{\includegraphics[width=0.31\textwidth,trim=20 25 20 12,clip]{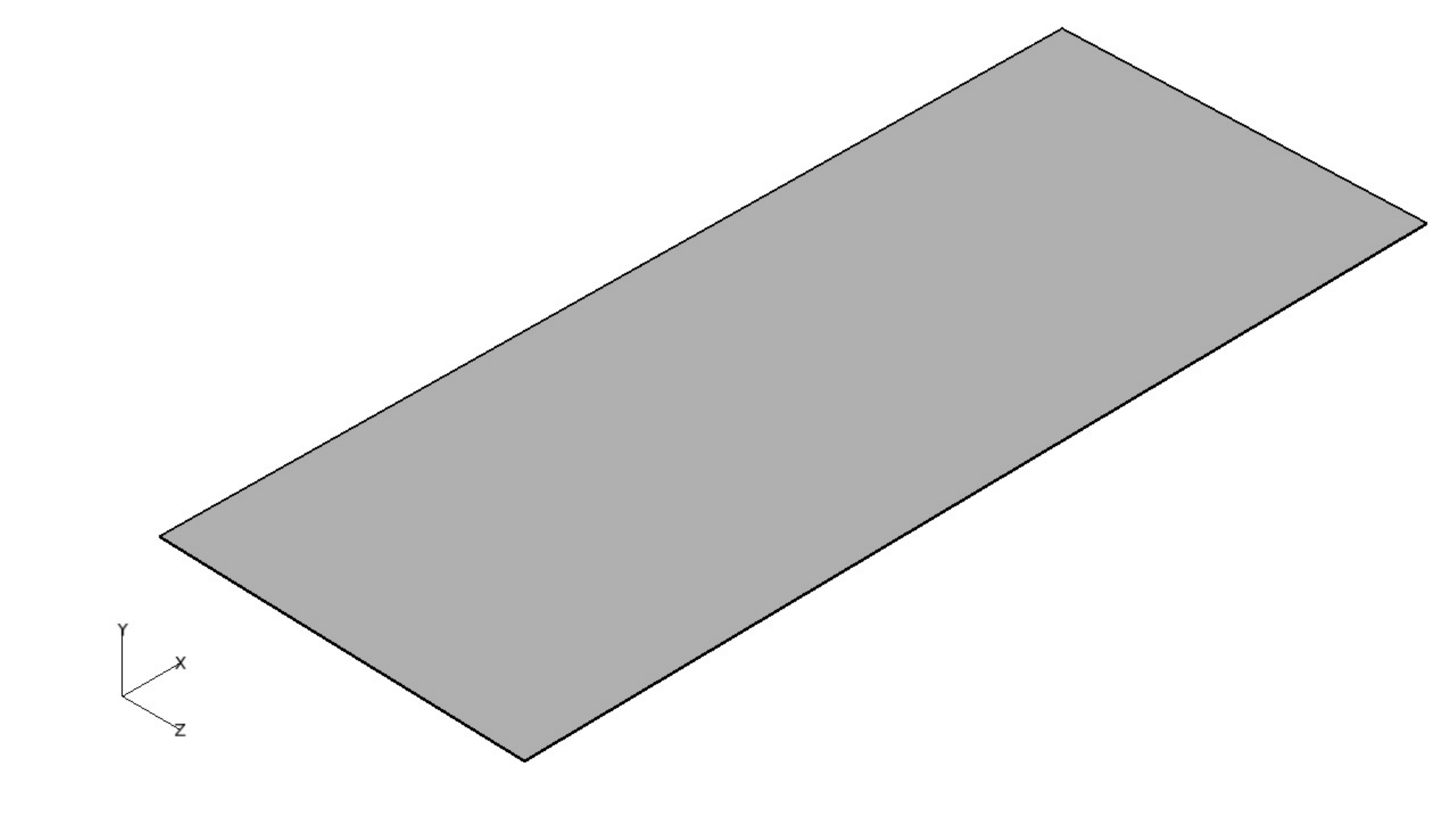}}
    \subfloat[ \label{fig:testcasesb}]{\includegraphics[width=0.31\textwidth,trim=20 25 20 20,clip]{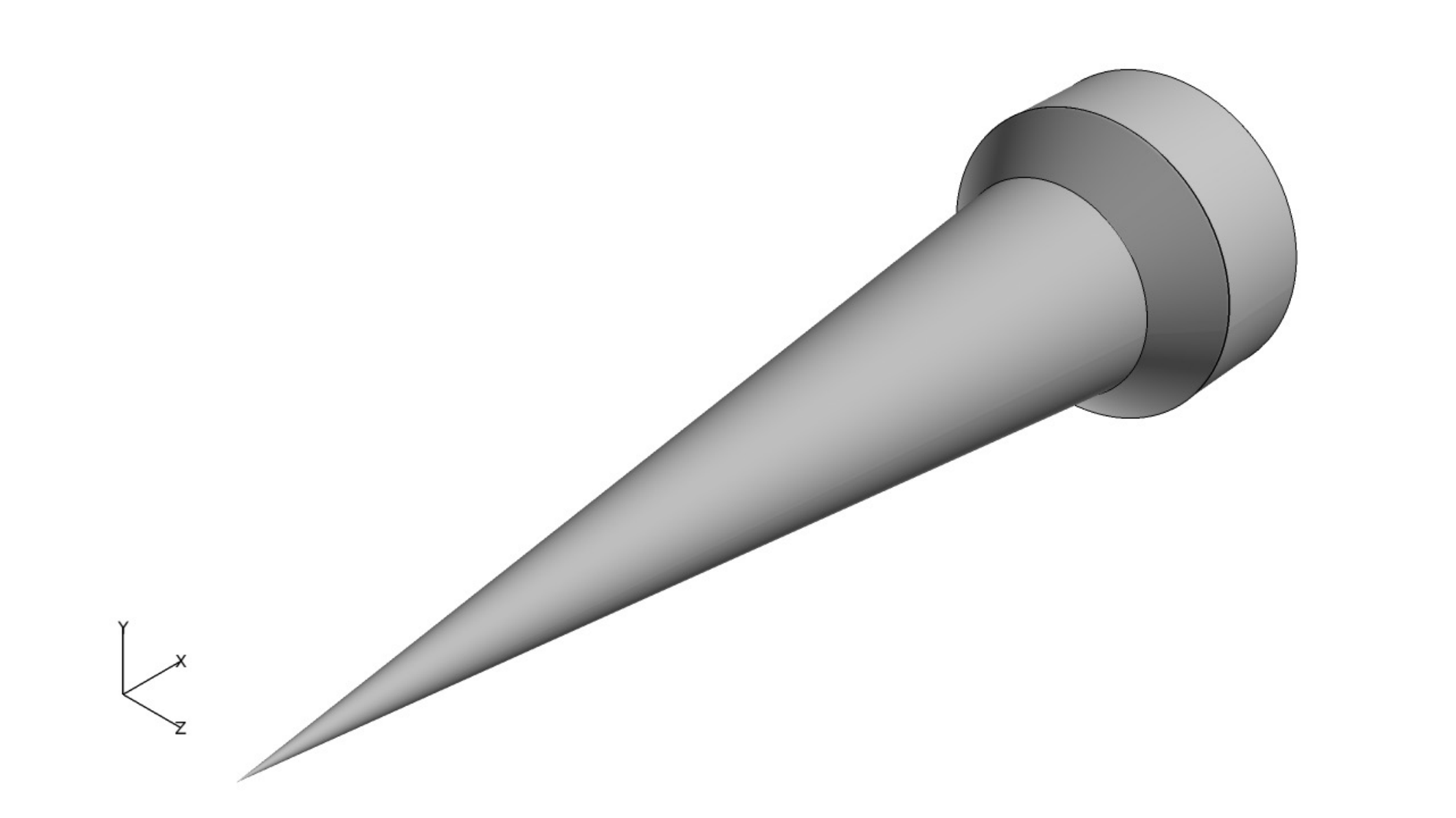}}
      \subfloat[\label{fig:testcasesd}]{\includegraphics[width=0.33\textwidth,trim=20 10 20 12,clip]{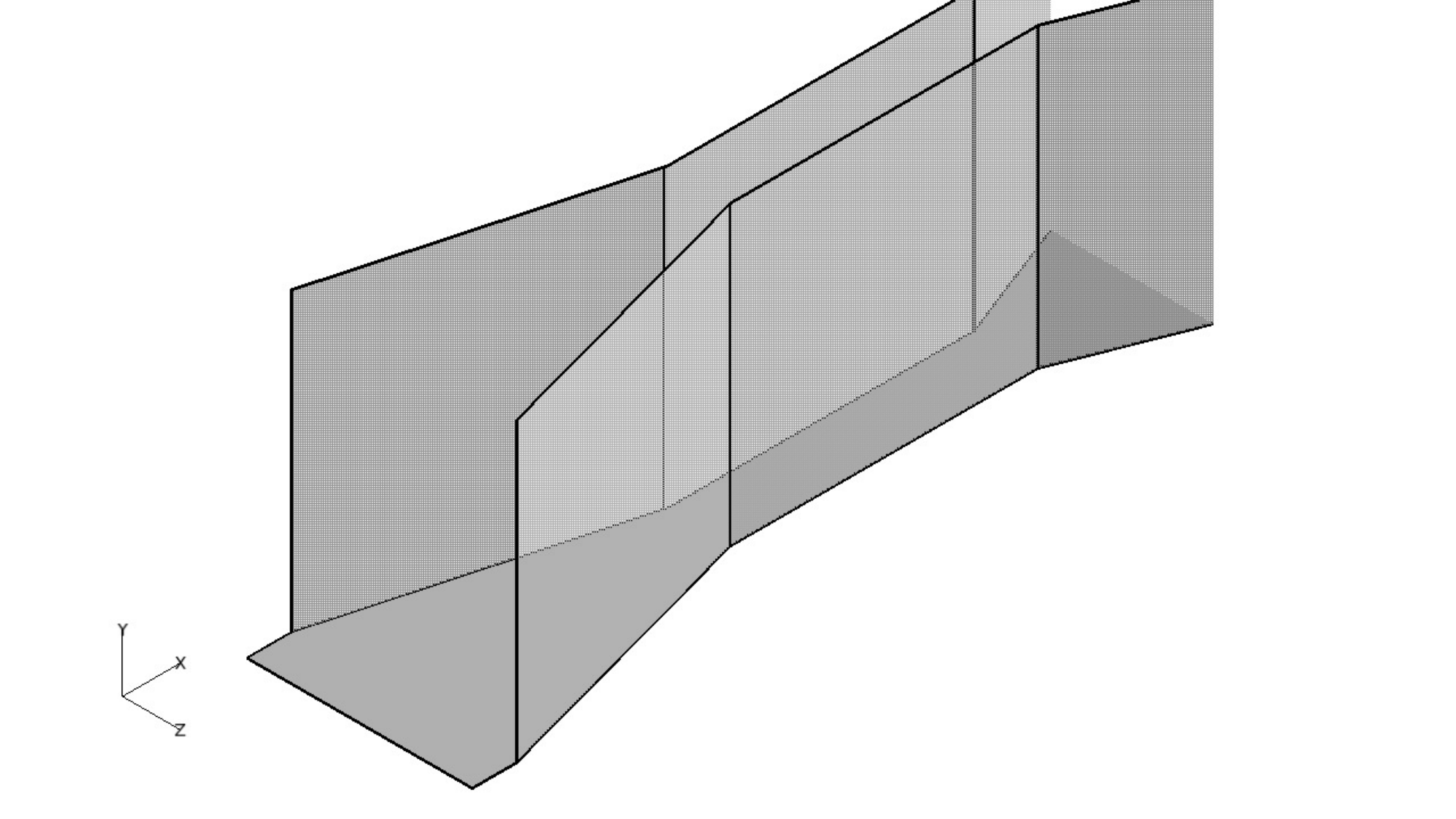}}
    \caption{Test configurations used in the present work: (a) Mach~$11.1$ flow over a flat plate (FPM11)~\citep{gnoffo2013uncertainty}, (b) Mach~$6.17$ flow over a long cone flare (LCFM6)~\citep{holden2018measurements} and (c) Mach~$8.3$ inflow to a double fin (DFM8)~\citep{kussoy1993hypersonic}. }
    \label{fig:testcases}
\end{figure}

%This section briefly describes all the test configurations used in the present study. 
The test cases are selected so that (i) they collectively exhibit a wide range of phenomena expected to occur in current and future high-speed vehicles, and (ii) have high-quality databases for validation in the form of benchmark DNS or experiments.
The configurations are presented in Fig.~\ref{fig:testcases}, arranged in increasing order of complexity. 

\subsection{Flat Plate (FPM11) \label{section:FPM11}}
%Mach 11. Cooled plate. Heat transfer and shear stress measurements. DNS and CUBRC data. Also serves as inflow condition for other subsequent test configurations.
The first test case shown in Fig.~\ref{fig:testcasesa} corresponds to a Mach~$11.1$ uniform flow over a cold flat plate at zero angle of attack. 
Accurate modeling of turbulent boundary layers subjected to wall cooling is critically important in the design of thermal protection systems. 
In addition, results from flat plate simulations are often required as inflow conditions for other hypersonic configurations of interest as shown later. 
These factors make the flow over a cooled flat plate an important step in assessing any turbulence model. 

The freestream conditions are $U_\infty=1780$ m/s, $\rho_\infty=0.09483$ kg/m$^3$ and  $T_\infty=64 $~K. 
The wall temperature $(T_w)$ is kept constant at $300$ K; these conditions are based on the measurements conducted as Calspan–University of Buffalo Research Center (CUBRC)~\citep{gnoffo2013uncertainty}.  
This test case has been studied extensively in literature~\citep{gnoffo2013uncertainty,rumsey2010compressibility,huang2020simulation} to assess the performance of turbulence models and their corrections.
%High-quality DNS data is also available for this case~\cite{huang2020simulation}.
The important parameters of interest are the shear stress and the heat transfer at the wall. 
This case is henceforth referred to as FPM11.

\subsection{Long Cone Flare (LCFM6) \label{section:LCF}}
The second test configuration used in the present study consists of a $7^\circ$ long sharp cone with a $40^\circ$ downstream flare as shown in Fig.~\ref{fig:testcasesb}. 
The long cone is representative of a hypersonic vehicle forebody~\citep{wadhams2008pre} and allows for the development of a fully turbulent flow prior to an SBLI generated by the higher angle flare. 

%\textit{Maybe add about how compression corner cases have been studied extensively in literature.}
Similar compression corner flows have been used extensively in literature to test and improve turbulence model predictions~\citep{xiao2007role,gnoffo2011uncertainty,wang2016modular,pathak2018phenomenological,raje2021anisotropic,zhang2021formulation,zhang2022application}.
This particular test case has been studied extensively for both laminar and turbulent conditions by %Holden~\emph{et al.}~
\cite{holden2018measurements}; %using the LENS facilities. 
%This database includes both standard tunnel conditions as well as a flight enthalpy case, providing the opportunity to consider the impact of high-temperature effects within the studied RANS models. 
the flow conditions used here correspond to ``Run 33'' in~\cite{holden2018measurements}.
 %Holden~\emph{et al.}. 
This test case comprises a Mach~$6.17$ uniform flow at $U_\infty=931.16$ m/s, $\rho_\infty = 0.07367$ kg/m$^3$ and $T_\infty=56.67$~K over the cone geometry. 
The wall temperature is kept constant at $297.7$~K. 
The key measurements are surface pressure and heat transfer. 
This case is referred to as LCFM6 in the manuscript.

\subsection{Double Fin Configuration (DFM8)}
%3D shock boundary layer interaction. Inflow of a scramjet. Shock generated at the fin interacts with the boundary layer on the plate at 90degrees
The final test case consists of two identical $15^\circ$ fins mounted on a flat plate as shown in Fig~\ref{fig:testcasesd}. %is considered with two identical $15^\circ$ fins mounted on a plate. 
The flow consists of an incoming Mach~$8.3$ boundary layer at $U_\infty=1483$~m/s, $\rho_\infty = 0.0186$~kg/m$^3$ and $T_\infty=80$~K that interacts with the two shocks and other derivative features generated by the two fin surfaces. 
The wall temperature is kept constant at $300$~K based on the measurements of %Kussoy \emph{et al.}~
\cite{kussoy1993hypersonic} which are used for validation.
We refer to this test case as DFM8 in the rest of the manuscript.

Such 3D SBLIs where the shocks due to one surface interact with a boundary layer on an adjacent surface are expected to be ubiquitous in hypersonic vehicle designs in both internal (scramjet flowpaths) and external (appendages) circumstances. 
%The DFM8 flow-field exhibits drastically different characteristics that are not observed in general axisymmetric/2D SBLIs.
The resulting flows of interest exhibit 3D flow separation, streamwise vortical structures, and complex shock/expansion systems, which involve efficiency degradation and even catastrophic unstart~\citep{gaitonde2015progress}. 
Such interactions also play a crucial role in mode transition, such as ramjet to scramjet in dual-mode designs.
Turbulence model  modifications designed for 2D configurations (such as FPM11) do not extend to flows containing peculiarly 3D features.
As elucidated later, the choice of DFM8 as a  target flowfield of interest is motivated by the fact that it contains such features.

%\begin{table}
%    \centering
%    \caption{Test Conditions for the test case DFM8 shown in Fig.~\ref{fig:testcasesd}.}
%    \label{table:DFconditions}
 %   \begin{tabular}{|c|c|c|c|c|c|c|c|} \hline
  %      $M_\infty$ & $U_\infty$ & $\rho_\infty$ & $T_\infty$ & $ \delta$ & $\theta$ & Re$_\delta$ & $T_w$\\ \hline
   %      8.28 & 1483 m/s & 0.0186 kg/m$^3$ & 80K & 3.25 cm & 0.083 cm & $1.7 \times 10^5$ & 300K \\ \hline
    %\end{tabular}
    
%\end{table}

\section{Results and Discussion}\label{section:results}
\subsection{Baseline Results}
\begin{figure}
\centering
    \subfloat[]{\includegraphics[width=0.47\textwidth,trim=90 255 100 240,clip]{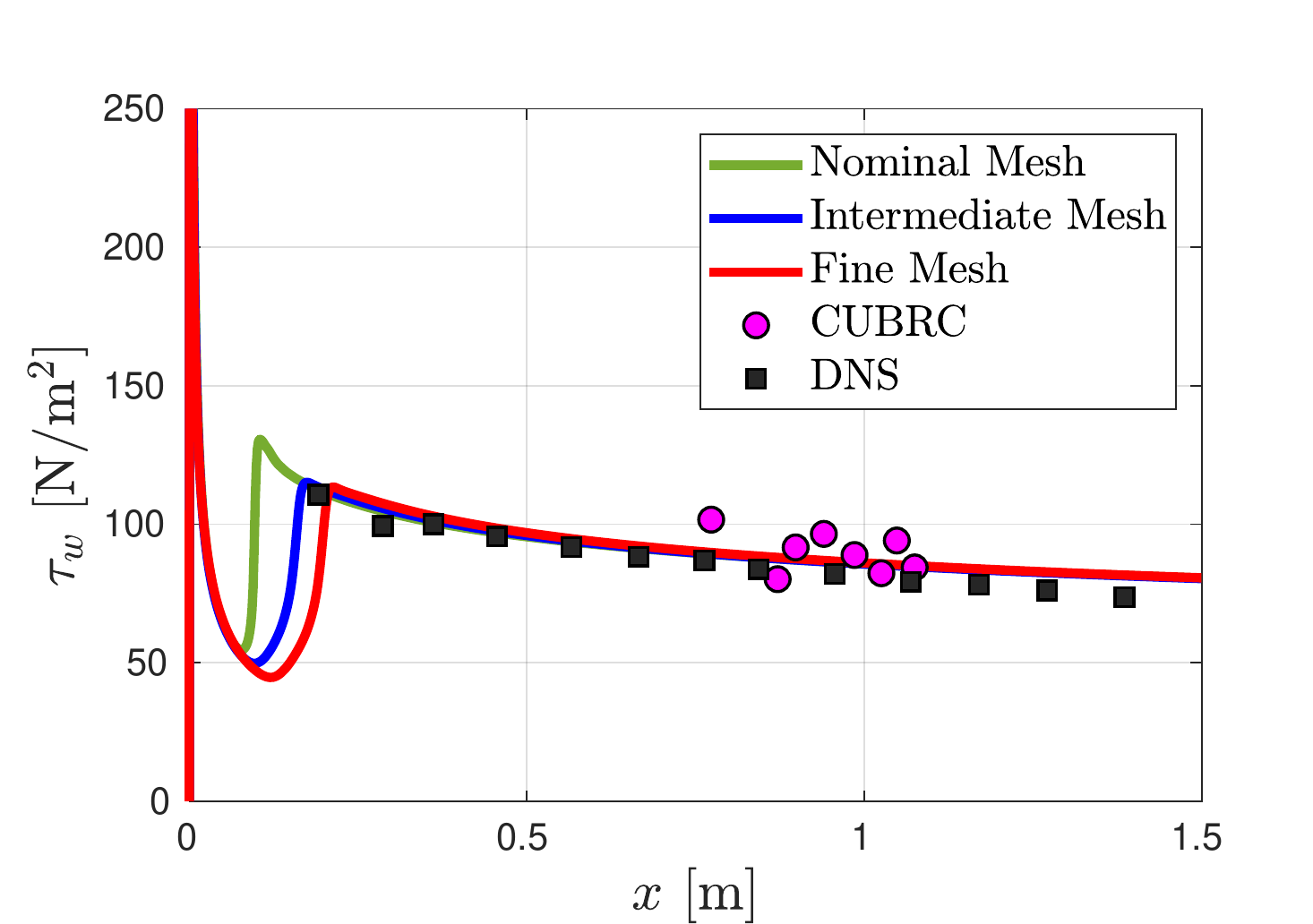}} \hspace{0.1in}
    \subfloat[]{\includegraphics[width=0.47\textwidth,trim=100 270 100 260,clip]{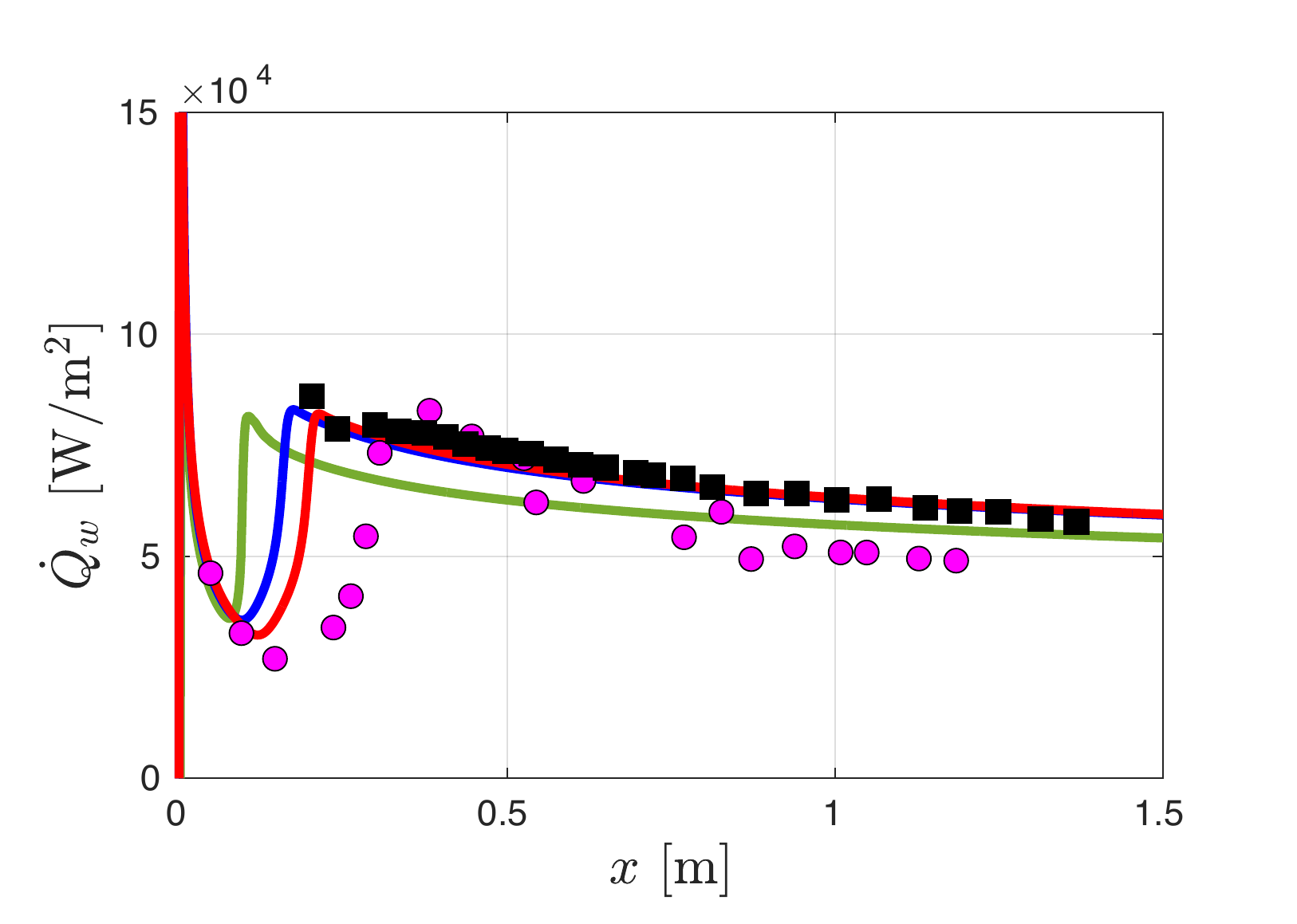}} \\
     \caption{FPM11: Baseline predictions for wall shear stress (left) and heat transfer (right).}
    \label{fig:FPM11_mesh}
\end{figure}

We first assess the performance of the baseline $k-\epsilon$ model for each of the three test cases.
As mentioned earlier in $\S$~\ref{sec:Intro}, errors in RANS predictions can occur due to %a wide range of factors. 
%These include (but are not limited to) 
grid resolution, inflow boundary conditions and turbulence model limitations. 
%Before investigating the turbulence model limitations, 
We first examine the effects of mesh resolution and the inflow turbulence on the FPM11 predictions.
Three meshes are employed, details of which are documented in Table~\ref{tab:MeshFPM11}.
\begin{table}
    \centering 
    \caption{FPM11: Grid size and spatial resolution ($
    \delta_o=0.01$ m).}
    \begin{tabular}{l c c c c}
    \hline \hline
      Mesh & $(n_x \times n_y)$ & $\Delta x/\delta_o$ & $\Delta t \delta_o/U_{\infty}$ & $\Delta y^+$ \\\hline
    Nominal & $625 \times 115$ & 0.24& 0.001 & 1.26 \\
    Intermediate & $1250 \times 173$ & 0.12& 0.001 & 0.83 \\
    Fine & $2500 \times 345$ & 0.06& 0.001 & 0.41 \\ \hline \hline
    \end{tabular}
    
    \label{tab:MeshFPM11}
\end{table}
The nominal mesh consists of $625$ and $115$ points in the streamwise $(x)$ and wall normal $(y)$ directions respectively.
The two finer meshes are obtained by subsequently refining the nominal mesh in the $x-$ and $y-$ directions as shown in Table~\ref{tab:MeshFPM11}.
The $\Delta y^+$ of the grid point adjacent to the wall at the plate center ranges from $1.26$ to $0.41$ for the three meshes.
The solutions are marched to steady state using a non-dimensional time-step $\Delta t=0.001 U_{\infty}/\delta_o$ where $\delta_o=0.01$ m.
%All three meshes have a $\Delta y^+ < 1$ at the first mesh point adjacent to the wall.
Figure~\ref{fig:FPM11_mesh} compares the baseline predictions for the wall shear stress ($\tau_w$) and heat transfer ($\dot{Q}_w$) on the three meshes with DNS results of %Huang \emph{et al.}~
\cite{huang2020simulation} and experimental measurements~\citep{gnoffo2013uncertainty}.
In general, the baseline $k-\epsilon$ model provides accurate estimates for both quantities once transition is obtained.
The transitional behavior of the predictions can be linked to %the relatively low Reynolds number of the experiment and 
the 
tendency of $k-\epsilon$ type models to remain laminar at low Reynolds and high Mach numbers~\citep{coakley1992turbulence}.
The values of the $\tau_w$ in turbulent regime is independent of the mesh used, whereas $\dot{Q}_w$ becomes mesh independent once 
%Special care is taken to ensure that all the results reported in this paper are mesh-independent and the boundary values of the turbulence variables so that the focus is on turbulence modeling limitations.
the $\Delta y^+$ value falls below unity. %for surface quantities holds true for this test case.
%; this is true for all the test cases studied in this paper.
Once $\Delta y^+ < 1$ is achieved, further refinement only affects the location of transition and not the surface quantities after transition.
It must be noted that such guidelines, although helpful in equilibrium situations, can be misleading in the presence of streamwise or spanwise gradients as the value of $\Delta y^+$ depends on the friction velocity on the surface.
For flows with separation, this guideline becomes even more inapt as the friction velocity vanishes at the separation and reattachment points.

\begin{figure}
\centering
    \subfloat[]{\includegraphics[width=0.47\textwidth,trim=90 270 90 260,clip]{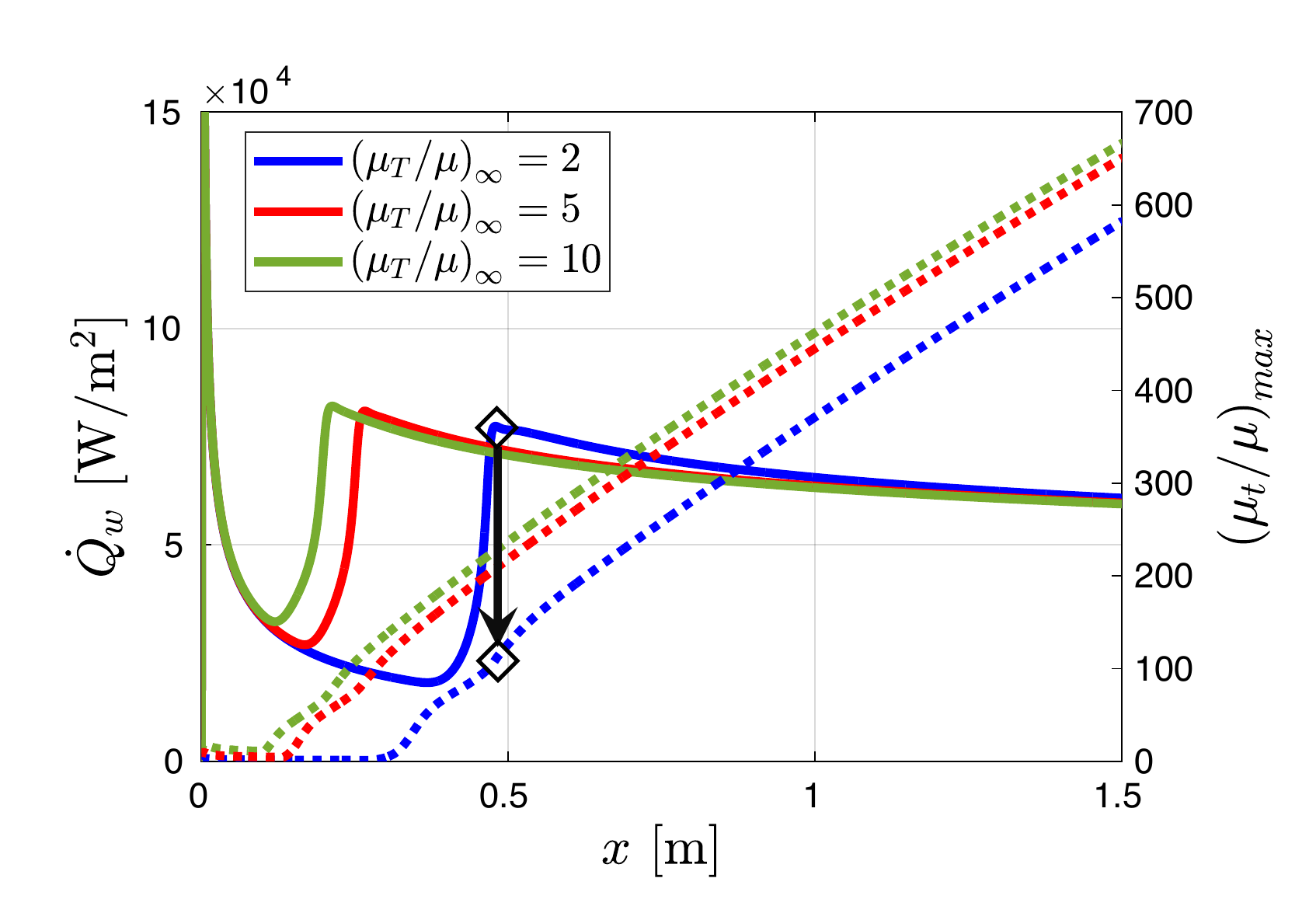}} \hspace{0.1in}
    \subfloat[]{\includegraphics[width=0.47\textwidth,trim=90 270 90 260,clip]{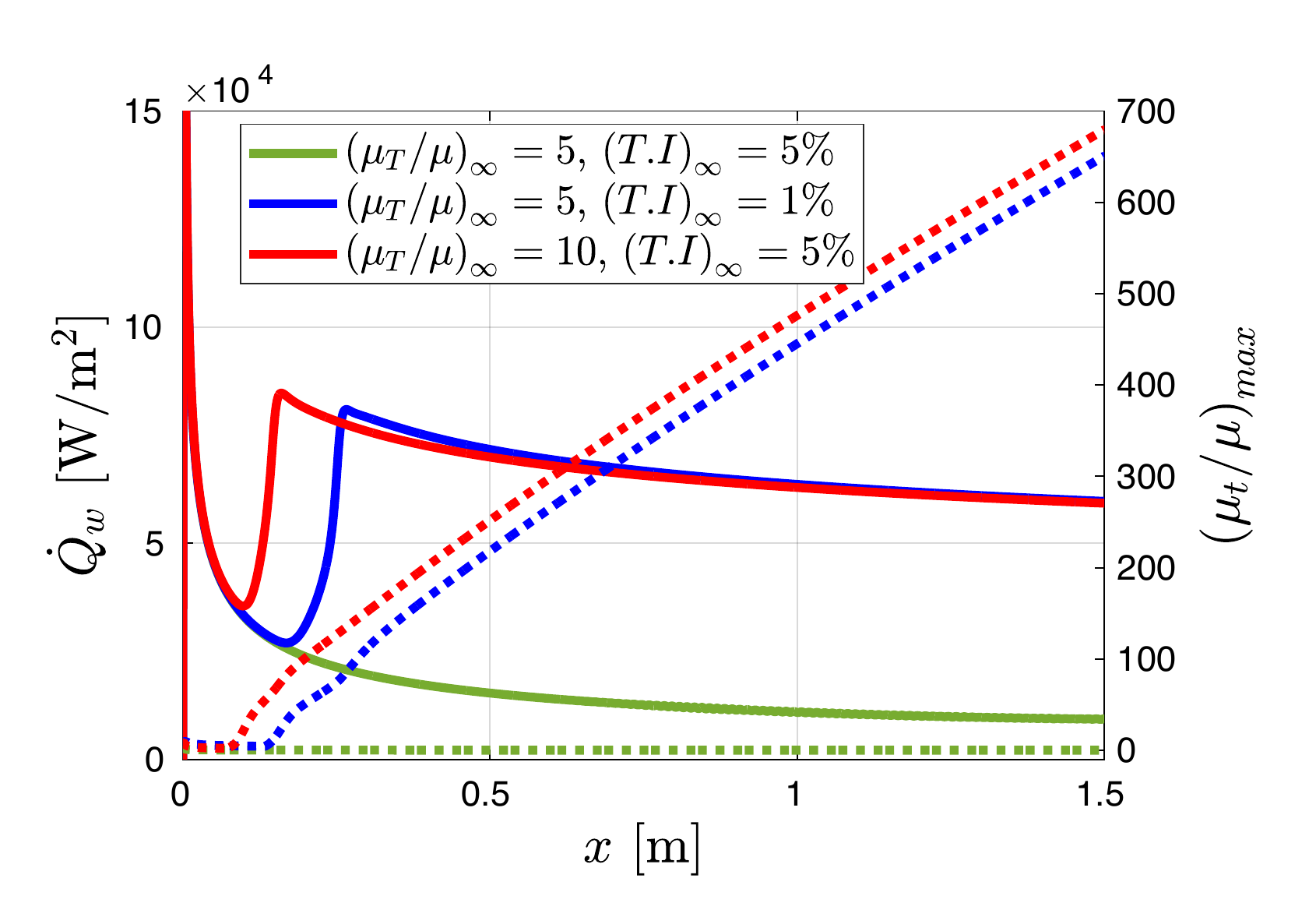}} \\
    \caption{FPM11: Effect of inflow eddy viscosity (left) and turbulence intensity (right) on wall heat flux. The right axis shows the evolution of maximum eddy viscosity along the length of the plate. }
    \label{fig:FPM11_eddyratio}
\end{figure}

We now examine the effect of freestream turbulence on these predictions. 
In most RANS codes, the freestream turbulence is represented by specifying a turbulence intensity and an eddy viscosity ratio $(\left(\mu_t/\mu\right)_\infty)$ at the inlet boundary. 
Figure~\ref{fig:FPM11_eddyratio} shows the effect of varying these inflow turbulence values on FPM11 heat transfer predictions on the intermediate mesh. 
The solid colored lines in Fig.~\ref{fig:FPM11_eddyratio}a correspond to different inlet eddy viscosity ratios at an inflow turbulence intensity of $1\%$, whereas the dotted lines indicate the evolution of $(\mu_t/\mu)_{\text{max}}$ %the max eddy viscosity ratio 
along the plate length using the right $y-$axis. %, whereas the dashed lines represent an inflow turbulence intensity of $5\%$ for the respective inlet viscosity ratios. 
As expected, the transition location moves upstream when an increase in inflow turbulent viscosity.
The maximum eddy viscosity experiences a sudden jump at the transition location as marked by the downward arrow in Fig.~\ref{fig:FPM11_eddyratio}a for $(\mu_t/\mu)_\infty=2$.
As a general guideline, transition is achieved once the maximum eddy viscosity in the boundary layer crosses above 100; this guideline is used throughout this investigation to test for fully turbulent boundary layer in the subsequent test cases.
%A turbulent viscosity ratio of five is found to be appropriate to match the transition location with the experiments.
Figure~\ref{fig:FPM11_eddyratio}b shows the effect of increasing turbulent intensity on the $k-\epsilon$ predictions. 
For a constant $(\mu_t/\mu)_\infty$ value of 5, an increase in freestream turbulence intensity from $1\%$ to $5\%$ inhibits boundary layer transition and the flow remains laminar for the entire length of the plate.
The maximum eddy viscosity in the laminar boundary layer barely rises above the $(\mu_t/\mu)_\infty$ value.
%This is especially true for lower inflow turbulent viscosity ratios. 
%Laminar values of heat transfer rates show good agreement with the experimental measurements in the laminar regime as shown earlier in Fig.~\ref{fig:FPM11_mesh}b. 
Since the present focus is on predicting the turbulent quantities, this behavior is undesirable and
%This undesirable behavior 
can be circumvented by selecting a higher inlet eddy viscosity ratio of 10, which results in an earlier transition of the boundary layer despite increased turbulence intensity. %as shown in Fig.~\ref{fig:FPM11_eddyratio}b.
%$(\mu_t/\mu)_\infty=10$ results in an earlier transition of the boundary layer despite increased turbulence intensity.
Based on these observations, for flow over a cooled flat plate with no pressure gradients, the baseline $k-\epsilon$ model (without any corrections) is sufficient to provide accurate shear stress and heat transfer predictions %based on comparisons with DNS and experimental measurements, 
provided an adequately high inflow turbulent viscosity is specified. 
%Since the flat plate boundary layer simulations are required as an inflow to other test cases, further analysis is based on $k-\epsilon$ model.

\begin{figure}
    \centering
    \includegraphics[width=0.8\textwidth]{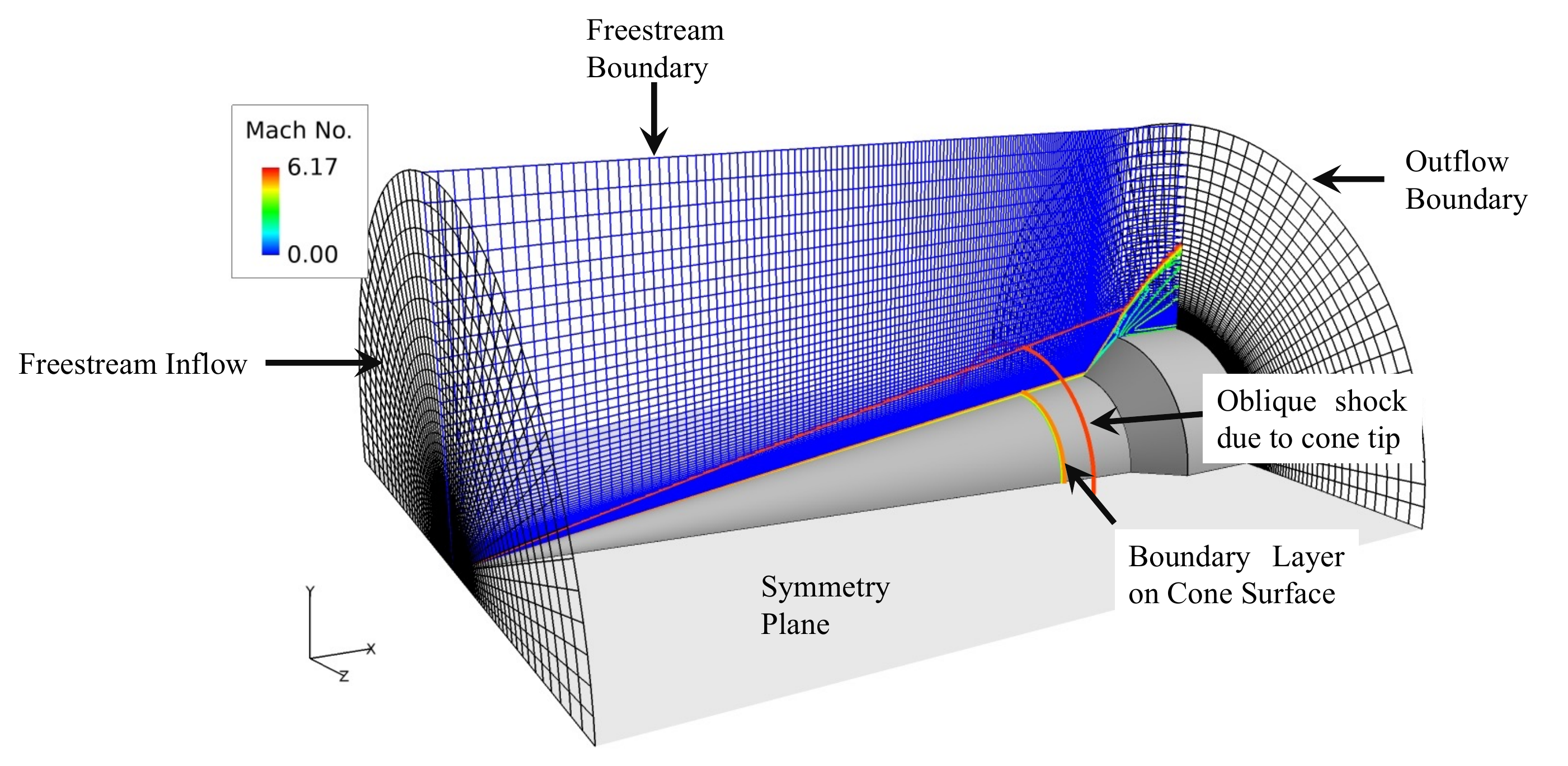}
   \caption{LCFM6: Schematic showing the computational domain, mesh distribution and boundary conditions.}
  \label{fig:LCF}
\end{figure}

\begin{figure}
\centering
 \subfloat[]{\includegraphics[width=0.47\textwidth,trim=90 250 90 250,clip]{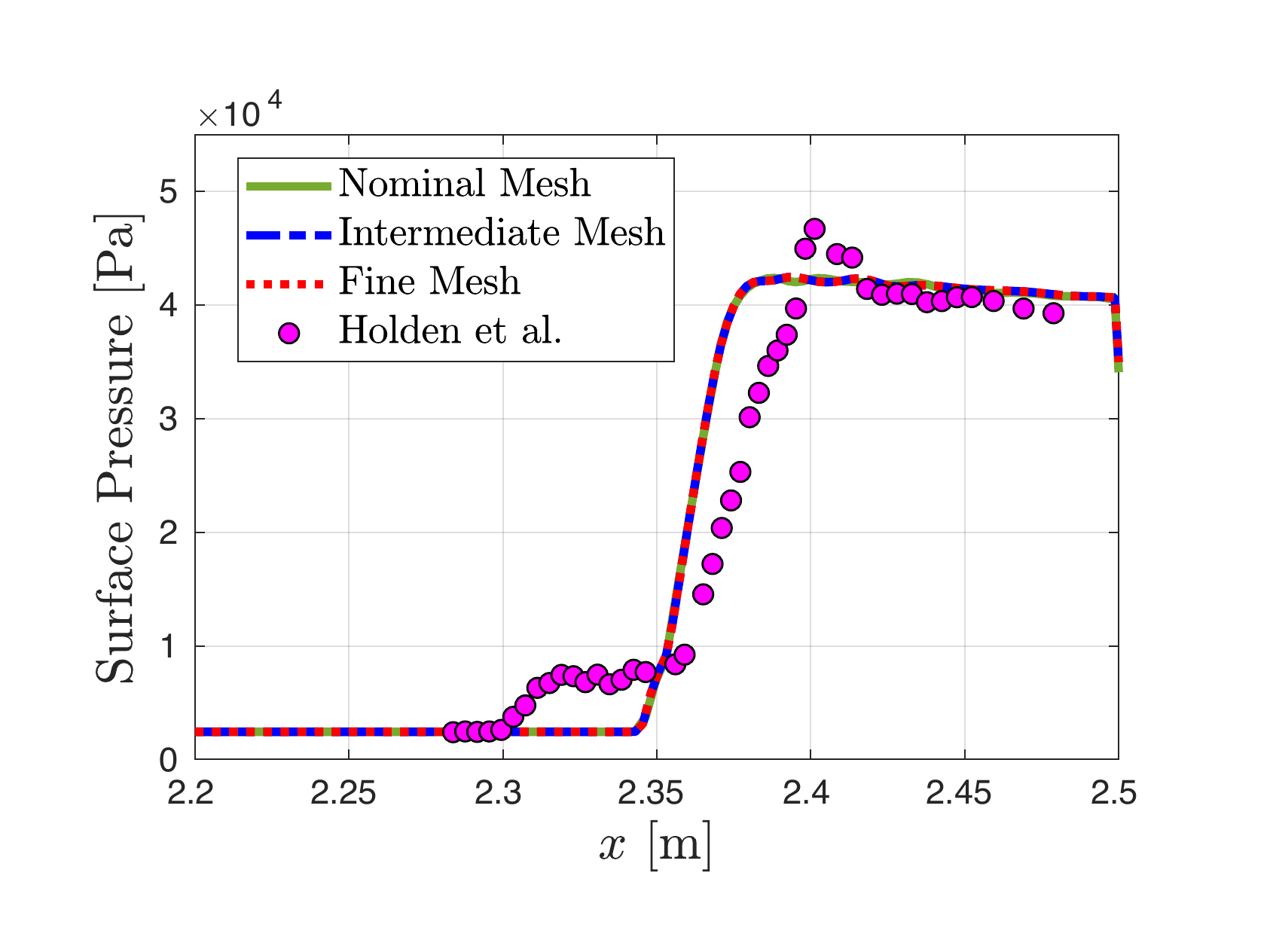}} \hspace{0.1in}
    \subfloat[]{\includegraphics[width=0.47\textwidth,trim=90 250 90 260,clip]{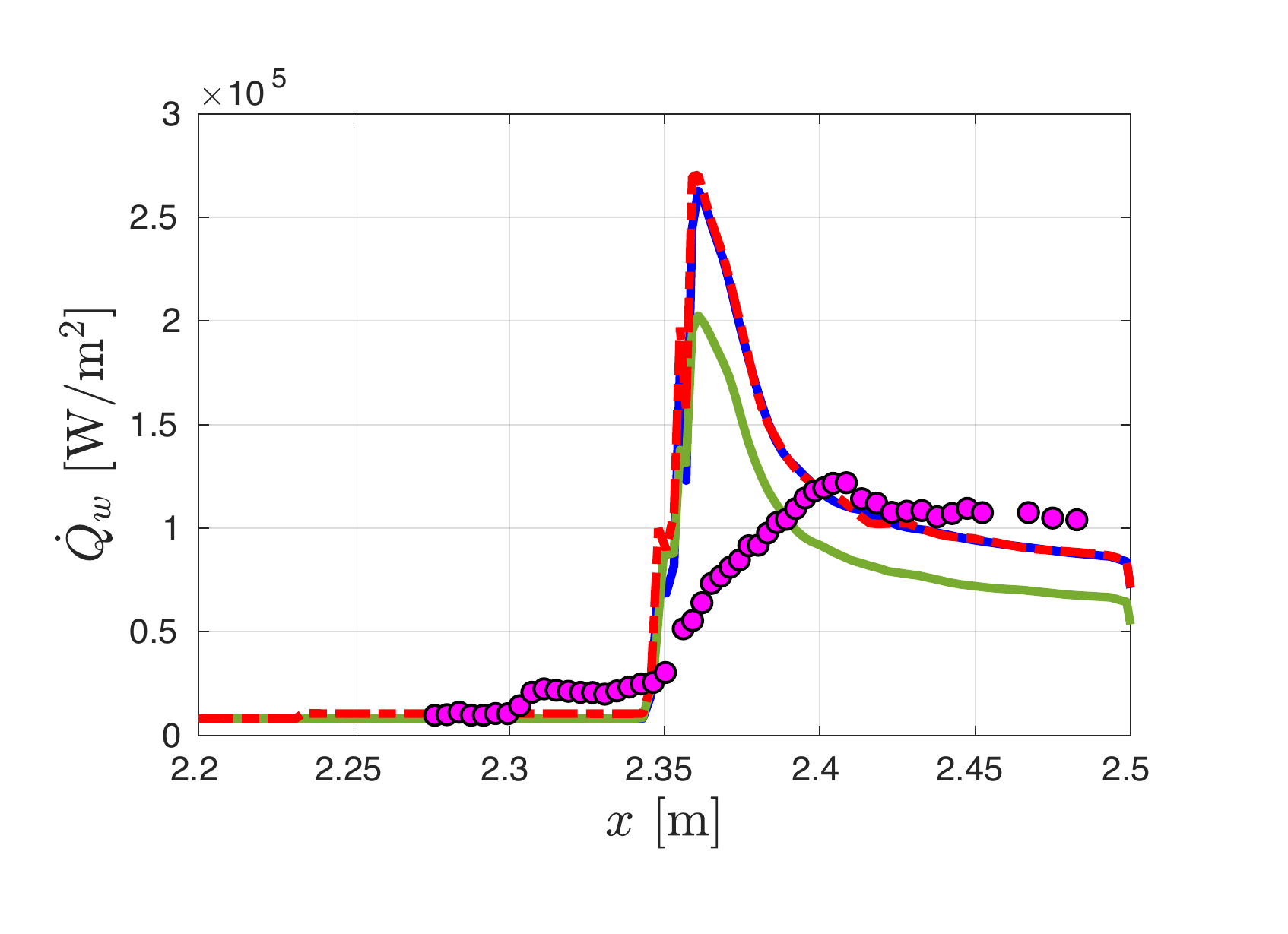}} \\
 
    \caption{LCFM6: Baseline predictions of surface pressure (left) and heat transfer (right). }
    \label{fig:LCFbaseline}
\end{figure}

Figure~\ref{fig:LCF} presents %the results for the LCFM6 test case. 
a schematic of the computational domain LCFM6 test case. %is presented in Fig.~\ref{fig:LCFvarya}. 
Due to the axisymmetric nature of the problem, a cylindrical coordinate system is chosen and only half the cone geometry is simulated using symmetry boundary conditions as shown.
%As stated earlier
%The flow-field is dominated by two oblique shocks due to the cone tip and the flare.
%Based on the previous FPM11 results, 
In order to ensure a turbulent boundary layer on the cone surface prior to the SBLI at the flare, a $(\mu_t/\mu)_\infty=10$ at $1\%$ turbulence intensity is specified at the inflow based on previous FPM11 results.
%As stated previously, the long cone geometry allows for a turbulent boundary layer to develop before the 
%The flow-field consists of an oblique shock at the cone tip followed by a boundary layer growth. The turbulent boundary layer develo
The effect of mesh resolution is tested using three meshes, details of which are documented in Table~\ref{tab:MeshLCF}. 
\begin{table}
    \centering 
    \caption{LCFM6: Grid size and spatial resolution ($
    \delta_o=0.0254$ m).}
    \begin{tabular}{l c c c c c c}
    \hline \hline
      Mesh & $(n_x \times n_r \times n_{\theta})$ & $\left(\Delta x/\delta_o\right)_{min}$ & $\left(\Delta x/\delta_o\right)_{\text{max}}$ & $\Delta \theta$ & $\Delta d^+$ & $\Delta t \delta_o/U_{\infty}$ \\\hline
    Nominal & $331 \times 121 \times 41$ & 0.12& 0.912 & $4.5^\circ$ & 0.8 & 0.001\\
    Intermediate & $481 \times 151 \times 51$ & 0.08& 0.631& $3.6^\circ$ & 0.4 & 0.001\\
    Fine & $531 \times 181 \times 61$ & 0.07& 0.508 & $3^\circ$ & 0.3 & 0.0005\\ \hline \hline
    \end{tabular}
    
    \label{tab:MeshLCF}
\end{table}
The nominal mesh consists of $331$, $121$ and $41$ points in the streamwise, radial and circumferential directions respectively.
The intermediate and fine grids are obtained by subsequently refining the nominal mesh along the three directions as shown.
Based on prior results from the FPM11 test case, the mesh point closest to the cone surface in the turbulent boundary layer upstream of the flare is located at $\Delta d^+ < 1$ (calculated at $x = 2.2$ m) for each of the three meshes. 
Here $d$ represents the normal distance from the wall.
%The streamwise spacing is clustered along the flare for all the three meshes.
Figure~\ref{fig:LCFbaseline} compares the resulting surface pressure and heat flux in the vicinity of the flare junction with experimental measurements for each of the three meshes.
%The background velocity contours are normalized by the freestream velocity.
The surface pressure predictions are independent of the mesh used, whereas the wall heat flux %on the other hand, due to the shock interaction 
requires a significantly finer mesh to converge than that required for the FPM11 test case ($\Delta d^+ < 0.4$ in the upstream attached boundary layer).
%heat transfer requires a finer mesh to be converged than the surface pressure.
Moreover, the $k-\epsilon$ model that provided very accurate results for the FPM11 test case, shows very little separation at the flare junction, resulting in an under-prediction of the peak surface pressure and a severe over-prediction $(\sim120\%)$ in the peak surface heat flux.
This is not unexpected, as it is well-known that the behavior of the $k-\epsilon$ model deteriorates even for incompressible adverse pressure gradient flows~\citep{menter1994two}. 
These deficiencies are addressed later in this section.
%Similar to the FPM11 test case, the surface heat flux requires a more finer mesh to become grid independent.

\begin{figure}
    \centering
    \includegraphics[width=0.75\textwidth,trim=0 35 0 0,clip]{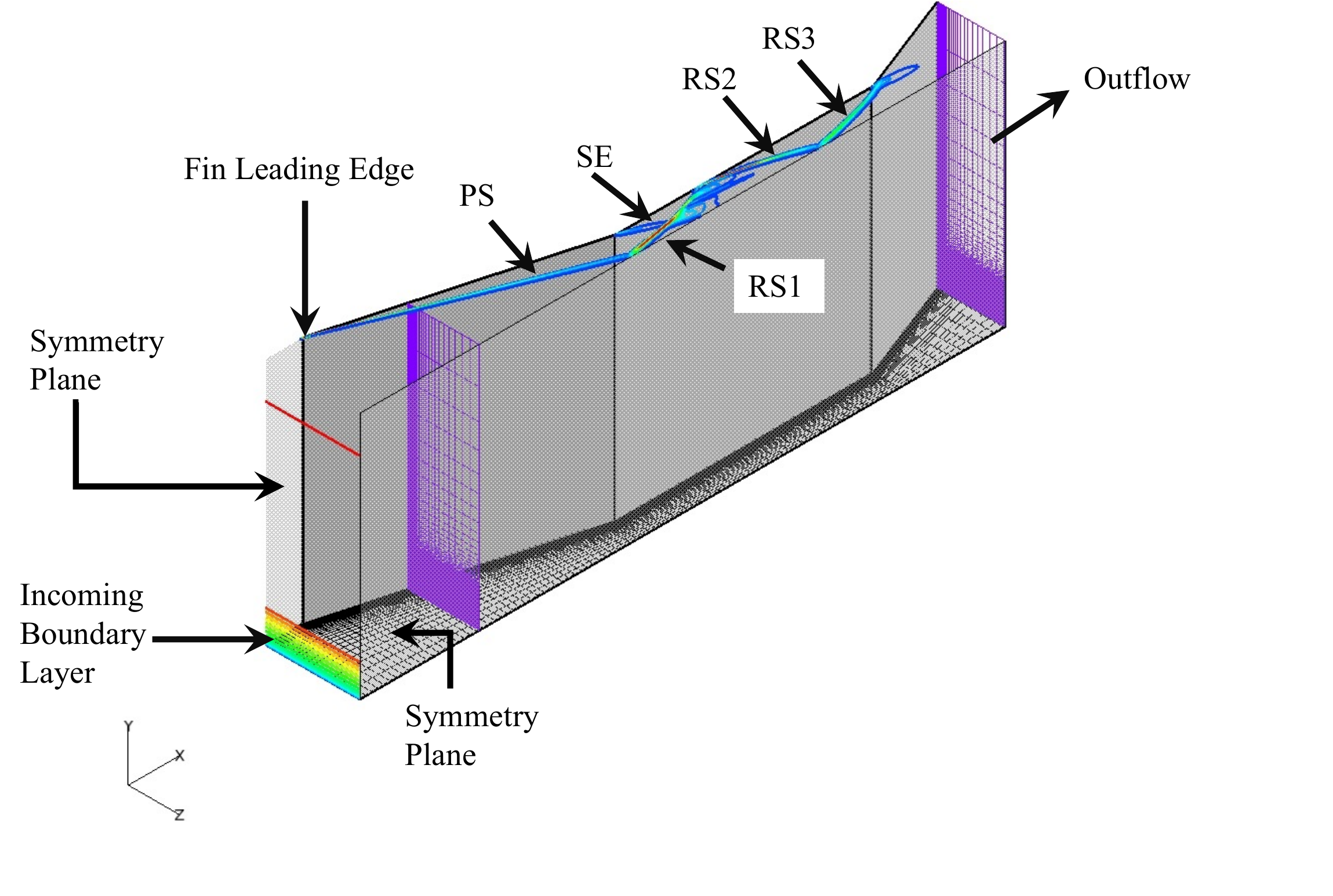} \\
    \caption{DFM8: Schematic showing the computational domain, boundary conditions and freestream shock-structure.}
    \label{fig:DFschematic}
\end{figure}

Figure~\ref{fig:DFschematic} shows a schematic of the computational domain for the DFM8 test case. 
The incoming flow directed along the $x$-axis with the $y$-direction being the normal to the plate and the $z-$direction representing the spanwise direction. 
The fin leading edge is placed at $x=0$.
For computational efficiency only half the experimental domain is simulated.
%The mesh consists of $273$, $173$ and $101$ points in the streamwise, wall-normal and spanwise directions respectively.
Symmetry boundary conditions are applied at two planes as shown: the plane bisecting the two fins and the plane upstream of the leading edge.
The freestream shock-structure comprising the primary shock (PS), the shoulder expansion (SE) and several reflected shocks (RS1 to RS3) are also shown for completeness. 
Three meshes are used to test for grid convergence; the details of each mesh are provided in Table~\ref{tab:MeshDFM8}.
\begin{table}
    \centering
    \caption{DFM8: Grid size and spatial resolution ($
    \delta_o=3.25$ cm). The $\Delta y^+$ value is calculated at the inflow plane.}
    \begin{tabular}{l c c c c c}
    \hline \hline
      Mesh & $(n_x \times n_y \times n_z)$ & $\left(\Delta x/\delta_o\right)_{min}$ & $\left(\Delta x/\delta_o\right)_{\text{max}}$ & $\Delta t \delta_o/U_{\infty}$ & $\Delta y^+$ \\\hline
    Nominal & $181 \times 115 \times 101$ & 0.081& 0.125 & 0.001 & 0.8\\
    Intermediate & $289 \times 173 \times 101$ & 0.051& 0.085& 0.001 & 0.4 \\
    Fine & $289 \times 173 \times 151$ & 0.051& 0.085& 0.0005 & 0.3 \\ \hline \hline
    \end{tabular}

    \label{tab:MeshDFM8}
\end{table}
These meshes collectively test the effect of streamwise, wall normal and spanwise spacing on $k-\epsilon$ predictions. 
Since it is already established that the $k-\epsilon$ model gives accurate predictions for flat plate boundary layer flows, the inflow boundary conditions are specified by matching the boundary layer momentum thickness of a precursor flat plate boundary layer simulation with the measurements of %Kussoy \emph{et al.}~
\cite{kussoy1993hypersonic} $(\theta=0.083 \text{cm})$;
a comparison of the inflow quantities with that of %Kussoy \emph{et al.} 
\cite{kussoy1993hypersonic} is provided in Table~\ref{tab:DFinflow} for completeness.
\begin{table}
    \centering
     \caption{Comparison of inflow quantities with measured values. The subscript `ref' denotes measured values at the inflow plane by %Kussoy \emph{et al.}~
     \cite{kussoy1993hypersonic}.}
    \begin{tabular}{c c c c c} \hline \hline
         $\delta$/$\delta_{\text{ref}}$ & $\theta$/$\theta_{\text{ref}}$ & $\tau_w$/$(\tau_w)_{\text{ref}}$ & $\dot{Q}_w/(\dot{Q}_w)_{\text{ref}}$ & $p/p_{\text{ref}}$ \\ \hline
        $1.003$ & $1$ & $0.942$ & $1.043$ & $1.078$\\ \hline \hline
    \end{tabular}
   
    \label{tab:DFinflow}
\end{table}
%marked in the figure.

\begin{figure}
\centering
%\subfloat[]{
 %   \includegraphics[width=0.75\textwidth,trim=0 35 0 0,clip]{Figures/DF_MeshBC.pdf}} \\
    \includegraphics[width=\textwidth,trim=0 0 0 0,clip]{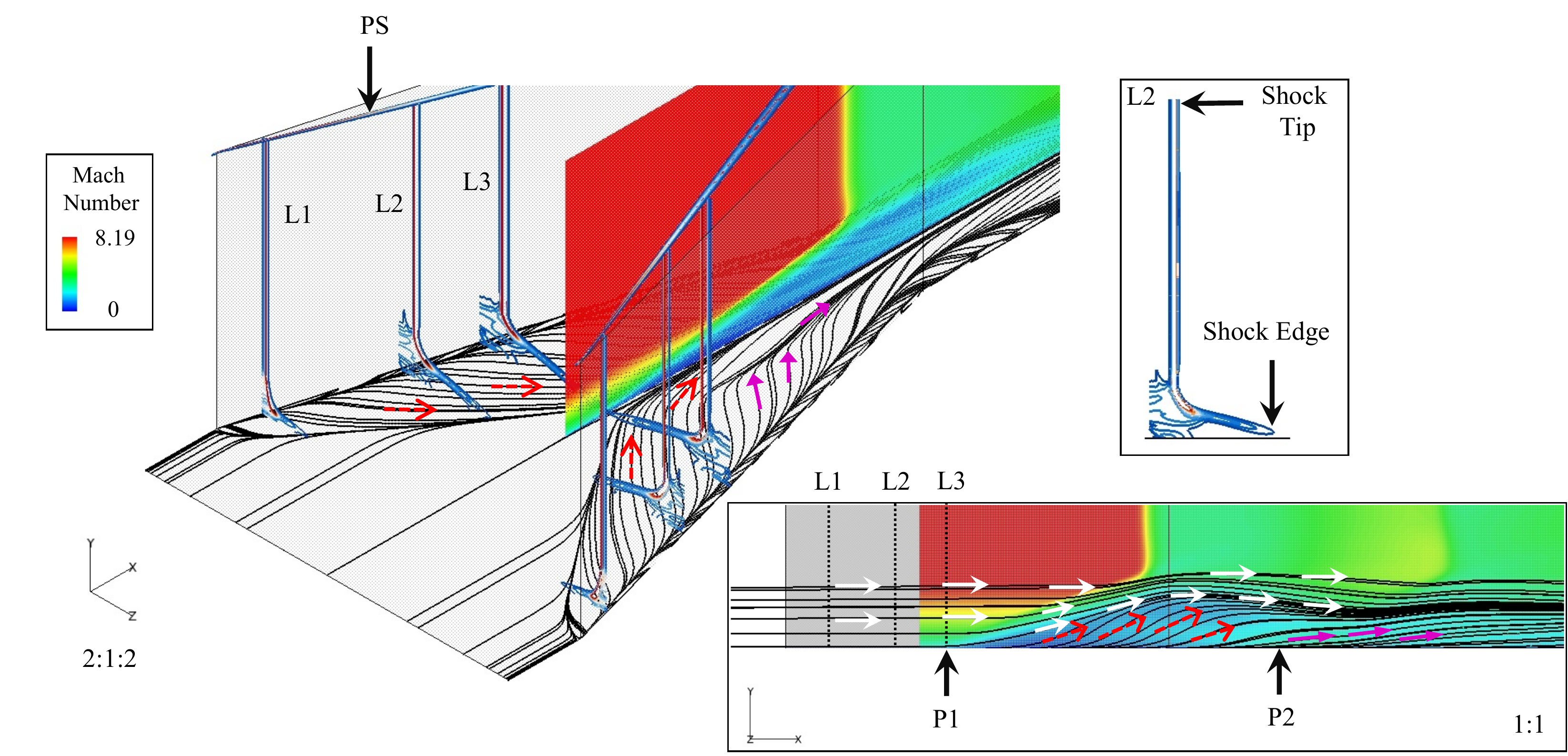}
    \caption{DFM8: Footprint of the flow using streamlines in the proximity of the bottom plate. 
    L1 to L3 denote three lamda shocks highlighted using pressure gradient contours. 
    Mach number contours are superimposed over the symmetry plane to highlight the flow separation. 
    The top inset highlights one of the lambda shocks whereas the bottom inset shows the streamlines in the symmetry plane.}
    \label{fig:DFFlow}
\end{figure}

Figure~\ref{fig:DFFlow} describes the 3D flow-field using streamlines, pressure gradient and Mach number contours %describes the footprint of flow-field of this test case using streamlines 
at an artificial aspect ratio of $(x : y : z) = (2 : 1: 2)$ to highlight different features under discussion that must be modeled.
%The streamlines highlight the footprint of the incoming boundary layer.
A more detailed description of the flow-field can be found in %Gaitonde \emph{et al.}~
\cite{gaitonde1995structure}.
The freestream primary shock extends all the way down to the bottom plate,
%The shock structure near the bottom plate is characterized by 
forming a succession of growing lambda shocks; three such shocks marked L1 to L3 are shown using pressure gradient contour lines. 
Each lambda shock is defined by its inviscid swept shock asymptote (shock tip) and the separation shock (shock edge) as marked in the top-right inset for shock L2.
%The freestream primary shock (marked as PS) is comprised of the tips of this lambda shock system as shown.
%The lamda shocks from each fin intersect
%The footprint of the flow near the bottom wall is highlighted using streamlines.
The incoming boundary layer separates along a line of coalescence formed by the lambda shock edges as shown. 
Separation is accompanied by a rapid increase in pressure which is maximum at the center symmetry plane. %flow along the centerline alk about how centerline is the last to separate.
%resulting in a separation of the incoming flow.
%The separation location is a function of the spanwise distance, 
%The incoming flow along the centerline is the last to separate at the intersection of the two lambda shocks from each fin. 
The bottom inset shows the flow structure in this plane %is examined in the bottom inset 
using streamlines superimposed over Mach number contours. %along the center symmetry plane.
The primary incoming flow (highlighted using white arrows) separates at the point P1 (the point of intersection of the two lambda shock edges from each fin) and does not reattach to the plate throughout the computational domain.
The region beneath the separated flow is occupied by the spanwise movement of the fluid from the fins towards the symmetry plane.
This %results in an 
open mean flow structure is absent in 2D/axisymmetric cases like the LCFM6, which exhibits a closed separation bubble at the flare junction.

\begin{figure}
\centering
%\subfloat[]{
 %   \includegraphics[width=0.75\textwidth,trim=0 35 0 0,clip]{Figures/DF_MeshBC.pdf}} \\
    %\subfloat[]{
    %\includegraphics[width=\textwidth,trim=0 0 0 0,clip]{Figures/DF_baselineflowv2.pdf}} \\
    \subfloat[ \label{fig:FPM11qw}]{\includegraphics[width=0.47\textwidth,trim=100 230 120 220,clip]{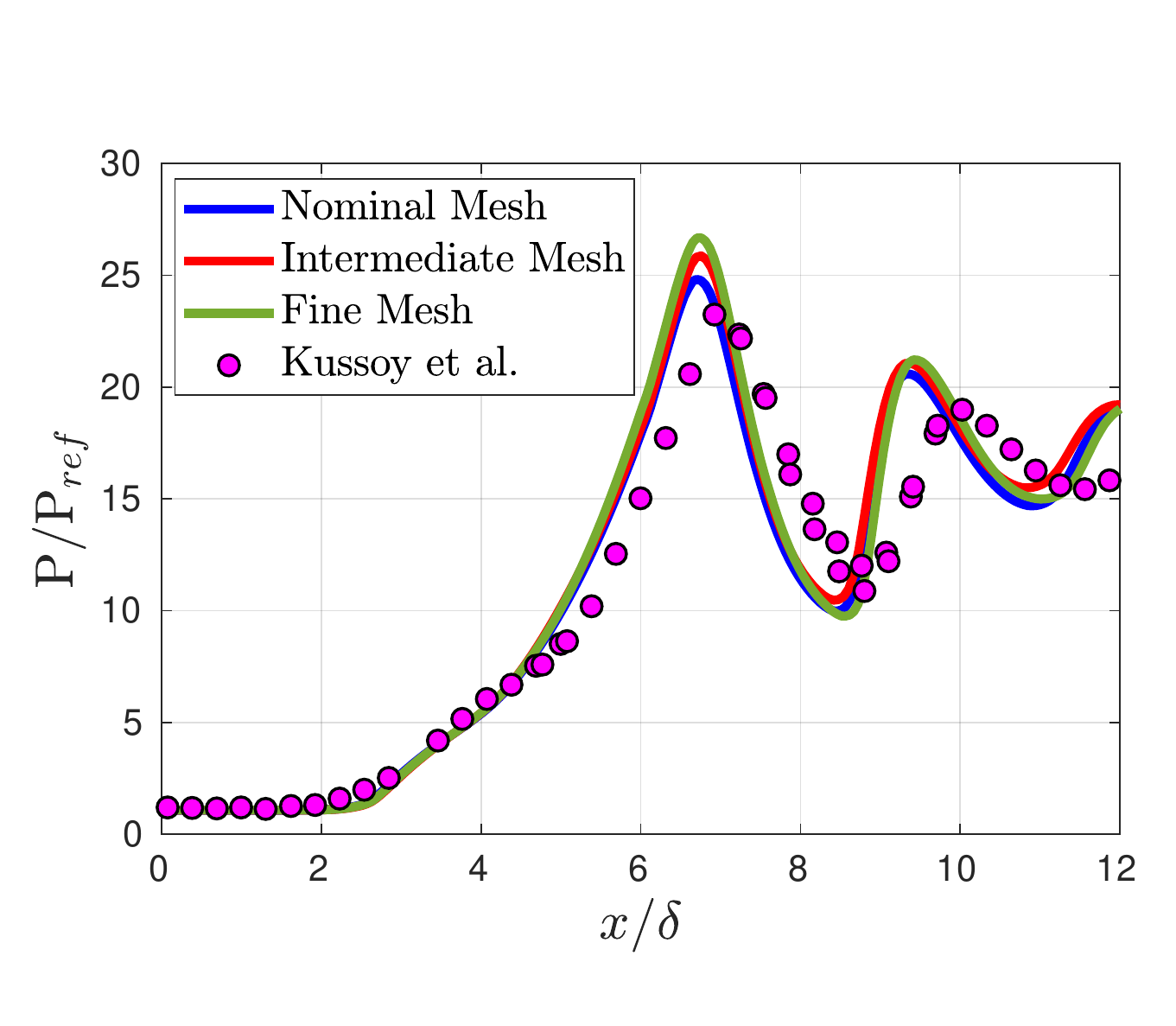}} \hspace{0.1in}
    \subfloat[ \label{fig:FPM11tauw}]{\includegraphics[width=0.47\textwidth,trim=120 250 120 260,clip]{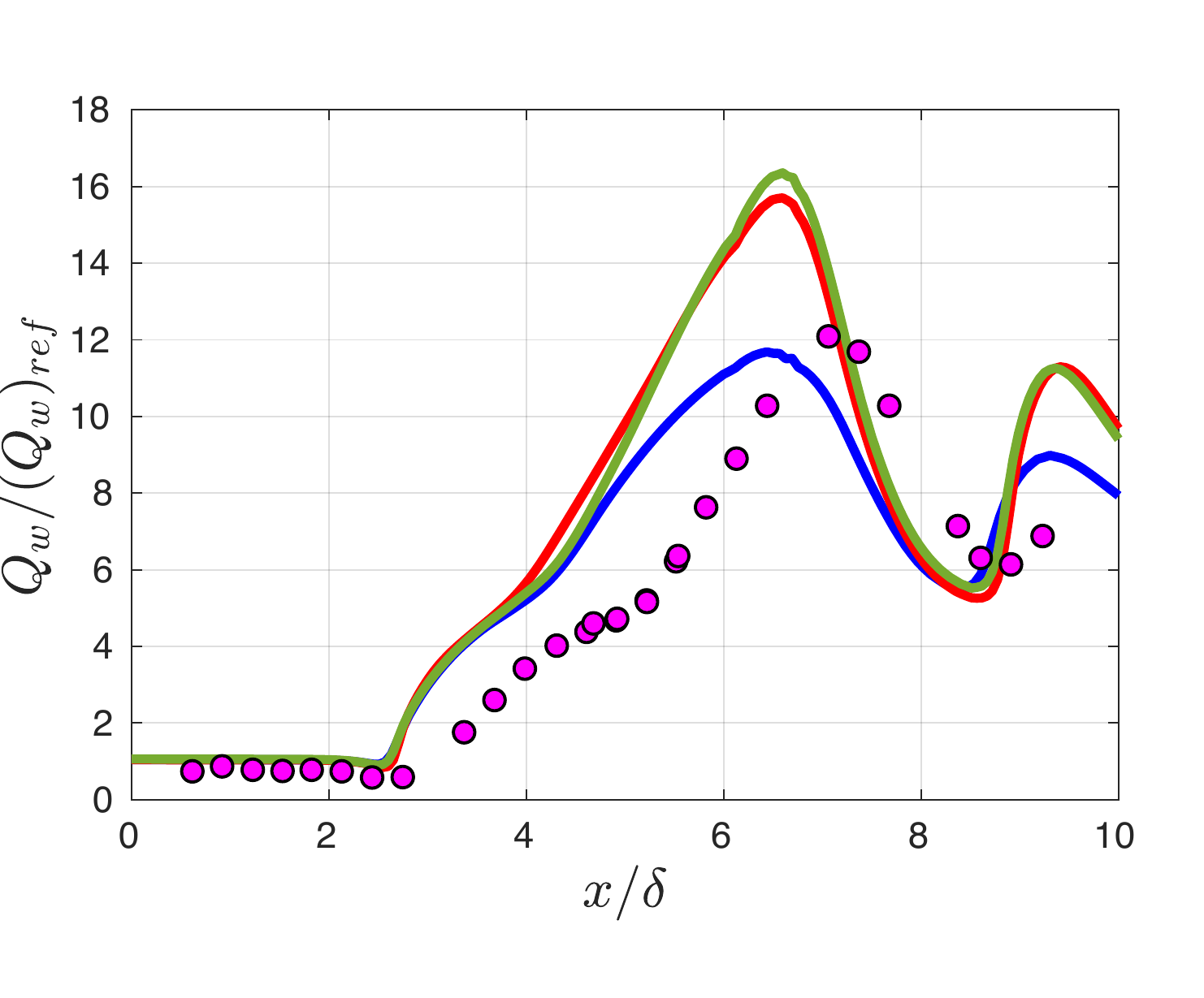}} \\
    %\subfloat[Van-Driest scaled velocity profile \label{fig:FPM11uvd}]{\includegraphics[width=0.5\textwidth]{Aviation_2021/Figures/uvd_inlet_FP_M11_v2.jpg}}
    %\subfloat[Normalized velocity profile \label{fig:FPM11u}]{\includegraphics[width=0.5\textwidth]{Aviation_2021/Figures/u_vs_ypls_inlet_FPM11_v2.jpg}} \\
    \caption{DFM8: Baseline predictions for wall pressure (left) and wall heat flux (right) along the center symmetry plane. }
    \label{fig:DFM8Mesh}
\end{figure}

Figure~\ref{fig:DFM8Mesh} shows the surface pressure and the heat flux predictions along $x$ due to this interaction at the centerline on each of the three meshes.
The pressure and heat transfer values are normalized by $430$ N/m$^2$ and $10{,}400$~W/m$^2$ which are the reference quantities used in experiments.
Similar to the LCFM6 test case, the surface heat flux requires a finer mesh for convergence in the incoming boundary layer than the surface pressure. 
Both quantities follow the same qualitative variation as the measured values.
%The heat transfer at the centerline also follows the same qualitative variation as pressure.
%The initial rise in heat transfer along with the peak value due to the upstream and downstream influence of the PS interactions are also significantly overpredicted by the $k-\epsilon$ model.
A key observation is the overprediction of
%As expected, 
%the baseline $k-\epsilon$ model overpredicts 
the rise in surface heat flux due to the flow separation along the centerline. 
However, the over-prediction is less severe ($\sim34 \%$) than the LCFM6 test case due to the absence of a closed recirculation region as discussed previously. %in the DFM8 flow. 
%In the case of surface pressure, 
The initial pressure rise at $x/\delta=2$ (point P1) due to the intersecting lambda shocks %upstream influence of the PS interaction 
is reasonably well predicted up to $x/\delta=5$, beyond which the baseline model predicts much sharper rates of increase and decay cycles than the experiment. 
%The initial pressure rise at $x/\delta=2$ due to the upstream influence of the PS interaction is reasonably well predicted up to $x/\delta=5$, beyond which the baseline model predicts much sharper rates of increase and decay cycles than the experiment.
%The peak surface heat flux value is over predicted by .

\subsection{Compressibility Correction}
\begin{figure}
\centering
    \subfloat[]{\includegraphics[width=0.33\textwidth,trim=120 270 100 270,clip]{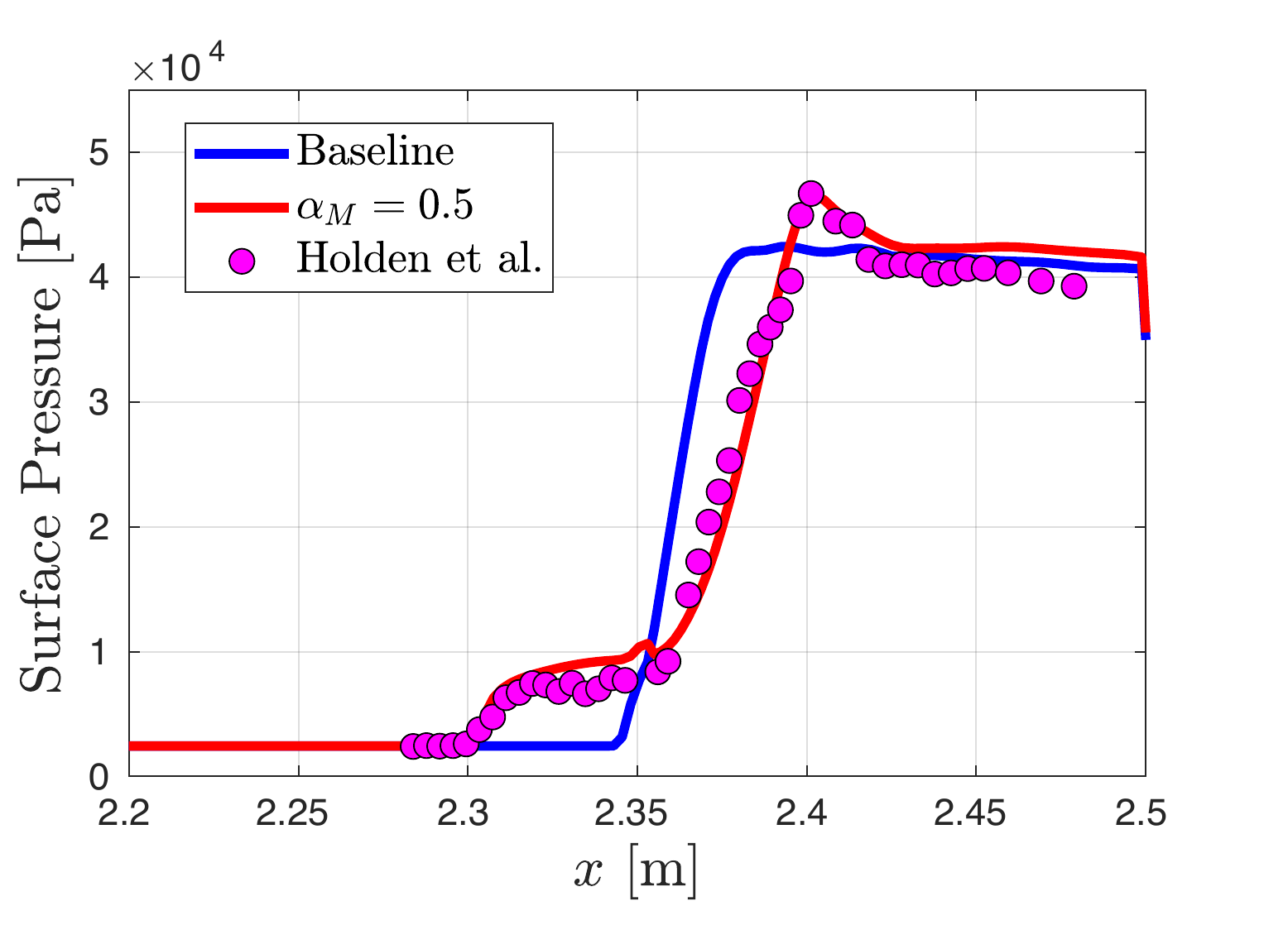}}
    \subfloat[]{\includegraphics[width=0.33\textwidth,trim=110 265 100 270,clip]{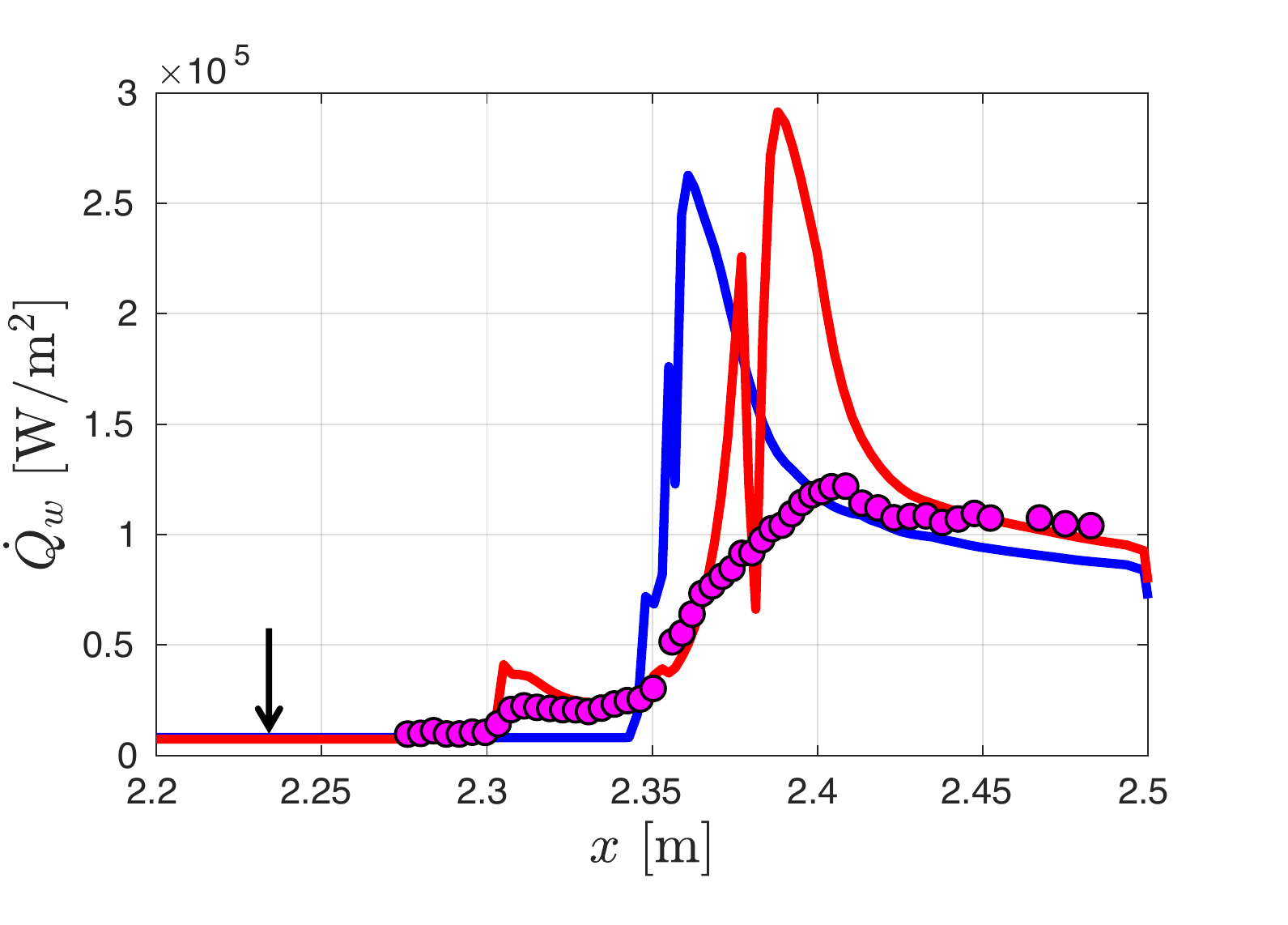}} 
     \subfloat[]{\includegraphics[width=0.33\textwidth,trim=110 270 100 260,clip]{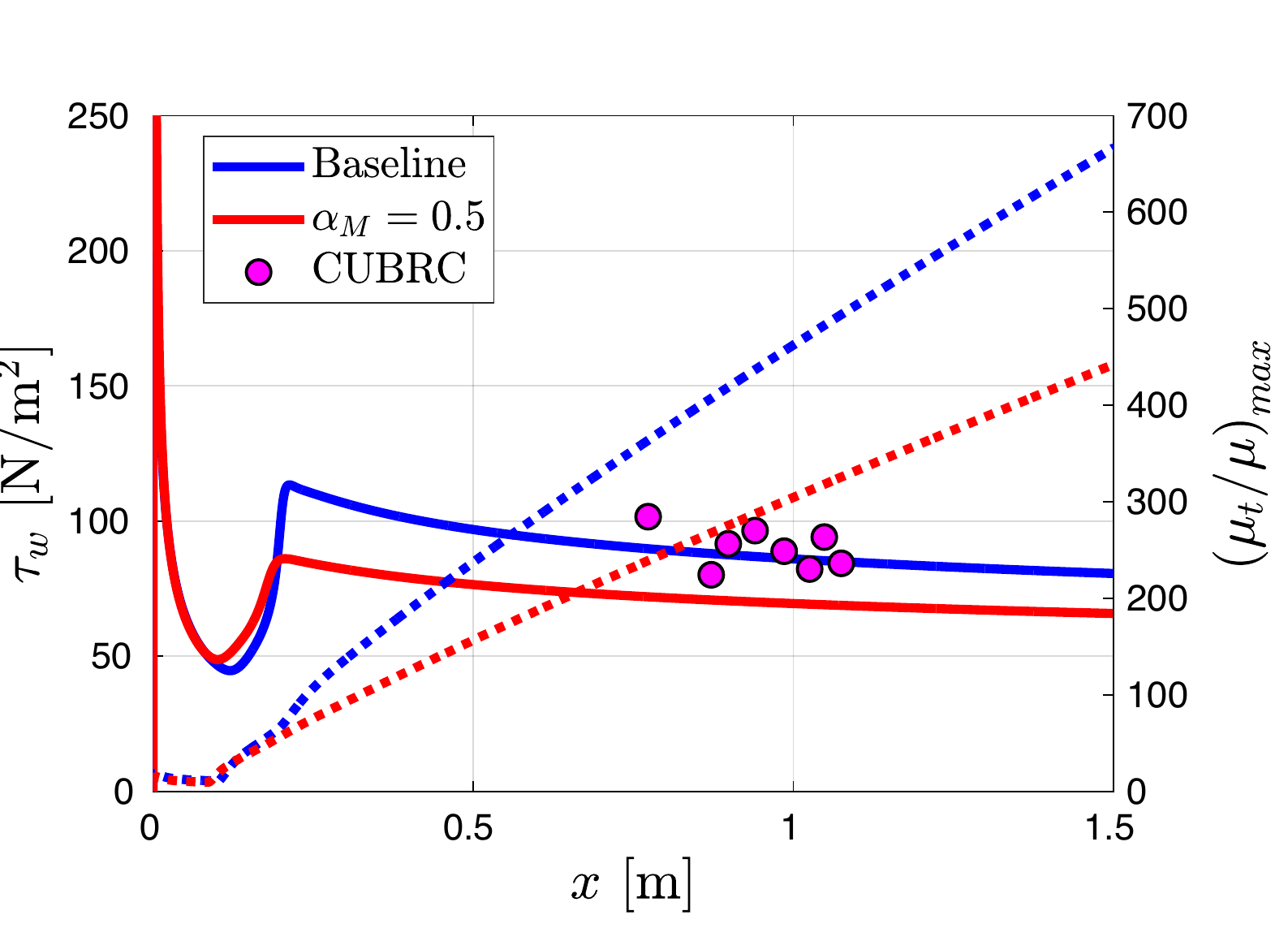}}
     \caption{Use of compressibility correction: (a) LCFM6 Surface Pressure, (b) LCFM6 Heat transfer, and (c) FPM11 wall shear stress.}
    \label{fig:CC}
\end{figure}

%Lack of separation at flare junction LCFM6. 
One of the key reasons for this degradation in accuracy for practical SBLI flows can be attributed to an over-amplification of turbulent kinetic energy immediately downstream of the shock~\citep{sinha2003modeling, zhang2017rans}. 
This results in a more energized boundary layer downstream of the shock than is physical; this delays, or in the case of LCFM6 completely inhibits separation.
One method %In order 
to improve the size of the separation bubble is to modify the coefficient $C_{12}$ in eqn.~\ref{eqn: source} to include the term $\alpha_M M_t^2$ as follows: 
\begin{equation}
    C_{12}=1+\alpha_M M_t^2+\frac{2 \nu}{\epsilon} \left[\left(\frac{\partial \hat{q}}{\partial x}\right)^2 + \left(\frac{\partial \hat{q}}{\partial y}\right)^2 + \left(\frac{\partial \hat{q}}{\partial z}\right)^2\right],
\end{equation}
where $M_t = \sqrt{2 k /(\gamma R T)}$ is the turbulent Mach number.
A value of $\alpha_M=0$ reverts to the baseline $k-\epsilon$ model whereas
an $\alpha_M=0.5$ is equivalent to Sarkar's compressibility correction~\citep{sarkar1991analysis}, which, although developed primarily for free shear layers to address the dilatation dissipation term, has shown success in similar separated flows~\citep{gaitonde2006cfd}. 
Figures~\ref{fig:CC}a and~\ref{fig:CC}b show the effect of employing this correction on the surface pressure and heat transfer %in the vicinity of the flare junction 
for the LCFM6 test case. %are shown with red lines in Figs.~\ref{fig:CC}a and~\ref{fig:CC}b respectively.
With the use of this compressibility correction, the surface pressure predictions show a very good match in both the separation bubble size and the peak value with the measurements.
This correction however, results in even higher values of the peak heat transfer ($\sim140 \%$ over-prediction) than the baseline, though the location of the peak heat transfer matches that of the experiment.

Nonetheless, despite the improvement in surface pressure, the compressibility correction can alter the upstream equilibrium boundary layer, for which model coefficients were obtained without the compressibility correction. 
To show the effect of including this correction for attached flows, Fig.~\ref{fig:CC}c displays results for the FPM11 test case with $\alpha_M=0.5$. 
The solid lines represent the wall shear stress along the plate whereas the dotted lines indicate the evolution of the maximum eddy viscosity using the right $y-$ axis. %compares the eddy viscosity profiles plotted at 100 mesh points upstream of the plate trailing edge, where the boundary layer is free of any potential influence of the outflow boundary condition. 
The use of the compressibility correction %for this case 
results in 
%an excess dissipation of 
a reduction of turbulent viscosity in the boundary layer;
this explains the reason for the increase in separation size with $\alpha_M=0.5$ for the LCFM6 test case, as shown previously.
However, as shown in Fig.~\ref{fig:CC}c, the reduction in eddy viscosity 
%due to enhanced dissipation however, 
degrades the performance of the $k-\epsilon$ model in predicting the wall shear stress for FPM11. %) as compared to the more accurate uncorrected model for the flat plate.
%The same trend is observed for the long cone region in which the boundary layer is expected to become turbulent before interacting with the flare. 
These observations highlight the underlying problem in turbulence modeling corrections where an improvement in one aspect is often accompanied by loss of accuracy in others. 
For the present investigation, it is therefore vital that modifications for one phenomenon not interfere with the accuracy of the baseline model for zero pressure gradient boundary layers.

\subsection{Correction 1: Production Limiting}
\begin{figure}
 \centering
    \subfloat[]{\includegraphics[width=0.7\textwidth]{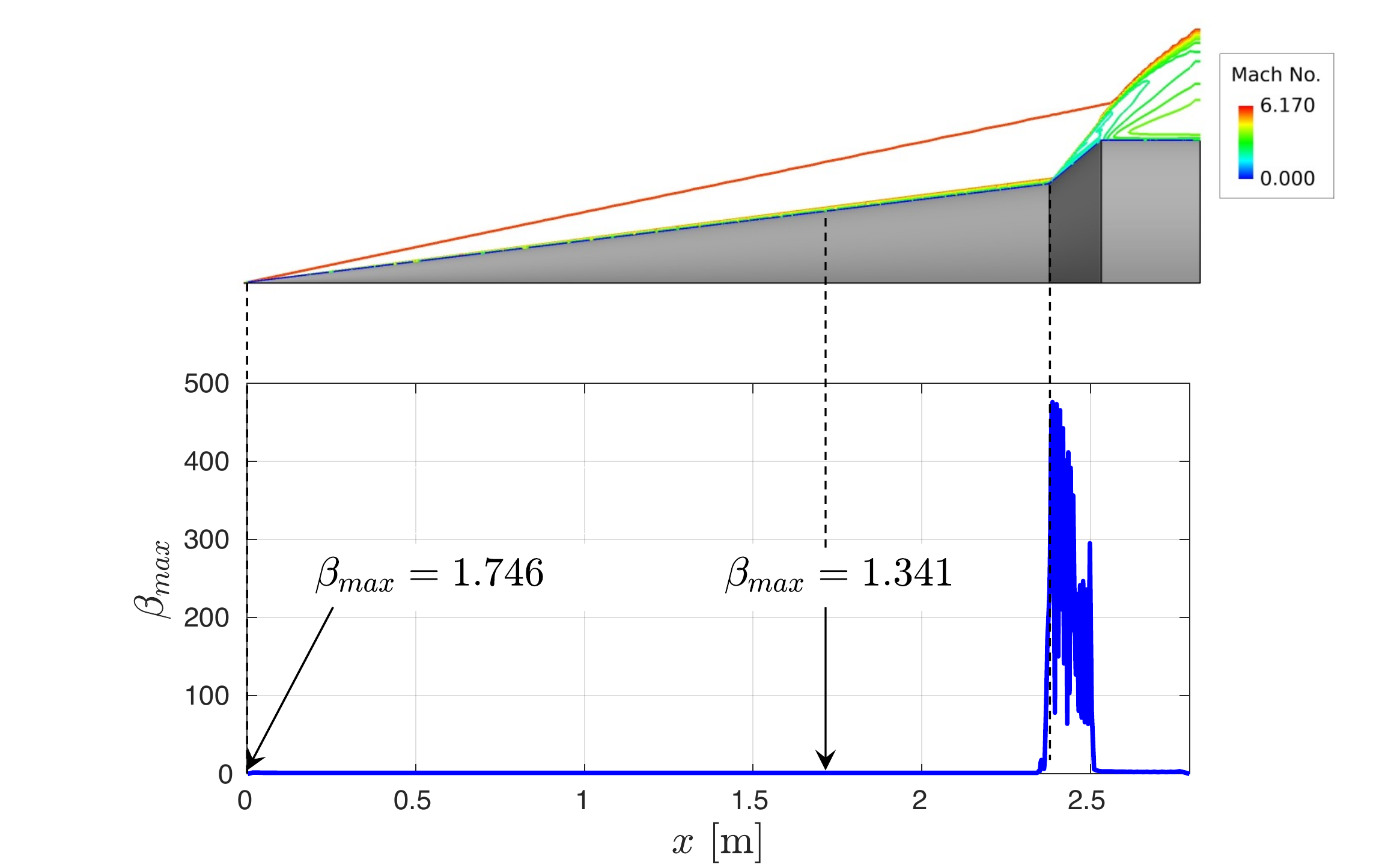}} \\
    \subfloat[]{\includegraphics[width=0.33\textwidth,trim=110 280 100 280,clip]{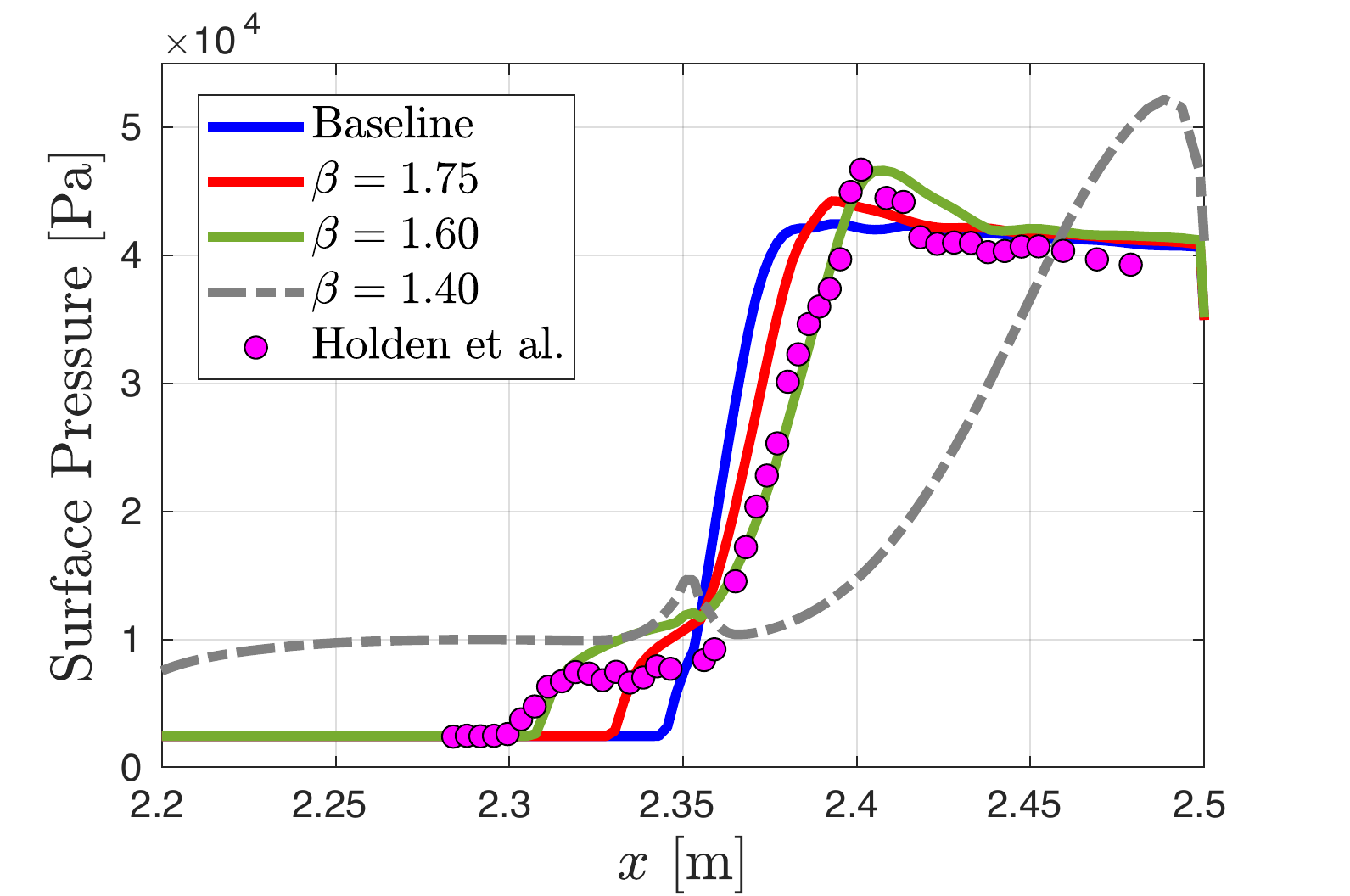}}
    \subfloat[]{\includegraphics[width=0.33\textwidth,trim=110 280 100 280,clip]{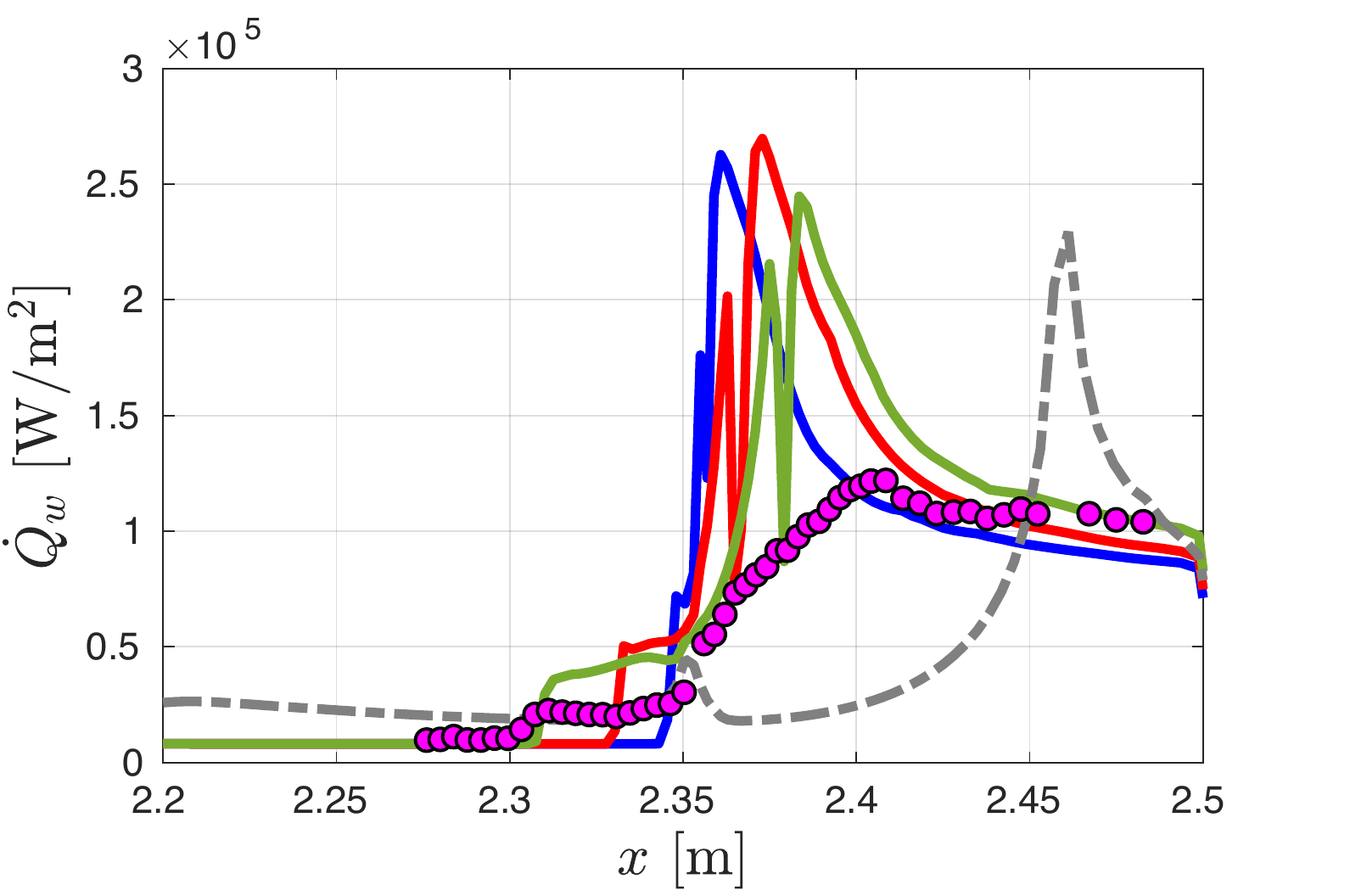}} 
    \subfloat[]{\includegraphics[width=0.33\textwidth,trim=110 280 100 280,clip]{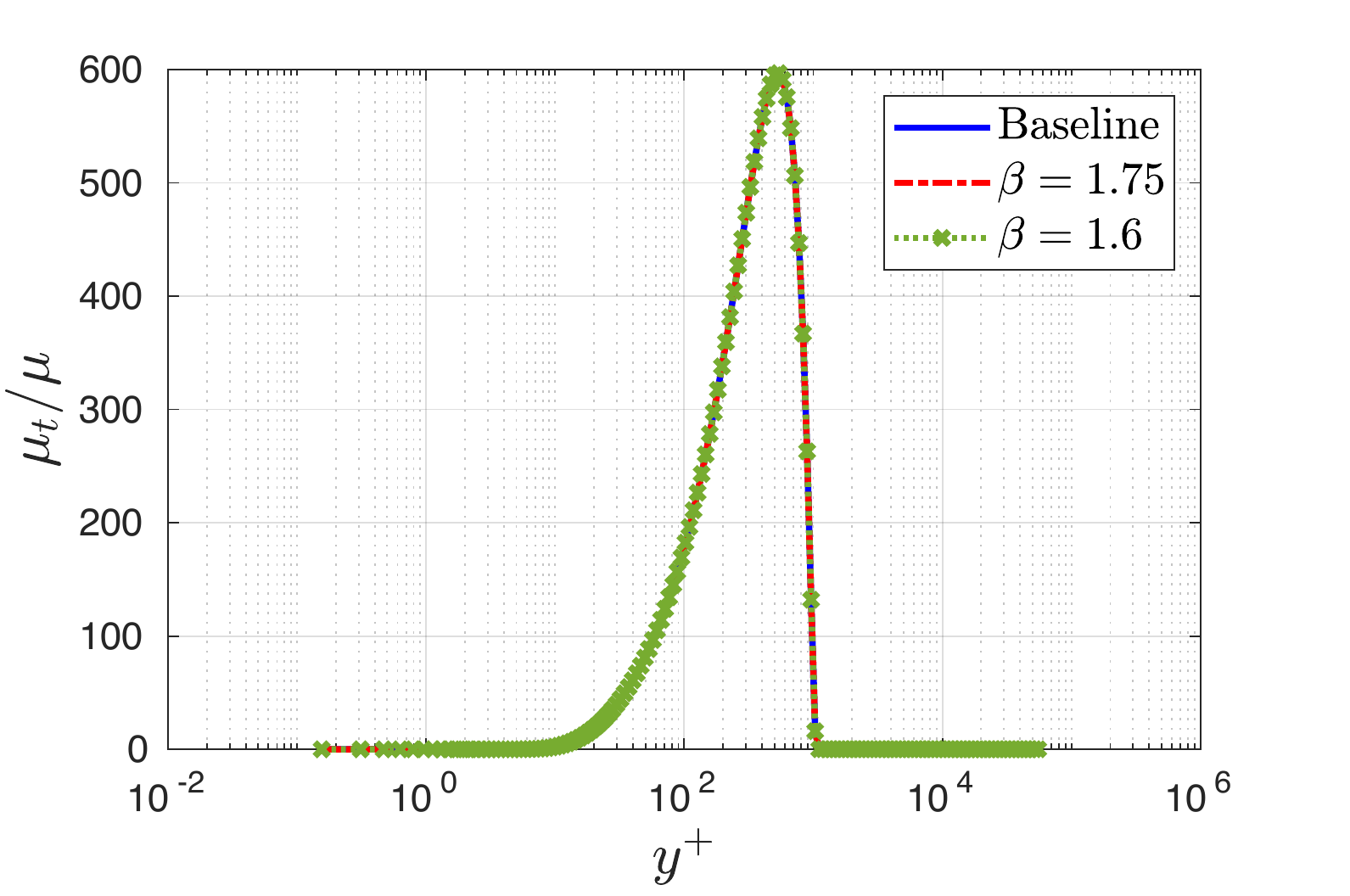}} 

    \caption{LCFM6: Variation of $\beta_{\text{max}}$ over the cone surface (top). The bottom rows shows the effect of user-defined $\beta$ on pressure (left), surface heat flux (middle) and eddy viscosity upstream of the flare (right). }
    \label{fig:LCF_Pk}
\end{figure}

Keeping this constraint in mind, we investigate the terms on the right hand side of the $k-$ equation. %study the turbulence production to dissipation ratio in the boundary layer upstream of the flare junction for the LCFM6 case. 
From eqn.~\ref{eqn: source}, the right hand side for the $k-$ equation can be expanded as
%\begin{equation}
%    \Psi_1 = \mu_t S^2 -\alpha_1 \rho k D - \left(1 + \frac{2 \nu}{\epsilon} \left(\frac{\partial k^{1/2}}{\partial x_k}\right)^2 \right)\rho \epsilon
%\end{equation}
\begin{equation}\label{eqn:RHSk}
    \Psi_1 = \mu_t S^2 - \rho \epsilon - \frac{2}{3}\rho k D - 2 \nu \left[\left(\frac{\partial \hat{q}}{\partial x}\right)^2 + \left(\frac{\partial \hat{q}}{\partial y}\right)^2 + \left(\frac{\partial \hat{q}}{\partial z}\right)^2\right].
\end{equation}
%where $P_k = \mu_t S^2 - (2/3)\rho k D$ is the turbulence production.
We define a non-dimensional parameter, $\beta$ given by
\begin{equation}
    \beta = \frac{\mu_t S^2}{\rho \epsilon},
\end{equation}
to examine the turbulence production to dissipation ratio for each test case.
%where, $\mu_t S^2$ can be regarded as a measure of turbulence production in the boundary layer.
Figure~\ref{fig:LCF_Pk}a plots the variation of the maximum value of $\beta$ with streamwise distance for the LCFM6 configuration.
After an initial rise due to the oblique shock at the cone tip ($x=0$), %Downstream of the initial oblique shock, 
the coefficient $\beta_{\text{max}}$ drops to $1.38$ immediately at~$x~=~0.1$~m, asymptotically approaching a value of $1.341$ in the fully turbulent boundary layer upstream of the flare.
The interaction of this boundary layer with the shock at the flare results in a significant increase ($\sim 354 \%$) in the $\beta_{\text{max}}$ value. 
Such high $\beta_{\text{max}}$ values represent a dominance of $k$ production over $\epsilon$, %which ultimately manifests as 
resulting in 
a more energized boundary layer downstream of the shock which inhibits separation.

This observation indicates that surface predictions can be improved by preventing this nonphysical amplification in turbulence production downstream of the shock.
A straightforward method to restrict this amplification is  %In order to improve the surface predictions, we 
to place an upper bound on the value of $\mu_t S^2$ %employ a limiter
based on a user-defined $\beta$ as follows:
\begin{equation} \label{eqn:PL}
    \Hat{P_s} = \text{min}\left( \mu_t S^2, \beta \rho \epsilon \right),
\end{equation}
where $\Hat{P_s}$ is the limited value which is substituted instead of $\mu_t S^2$ in eqn.~\ref{eqn:RHSk}. %, and $\beta$ is a user-defined parameter. 
The choice of $\beta$ dictates maximum value of turbulence production downstream of the shock. 
For example, a $\beta=\infty$ value implies no limitation on turbulence production and reverts back to the baseline model.
Note, however, that eqn.~\ref{eqn:PL} resembles the eddy viscosity limiting described in %Menter~
\cite{menter1994two} and %Kral \emph{et al.}~
\cite{kral1996application} to eliminate unrealistic buildup of eddy viscosity in stagnation regions of airfoils ($\beta=20$).
Despite the fundamentally different direction of the present effort with these earlier studies, the use of this ``production limiting'' provides a degree of unifying context. 

%Figure~\ref{fig:LCF_Pk}b shows the effect of incorporating this correction on the surface pressure predictions. %tests the sensitivity of the surface pressure to a pre-selected $\beta$ value. 
We take insights from the $\beta_{\text{max}}$ variation presented in Fig.~\ref{fig:LCF_Pk}a to select a suitable $\beta$ value. 
%Since the goal of this investigation is to improve A natural choice for the $beta$ 
It is evident that for any $\beta > 1.746$ in eqn.~\ref{eqn:PL}, the modified $k-\epsilon$ model behaves like its %uncorrected 
baseline variant upstream of the flare and therefore does not affect the attached boundary layer. %upstream of the flare.
%From Fig.~\ref{fig:LCF_Pk}b, it is observed that 
Figure~\ref{fig:LCF_Pk}b shows the effect of incorporating this correction on the surface pressure predictions. 
Specifying %use of
a pre-defined $\beta=1.75$ improves the surface pressure predictions compared to the baseline.
The separation bubble size and the peak pressure however, are still under-predicted;
this suggests the use of a lower $\beta$ to %The surface pressure predictions can be 
further improve predictions. %by using lower values of $\beta$ as shown.
Based on Fig.~\ref{fig:LCF_Pk}a, any $\beta < 1.746$ requires an additional geometric constraint of $x > x_0$ to avoid potential non-physical behavior near the oblique shock at the cone tip.
For example, %based on the $\beta_{\text{max}}$ variation in Fig.~\ref{fig:LCF_Pk}a, 
a $\beta = 1.40$, which is very close to the asymptotic $\beta_{\text{max}}=1.341$ in the attached boundary layer, can be employed for all $x > 0.1$~m. 
%The lowest permissible value of $\beta$ is 1.341 which requires 
%It must be noted, that any $1.38 < \beta < 1.746$ an additional geometric constraint of $x > 0.1$~m is required 
%to avoid potential non-physical behavior behavior near the oblique shock at the cone tip. %in the attached boundary layer.
%For example, a $\beta=1.38$ a geometrical constraint of $x > 0.1$~m to avoid potential nonphysical behavior near the oblique shock at the cone tip.
%A lower value of $\beta$ can be implemented using 
%Lower values of $\beta$ can be implemented 
%The threshold value of $\beta$ can be further reduced to $\sim 1.38$ with a geometrical constraint of $x > 0.1$~m to avoid potential nonphysical behavior due to the oblique shock at the cone tip.
%Figures~\ref{fig:LCF_Pk}b and ~\ref{fig:LCF_Pk}c show the resulting surface pressure and heat flux in the vicinity of the flare junction for different $\beta$ values.
%It is observed from Fig.~\ref{fig:LCF_Pk}b that 
%In general, 
%a decrease in $\beta$ increases the separation bubble size at the flare.
%Based on these trends, 
The use of $\beta=1.4$ overpredicts the separation bubble size, resulting in higher peak pressures than the measurements.
A $\beta$ value of 1.6 on the other hand, is found to be appropriate to accurately match the separation bubble size and peak pressure observed in the experiments.
%Accurate results for the separation bubble and peak press
%The use of $\beta=1.4$ overpredicts the separation bubble size, resulting in higher peak pressures than the measurements.
The peak heat transfer (shown in Fig.~\ref{fig:LCF_Pk}c), although smaller than the baseline model, remains overpredicted for $\beta=1.6$. 
Despite this disagreement, the new predictions are highly encouraging because, unlike the compressibility correction, the modifications do not alter the upstream boundary layer, for which the baseline $k-\epsilon$ predictions are accurate.
This is ensured by examining the eddy viscosity distribution upstream of the flare ($x = 2.2$ m) for the baseline and the modified variants ($\beta=1.6, 1.75$) in Fig.~\ref{fig:LCF_Pk}d.
The modified $k-\epsilon$ predictions shows identical eddy viscosity distribution in the attached boundary layer as the baseline, %; this ensures that 
thus preserving the accuracy of predictions in the upstream boundary layer. %are preserved.
%In order to ensure that these modifications do not interfere with the upstream boundary layer, Fig.~\ref{fig:LCF_Pk}d examines the eddy viscosity distribution upstream of the flare ($x = 2.2$ m) for the baseline and the modified variants ($\beta=1.6, 1.75$).
%The identical behavior of the eddy viscosity with the baseline makes the new predictions highly encouraging because, unlike the compressibility correction, the modifications do not alter the upstream boundary layer, for which the baseline $k-\epsilon$ predictions are accurate.
%Figure~\ref{fig:DFM8_Pk}d shows the eddy viscosity distribution at $x = 2.2$ m for two $\beta$ values. %
%The eddy viscosity variation in the boundary layer upstream of the flare ($x = 2.2$ m) for $\beta= 1.75$ and $1.6$ is identical to the baseline model as shown in Fig.~\ref{fig:LCF_Pk}. 

%For reference, the velocity profile is also plotted with dashed lines.

\begin{figure}
\centering
    \subfloat[ ]{\includegraphics[width=0.47\textwidth,trim=110 280 100 280,clip]{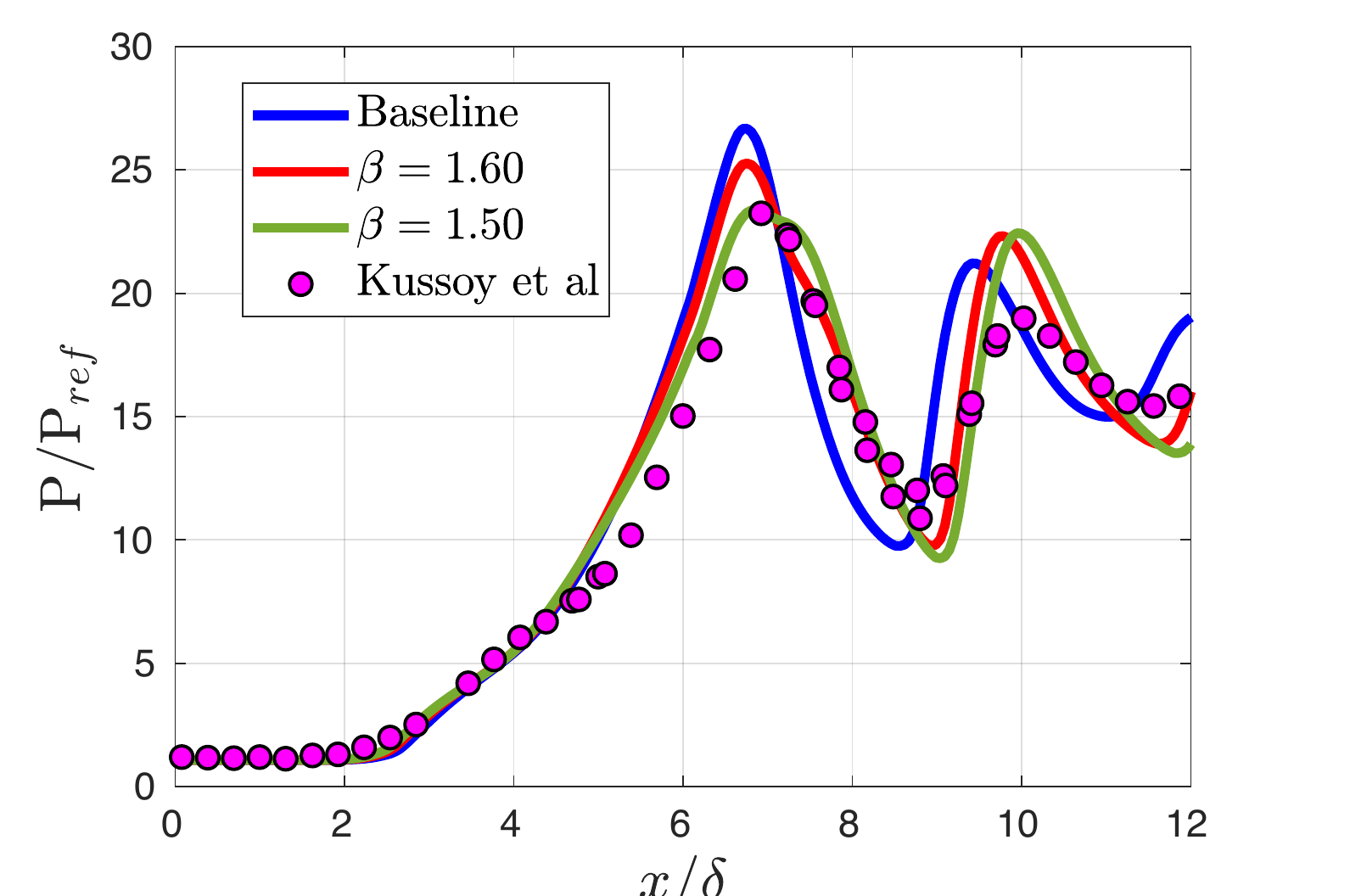}}\hspace{0.1in}
    \subfloat[ ]{\includegraphics[width=0.47\textwidth,trim=110 280 100 280,clip]{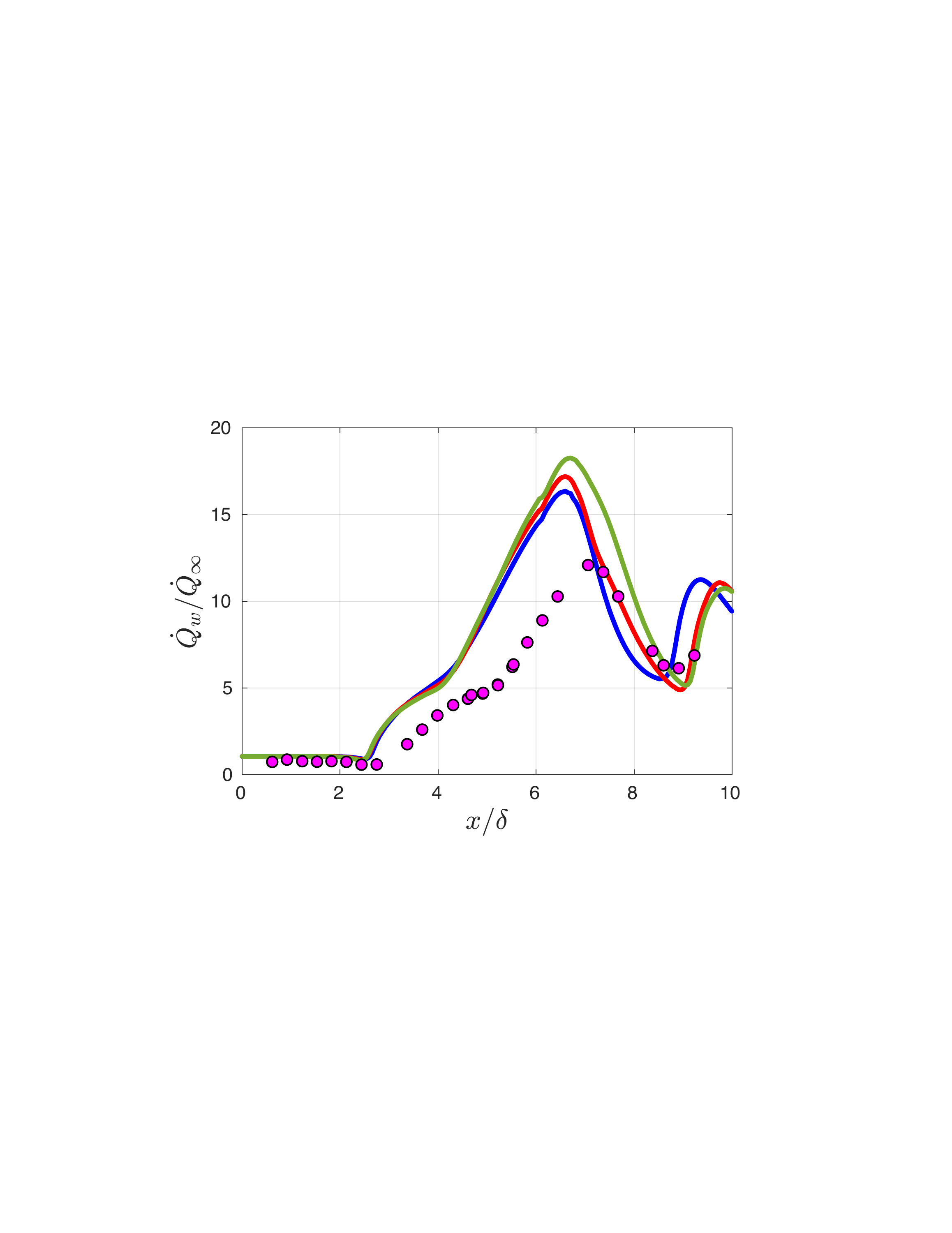}} 
    \caption{DFM8: Effect of user-defined $\beta$ on surface pressure (left) and surface heat transfer (right).}
    \label{fig:DFM8_Pk}
\end{figure}

Figure~\ref{fig:DFM8_Pk} shows the effect of incorporating this correction on the centerline pressure and heat transfer for the DFM8 test case. 
The threshold values of the parameter $\beta$ is calculated to be 1.381 in the inflow boundary layer.
Since the inflow is specified from a precursor boundary layer simulation, a geometrical constraint  is not necessary for this test case.
The use of $\beta=1.5$ is observed to accurately capture the mean surface pressure along the centerline, including the pressure rise and peak pressure due to the PS interaction $(2 \leq x/\delta \leq 6.5)$, the relief effect due to SE $(6.5 \leq x/\delta \leq 9)$ and the subsequent rise due the upstream influence of the RS2 interactions, with only modest disagreement between the secondary maxima and minima. %
%found to accurately match the peak pressure measured in the experiments.
Similar to the LCFM6 calculations, the peak heat transfer along the centerline remains over-predicted.

\subsection{Correction 2: Length-scale Limiting}
%The over-prediction in heat transfer indicates that the baseline model does not sufficiently capture the heat convected away from the wall by the fluid; this results in an increased heat transfer at the wall downstream of the SBLI.
To improve heat transfer predictions, we take recourse to %Reynolds~
\cite{reynolds1980modeling}, %and Morel and Mansour~\cite{morel1982modeling} 
who examined the role of turbulent dissipation in flows inside internal combustion engines.
%flows undergoing rapid compression.
Based on the conservation of angular momentum, he argued that for flows undergoing rapid compression, $\epsilon$ is too slow to be of significance and could be neglected.
%This idea was further generalized %in terms of the turbulent length scale ($l$) 
%by 
%Morel and Mansour~
\cite{morel1982modeling} expanded on Reynolds' findings by modifying the 
$\epsilon$ equation %to correctly reproduce the trends expected 
%formulating the problem in terms of a turbulence length scale ($l$) 
to ensure a reduction in the turbulence length scale %by placing a constraint on the 
with an increase in density. %based on the type of compression.
%argued that $l$ and $k^{1/2}$ provided an improved description of turbulent motions.
This idea of length-scale limiting was extended to shock-induced compression by %Vuong and Coakley~
\cite{vuong1987modeling}, who placed an upper bound on the turbulence length-scale  %extended this idea to aerodynamic flows undergoing compression due to shocks by suggesting a length-scale limiter . %containing flows %undergoing compression due to shocks and
%to modify $\epsilon$ as follows
%\begin{equation}
%    \epsilon = \frac{k^{(3/2)}}{\text{min}\left( \frac{k^{(3/2)}}{\epsilon}, \lambda d\right)},
%\end{equation}
%where $d$ is the normal distance from the wall and $\lambda$ is a user-defined parameter. 
%who 
%and suggested 
%the use of length-scale limiting 
to improve heat transfer for separated flows in the reattachment regime.
%Following Vuong and Coakley~\cite{vuong1987modeling}, the turbulent length-scale can be constrained as
%This is incorporated in the $\epsilon$ equation by defining a length-scale, $l$ as follows
%\begin{equation}\label{eqn:lengthscale}
%    l = \text{min}\left[ \frac{k^{(3/2)}}{\epsilon}, \lambda d\right],
%\end{equation}
%where $d$ is the normal distance from the wall.
%This constraint places an upper bound of the turbulence length-scale which can is used to calculate $\epsilon$ as follows
%\begin{equation}\label{eqn:eps}
%    \epsilon = \frac{k^{(3/2)}}{l}.
%\end{equation}
%The use of $2.5 d$ as an upper bound for $l$ is based on Bradshaw's assumption between the wall shear stress and the turbulence kinetic energy. 
%Although this ``length-scale limiting" has shown some success in improving heat transfer predictions for shock-induced separation, the effectiveness of this correction has been observed to vary from case to case~\cite{}.

\begin{figure}
\centering
\subfloat[]{\includegraphics[width=0.7\textwidth]{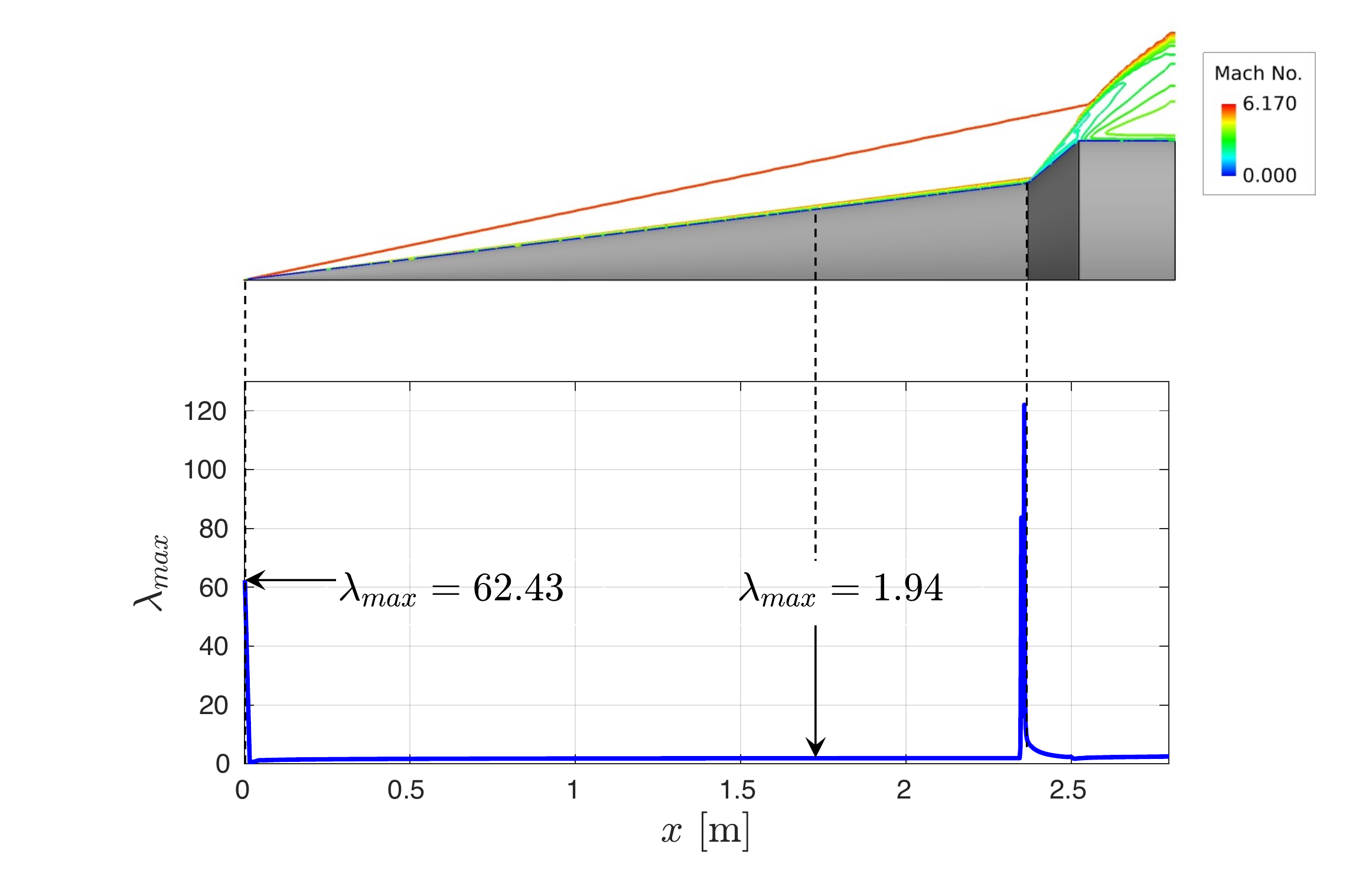} }\\
    \subfloat[ \label{fig:FPM11qw}]{\includegraphics[width=0.33\textwidth,trim=110 280 100 285,clip]{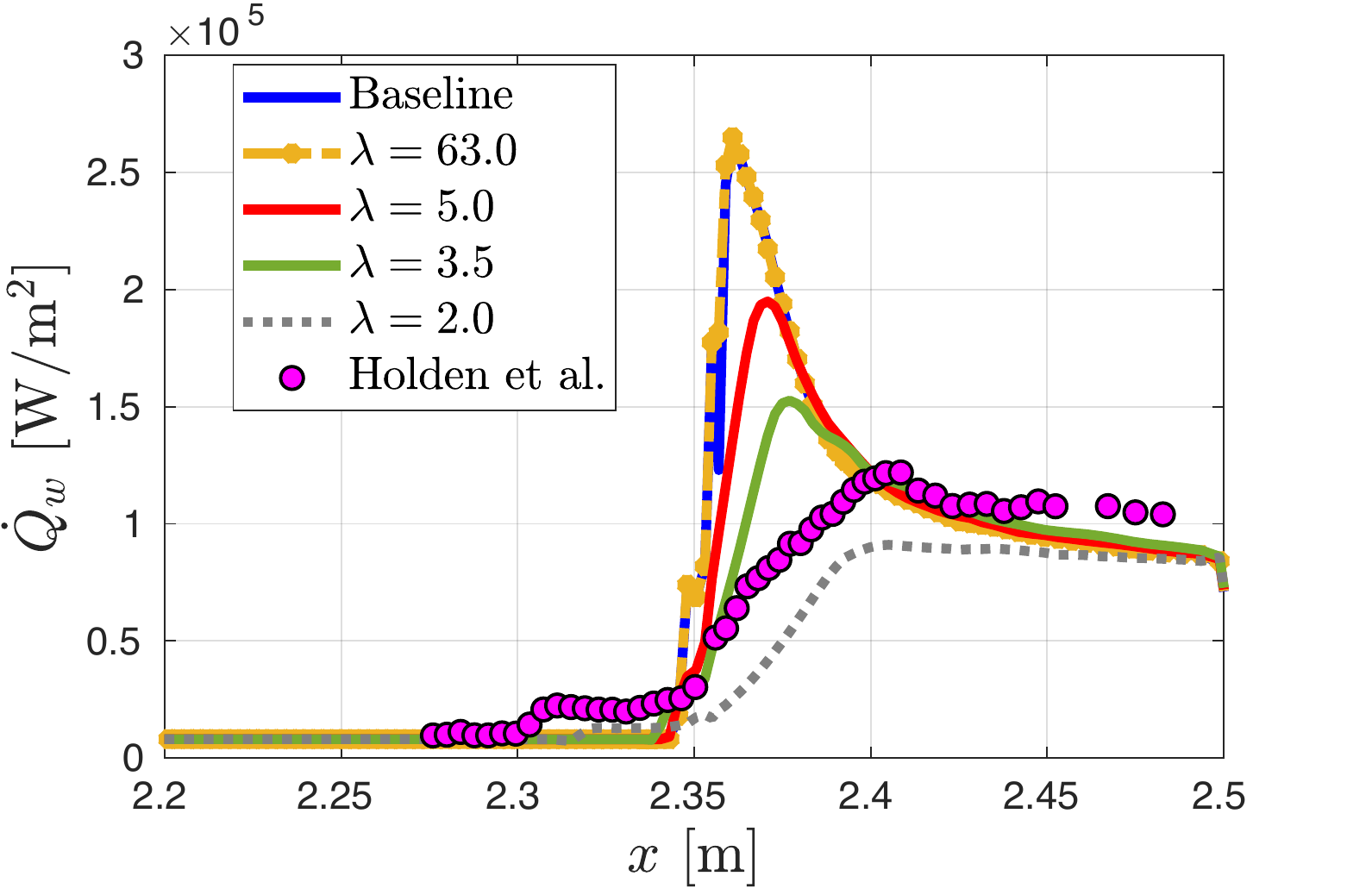}}
    \subfloat[ \label{fig:FPM11tauw}]{\includegraphics[width=0.33\textwidth,trim=110 280 100 280,clip]{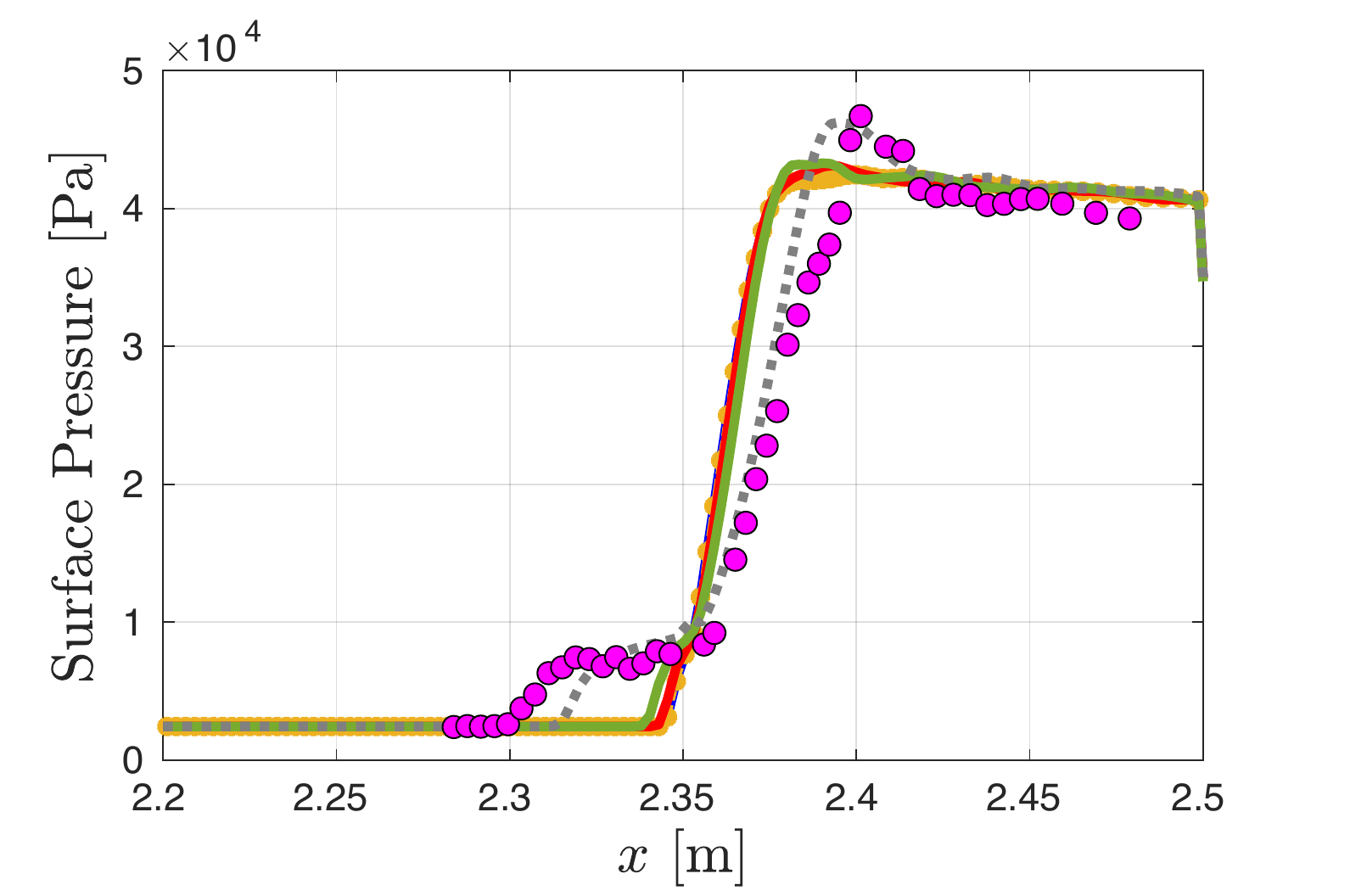}} 
    \subfloat[ \label{fig:FPM11tauw}]{\includegraphics[width=0.33\textwidth,trim=110 280 100 280,clip]{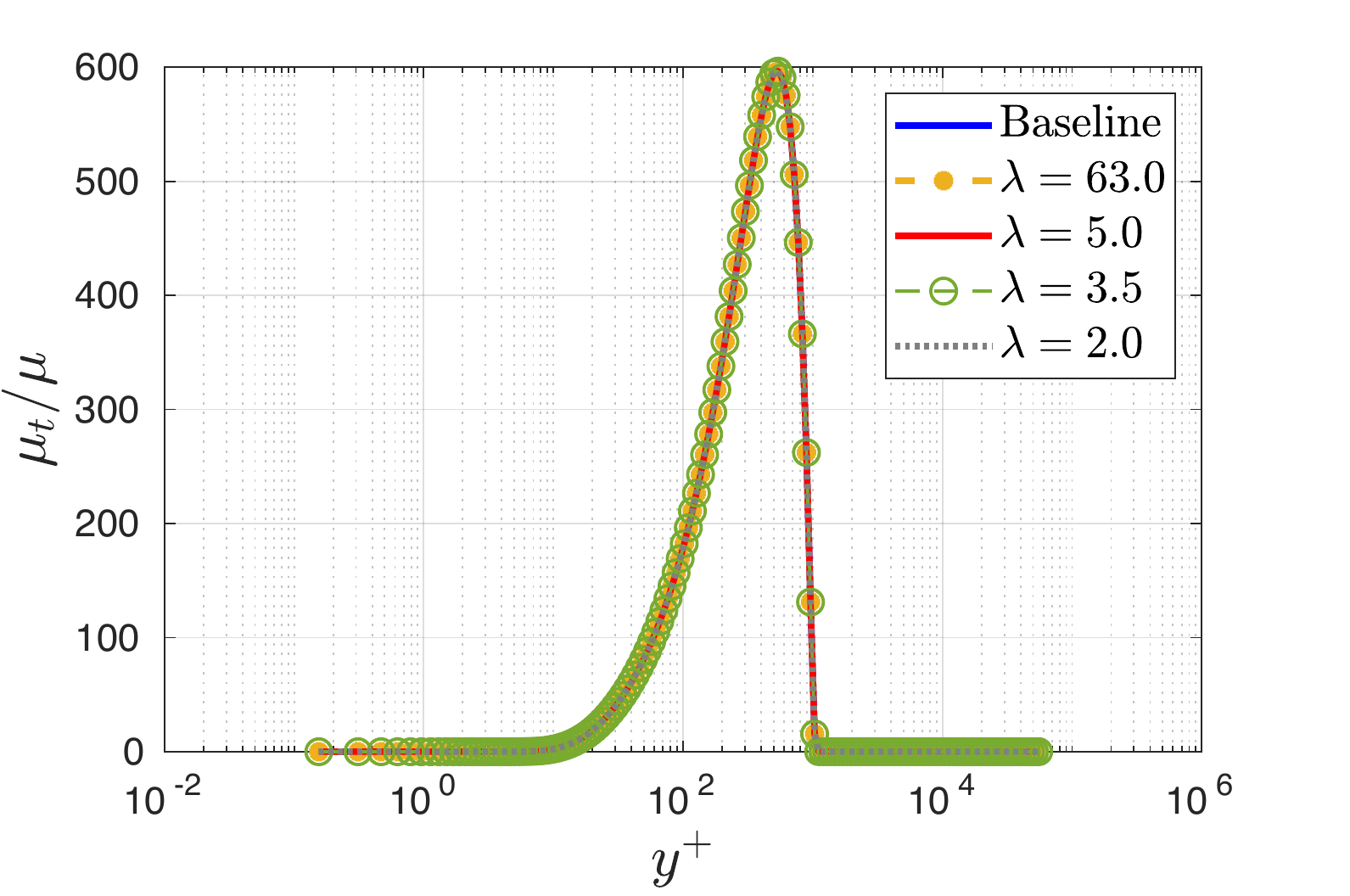}} 
    
    %\subfloat[Van-Driest scaled velocity profile \label{fig:FPM11uvd}]{\includegraphics[width=0.5\textwidth]{Aviation_2021/Figures/uvd_inlet_FP_M11_v2.jpg}}
    %\subfloat[Normalized velocity profile \label{fig:FPM11u}]{\includegraphics[width=0.5\textwidth]{Aviation_2021/Figures/u_vs_ypls_inlet_FPM11_v2.jpg}} \\
    \caption{LCFM6: Variation of $\lambda_{\text{max}}$ over the cone surface (top). The bottom rows shows the effect of user-defined $\lambda$ on pressure (left), surface heat flux (middle) and eddy viscosity upstream of the flare (right). }
    \label{fig:LCF_LL}
\end{figure}

%Eqn~\ref{eqn:lengthscale} prevents the turbulence length-scale from amplifying more that $2.5$ times
In the present investigation, we take a more generalized approach to Vuong and Coakley's length-scale limiting by defining a non-dimensional parameter $\lambda$ given by 
\begin{equation}\label{eqn:lambda}
    \lambda = \frac{k^{3/2}/\epsilon}{d }.
\end{equation}
%where $d$ represents the distance from the wall.
Figure~\ref{fig:LCF_LL} shows the distribution of the maximum value of $\lambda$ as a function of $x$ for the LCFM6 baseline calculations. 
%A $\lambda_{\text{max}}$ value of $1.94$ is observed for the attached boundary layer upstream of the flare.
Similar to the $\beta_{\text{max}}$ variation in Fig.~\ref{fig:LCF_Pk}a, the oblique shocks at the cone tip and the flare are observed to amplify the $\lambda_{\text{max}}$ value. 
%The oblique shock at the cone tip results in $\lambda_{\text{max}}=62.43$ which 
After an initial value of $62.43$ due to the oblique shock at the cone tip, the $\lambda_{\text{max}}$ value drops to zero at the very next axial location, approaching a value of $1.94$ in the attached boundary layer before being significantly amplified at the flare.
%From eqn.~\ref{eqn:lambda}, an amplification in $\lambda_{\text{max}}$ indicates a reduction in $\epsilon$ value; this reduces %which is insufficient to capture
%does not sufficiently capture 
%the heat convected away from the wall by the fluid resulting in an increased heat transfer at the wall.
This observation indicates that %suggests that %amplification of $\lambda$ can be controlled %suggests that 
wall heat flux predictions can be improved
%This indicates a value of $\lambda > 1.95$ in order to leave the upstream boundary layer unaltered.
%This is 
%incorporated in the $\epsilon$ equation 
by limiting this amplification of $\lambda$ at the flare; this is achieved by placing an upper-bound on the turbulence length-scale as follows %. 
%preventing a drastic reduction in $\epsilon$ prediction downstream of the shock.
%This is incorporated in the $\epsilon$ equation by defining a length-scale, $l$ as follows
\begin{equation}\label{eqn:LL}
    l = \text{min}\left[ \frac{k^{(3/2)}}{\epsilon}, \lambda d\right],
\end{equation}
where $\lambda$ is a user-defined parameter. 
This length-scale is used to calculate $\epsilon$ as shown below
\begin{equation}\label{eqn:eps}
    \epsilon = \frac{k^{(3/2)}}{l}.
\end{equation}
%This is implemented by means of a user-defined $\lambda$ as follows \begin{equation}
%    \epsilon = \max \left[\epsilon, \frac{k^{(3/2)}}{\lambda d}\right],
%\end{equation}
%where the parameter $\lambda$ can be used to increase the $\epsilon$ value downstream of the shock.

Based on the trends in Fig.~\ref{fig:LCF_LL}a, the use of any $\lambda > 1.94$ (for all $x>0$) results in an $\epsilon$ value which %calculated from eqn.~\ref{eqn:eps} 
is identical to the baseline model upstream of the flare; this preserves the accuracy of predictions in the attached boundary layer.
%Note that, $\lambda=2.5$ is equivalent to Vuong and Coakley's~\cite{vuong1987modeling} correction.
Figures~\ref{fig:LCF_LL}b and~\ref{fig:LCF_LL}c show the heat transfer and surface pressure predictions for the LCFM6 test case at different values of $\lambda$.
Contrary to production-limiting, the use of length-scale limiting has a significant effect on the heat transfer predictions.
In general, the peak heat flux at the wall decreases with a decrease in $\lambda$ value.
The surface pressure is mostly unaffected except at very low values of $\lambda$.
%Moreover, this improvement in heat transfer is achieved without affecting the upstream boundary layer as evident from the eddy viscosity distribution at $x=2.2$ m in Fig.~\ref{fig:LCF_LL}d.
Figure~\ref{fig:LCF_LL}d shows the eddy viscosity distribution in the attached boundary layer at $x=2.2$ m for all $\lambda$ values tested.
The perfect overlap of the length-scale limited simulations with the baseline indicates that the improvements in heat transfer are obtained without altering the upstream boundary layer. 

\begin{figure}
\centering
 
    \subfloat[ \label{fig:FPM11tauw}]{\includegraphics[width=0.47\textwidth,trim=100 250 100 270,clip]{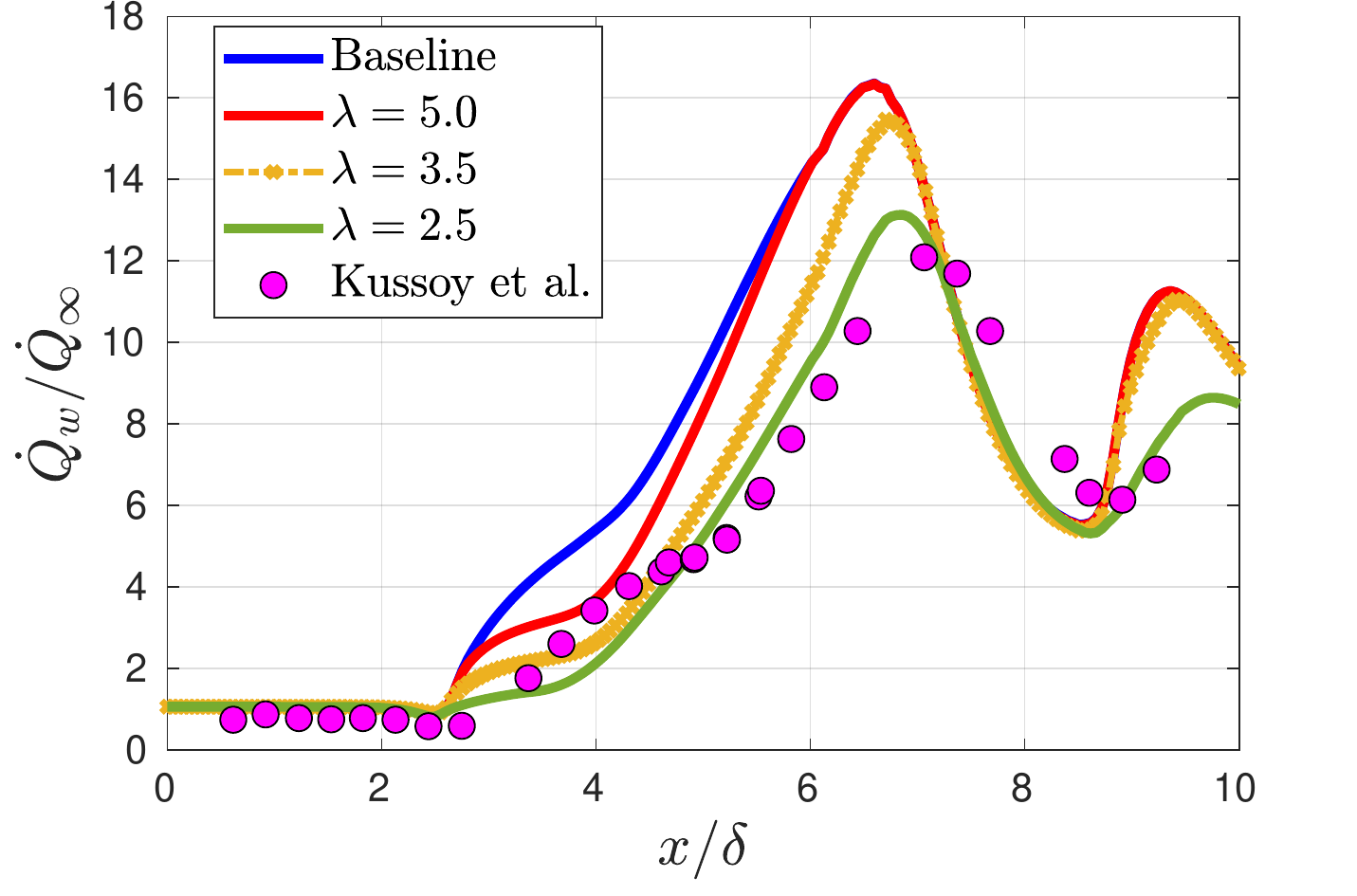}} \hspace{0.1in}
   \subfloat[ \label{fig:FPM11qw}]{\includegraphics[width=0.47\textwidth,trim=100 250 100 270,clip]{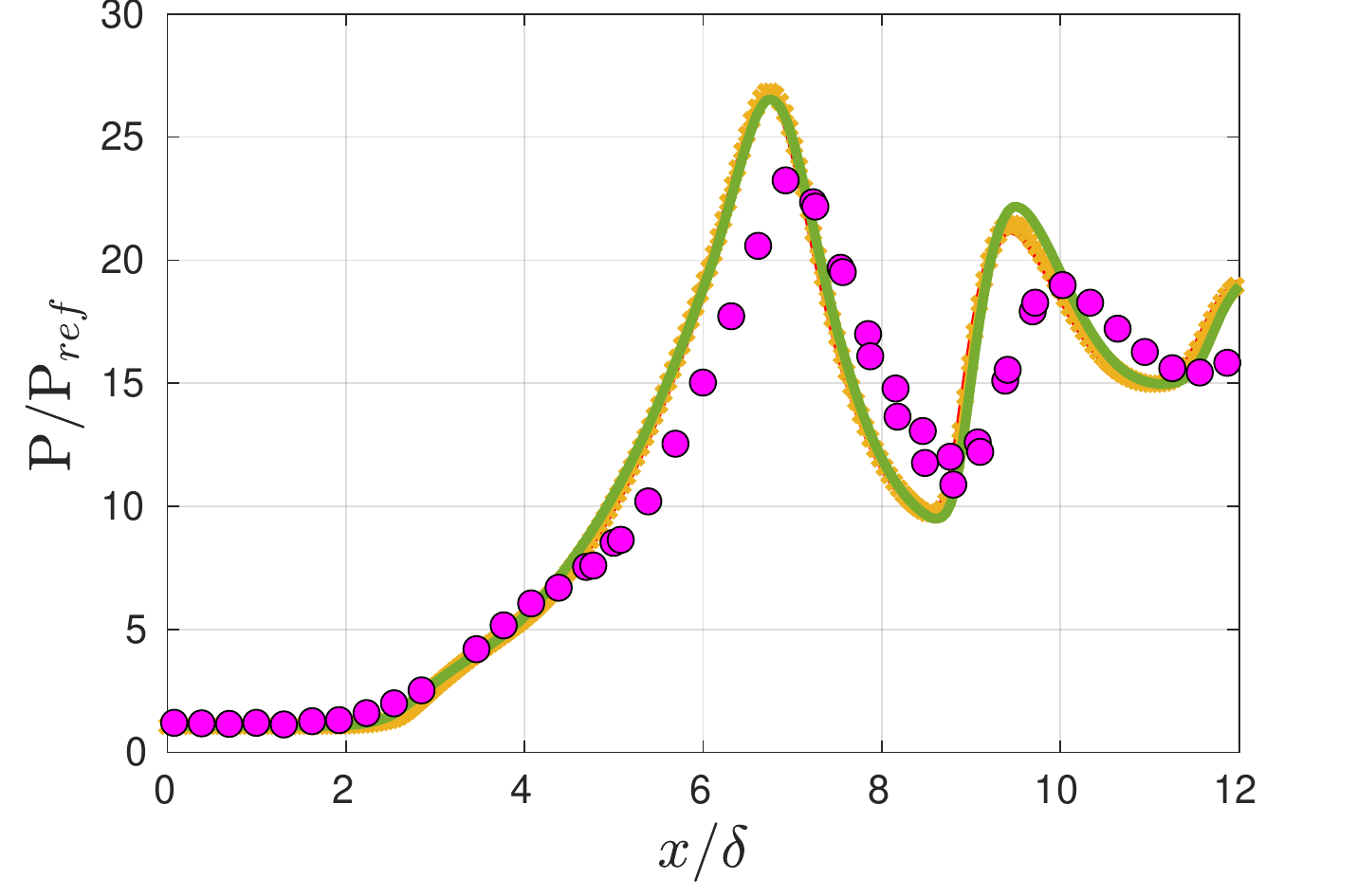}}
    \caption{DFM8: Effect of length-scale limiting on surface heat transfer (left) and surface pressure (right).}
    \label{fig:DFM8_LL}
\end{figure}

Figure~\ref{fig:DFM8_LL} shows the effect of incorporating this correction on the heat flux and surface pressure predictions for the DFM8 test case.
Based on a similar analysis as in Fig.~\ref{fig:LCF_LL}a, the threshold value of $\lambda$ is found to be $1.975$ in the inflow boundary layer when %.
%The threshold value of $\lambda$ is calculated 
the distance $d$ is calculated from the bottom plate $(y\text{-coordinate})$.
A similar trend is observed, where a decrease in $\lambda$ value decreases the peak heat transfer.
The use of $\lambda=2.5$ significantly improves heat transfer predictions, including peak heat transfer and the initial rate of increase along the centerline.
Similar to the LCFM6 case, this correction has no effect on the surface pressure predictions.
%This behavior suggests the use of mixed production and length-scale limiting to improve the overall $k-\epsilon$ predictions. 
These observations are leveraged next to improve the overall $k-\epsilon$ predictions for the two test cases.

\subsection{Combined Results: Production and Length-scale Limiting}
%This results in a more energized boundary layer which does not separate
The previous subsections have shown that the production and length-scale limiting corrections can independently improve surface pressure and wall heat flux without affecting the attached boundary layer. %for both LCFM6 and DFM8 test cases.
This subsection attempts to combine the two corrections %behavior suggests combining the use of mixed production and length-scale limiting 
to improve the overall $k-\epsilon$ predictions. 

\begin{figure}
    \centering
    \subfloat[]{\includegraphics[width=0.47\textwidth,trim=100 250 100 270,clip]{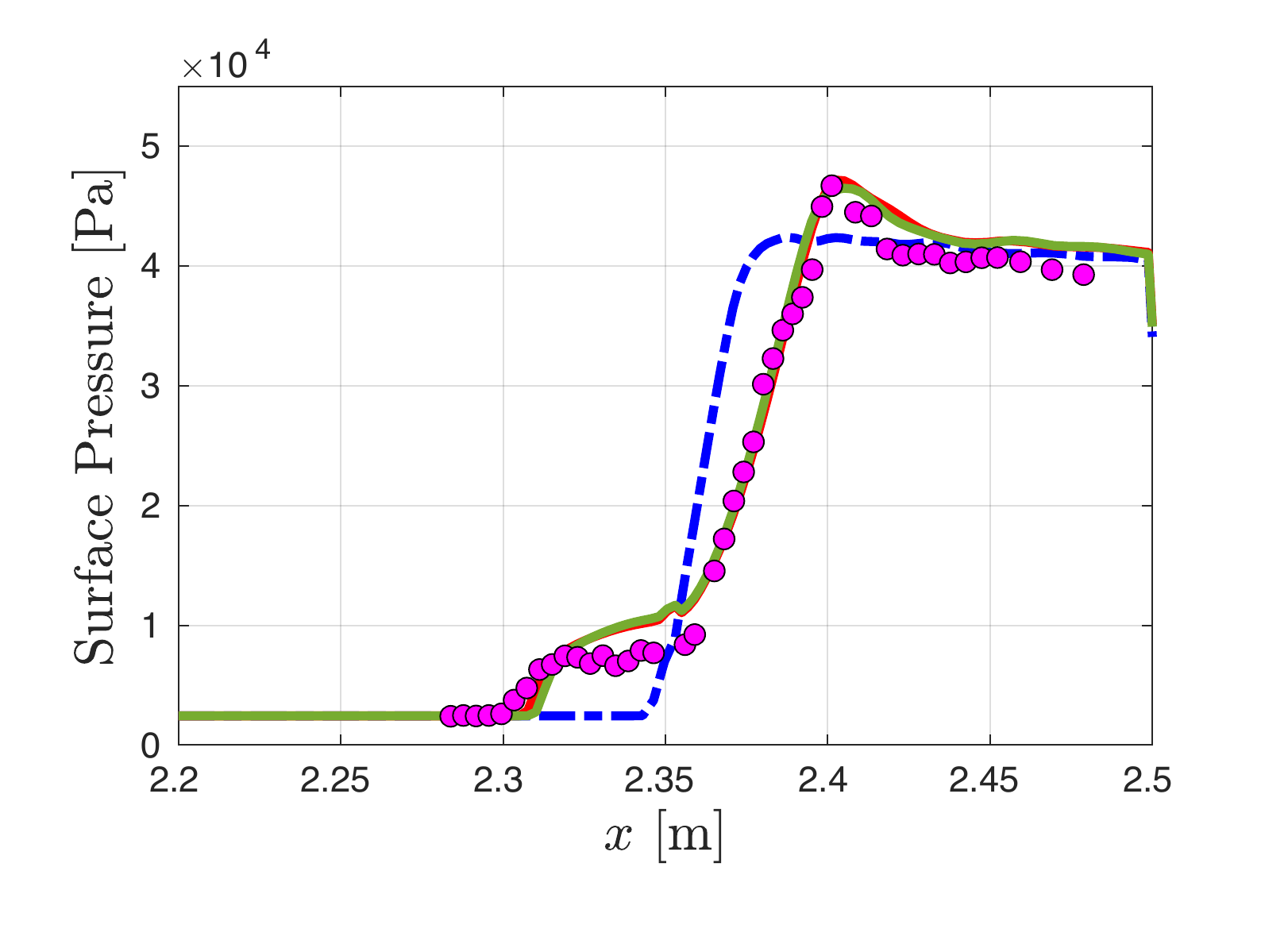}}\hspace{0.1in}
    \subfloat[]{\includegraphics[width=0.47\textwidth,trim=100 250 100 270,clip]{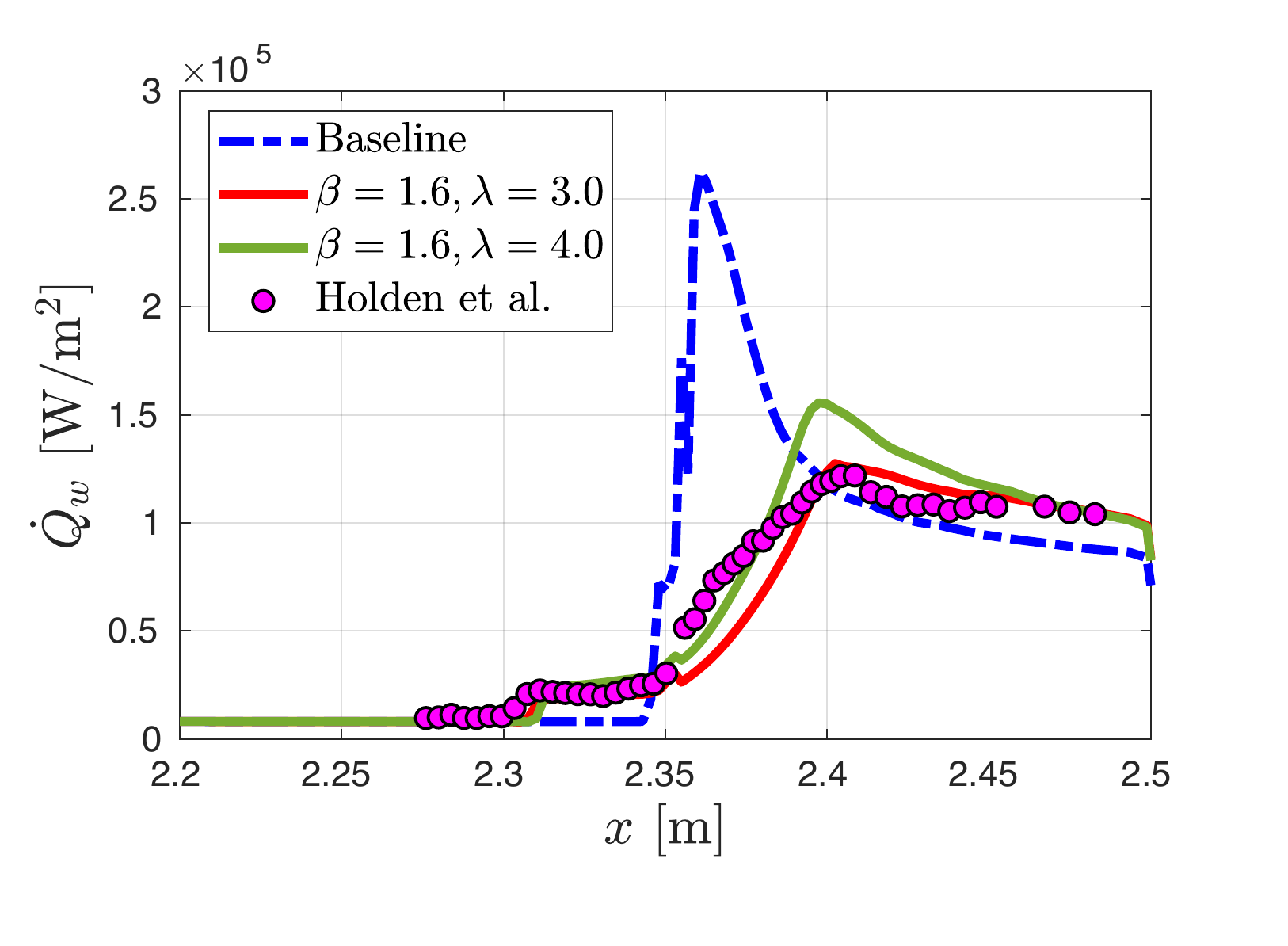}}
    \caption{LCFM6: Predictions of surface pressure (left) and wall heat flux (right) with combined production and length-scale limiting.}
    \label{fig:LCFBest}
\end{figure}

We first focus on the LCFM6 test case.
Figure~\ref{fig:LCFBest} shows the surface pressure and heat transfer predictions with $\left( \beta, \lambda \right) = \left( 1.6, 3.0 \right)$ and $\left( 1.6, 4.0 \right)$. %for the LCFM6 test case.
Note that, $\beta=1.6$ showed the best pressure predictions but significantly over-predicted the heat transfer $(\sim 100\%)$ for this test case (Fig.~\ref{fig:LCF_Pk}).
When combined with length-scale limiting, the modified $k-\epsilon$ model accurately capture the separation bubble size and the reattachment peak pressure observed in measurements irrespective of the $\lambda$ value;
this observation indicates that the two corrections do not conflict with one another for this test case.
For surface heat transfer, the modified $k-\epsilon$ predictions improve considerably compared to the baseline.
The use of $(\beta,\lambda)=(1.6, 3.0)$ accurately matches the rate of increase in $\dot{Q}_w$ until $x=2.37$ m with a $\sim27 \%$ over-prediction in the peak value.
In contrast, the use of $(\beta,\lambda)=(1.6, 3.0)$ accurately predicts the peak heat flux with an average $\sim35 \%$ under-prediction in the region $2.35 \text{ m}\leq x < 2.39 \text{ m}$. 
In both cases, the location of the peak heat transfer is correctly predicted.
These results are highly encouraging as these improvements are obtained without altering the upstream attached boundary layer.
Moreover, since the terms required to implement these corrections are already present in the baseline model, these modifications are straightforward to implement in existing codes and do not incur any additional computational cost.
%\textit{@@Cost of simulations@@}.
As an example, both the baseline and modified ($\beta=1.6 {,}$ $\lambda=3.0$) calculations for the LCFM6 test case required approximately $1{,}900$ CPU hours with 192 Intel\textsuperscript{\textregistered} Xeon\textsuperscript{\textregistered} 8268s Skylake ($2.9$ GHz) processors ($\sim 10$ hours of wall-clock time) on the intermediate mesh (Table~\ref{tab:MeshLCF}).

%The results thus suggest that a more generalized formulation for $\beta$ and $\lambda$ values based on shock angle and flow conditions can further improve heat transfer predictions. 
%This is being pursued by the authors and will be presented in the future.

\begin{figure}
    \centering
    \subfloat[]{\includegraphics[width=0.47\textwidth,trim=100 250 100 270,clip]{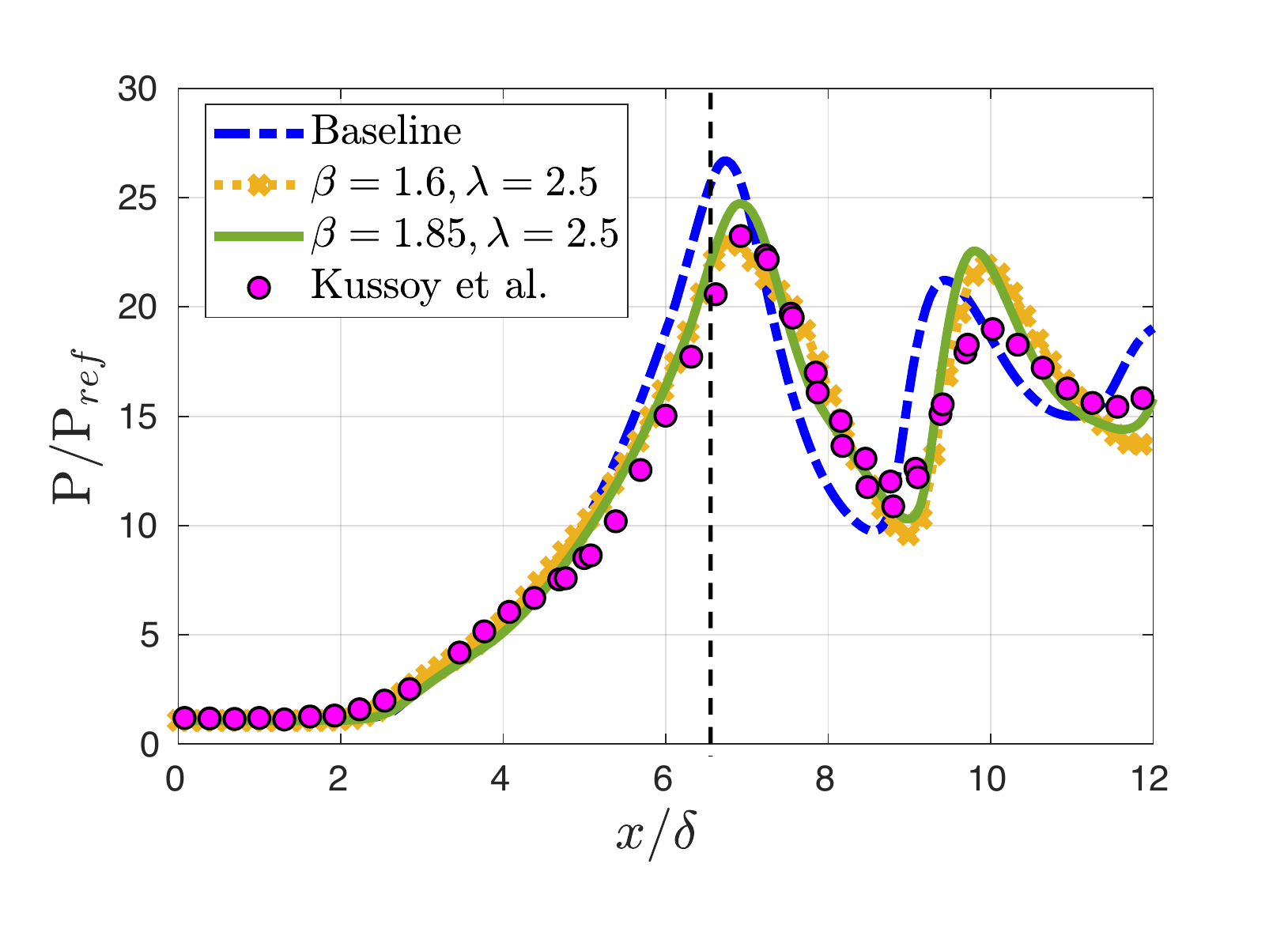}}\hspace{0.1in}
    \subfloat[]{\includegraphics[width=0.47\textwidth,trim=100 250 100 270,clip]{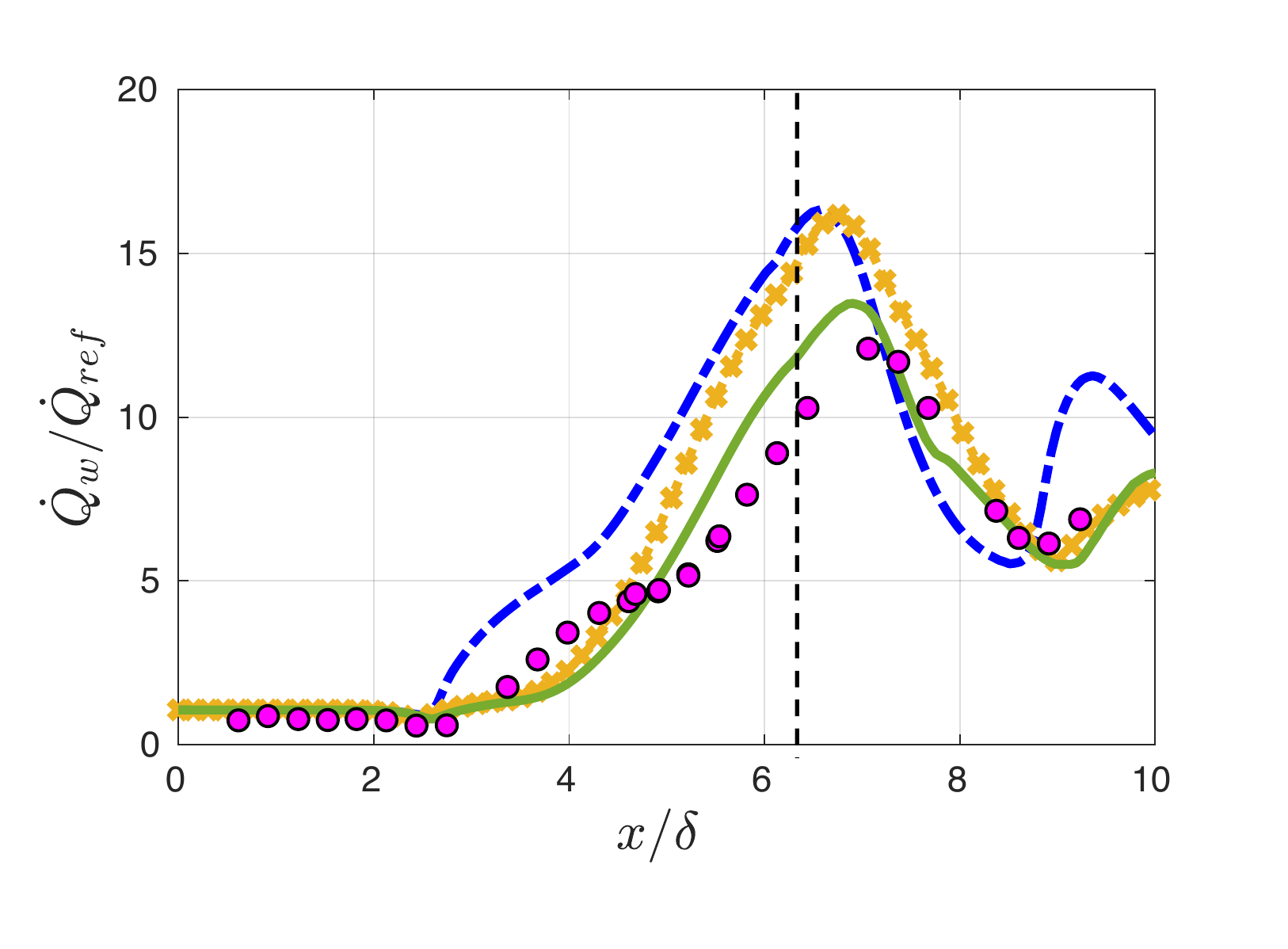}} \\
    %\subfloat[]{\includegraphics[width=0.45\textwidth,trim=100 250 100 270,clip]{Figures/DF_Press_SpanwiseFinal_v2.pdf}}
    %\subfloat[]{\includegraphics[width=0.45\textwidth,trim=100 250 100 270,clip]{Figures/DF_Qw_spanwiseFinal_v2.pdf}}
    \caption{DFM8: Predictions of surface pressure (left) and wall heat flux (right) with combined production and length-scale limiting along the centerline. }
    %The bottom row shows the spanwise predictions at the plane marked with dotted lines in the top row.}
    \label{fig:DFBest}
\end{figure}

We now focus our attention to the more complicated DFM8 test case. Figure~\ref{fig:DFBest} shows the predicted surface pressure and heat transfer rates along the centerline for the DFM8 test case with $(\beta,\lambda)=(1.6, 2.5)$ and $(1.85, 2.5)$.
%The pressure and heat transfer values are normalized by $430$ N/m$^2$ and $10{,}400$~W/m$^2$ which are the reference quantities used in experiments.
%Based on these figures, several observations can be made.
%Several observations may be made to assimilate the results.
Note that, $\beta=1.5$ and $\lambda=2.5$ provided the best surface pressure and heat transfer predictions when applied independently (Figs.~\ref{fig:LCF_Pk}a and~\ref{fig:LCF_LL}a).
When combined, the use of %both production and length-scale limiting with 
$(\beta,\lambda)=(1.6, 2.5)$ is observed to accurately capture the mean surface pressure along the centerline, including the initial pressure rise and peak pressure, the relief effect downstream of the PS and the subsequent rise due the upstream influence of the RS2 interactions, with only modest disagreement at the secondary maxima.
This correction combination however, only improves the surface heat flux in the $(2 \leq x/\delta \leq 5)$ range. 
The peak heat flux, similar to the baseline is over-predicted by $\sim 34\%$.
%This correction combination does not, however, reproduce the accurate heat transfer predictions of the $k-\epsilon$ model with only length-scale limiting, \textit{i.e.}, $\left(\beta,\lambda\right)=\left(\infty,2.5\right)$.
A further reduction in the value of $\lambda$ at $\beta=1.6$ does not reduce the over-prediction in peak heat transfer rates.
In fact, the best overall predictions are obtained by choosing a higher $\beta$ value of $1.85$ while keeping $\lambda=2.5$ as shown. 
The peak heat transfer now shows only a $\sim 12 \%$ over-prediction from the measured values without sacrificing the accuracy of surface pressure.

\begin{figure}
    \centering
    \subfloat[]{\includegraphics[width=0.65\textwidth,trim=20 100 10 10,clip]{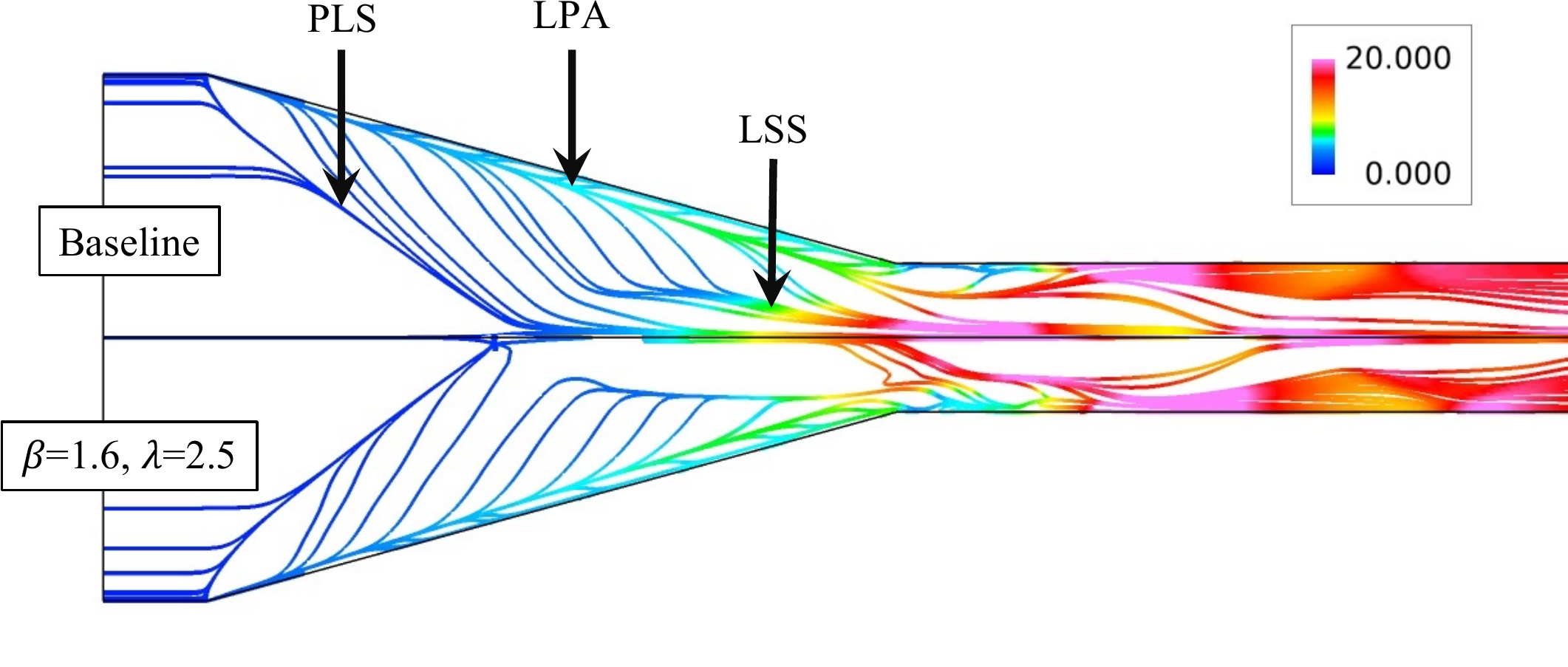}} \\
    \subfloat[]{\includegraphics[width=0.65\textwidth,trim=5 100 10 10,clip]{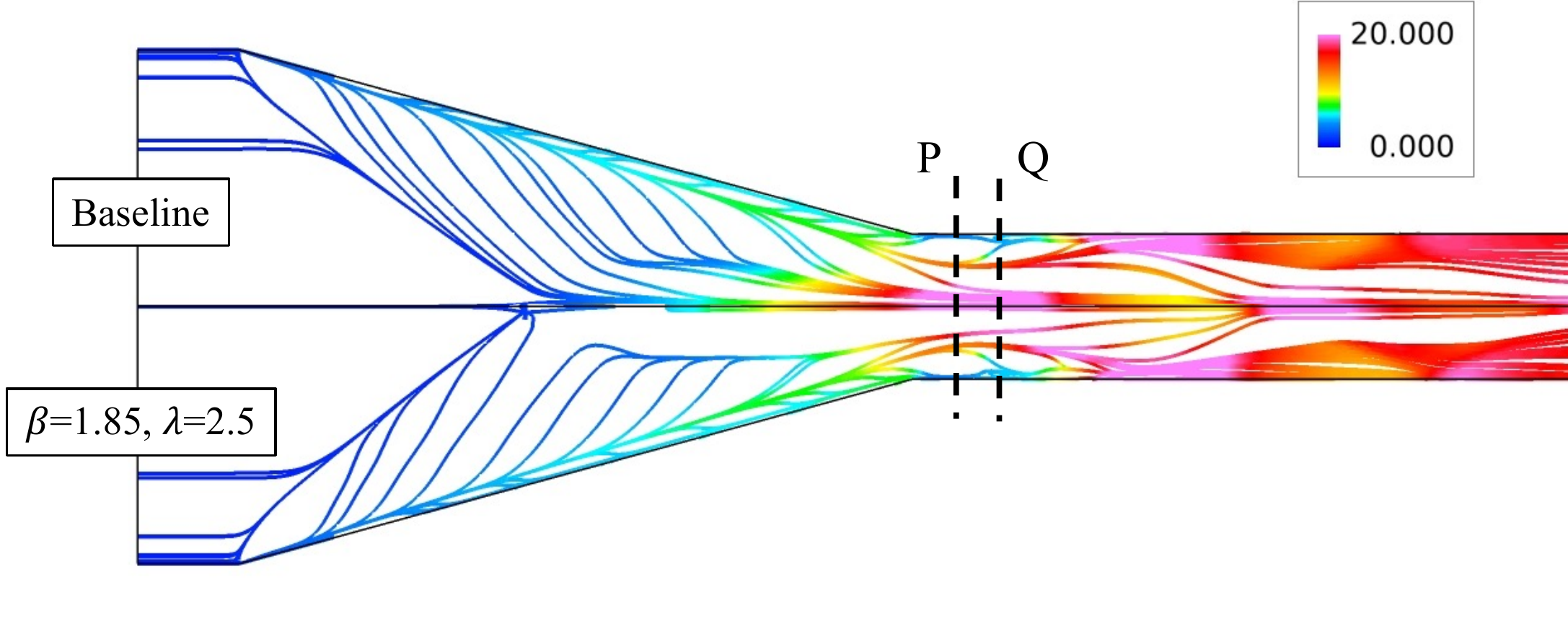}}
    \caption{DFM8: Comparison of the flow's footprint with baseline for (a)~$(\beta,\lambda)=(1.6,2.5)$, and (b)~$(\beta,\lambda)=(1.85,2.5)$.}
    \label{fig:DF_stream}
\end{figure}

%Improved predictions are obtained by choosing a higher $\beta$ value of 1.85 while keeping $\lambda=2.5$. 
%The peak heat transfer now shows only a $\sim 12 \%$ over-prediction from the measured values without sacrificing the accuracy of surface pressure. %use of these values accurately reproduces the accurate heat transfer prediction
This deviation from the expected behavior indicates that the two modifications interfere with one another for the more complicated 3D case; this warrants a closer look at the flow-field for the two calculations. %@@ New results show 12 percent over prediction in peak heat transfer@@
Figure~\ref{fig:DF_stream} compares the streamlines (colored with P/P$_{\text{ref}}$) in the proximity of the bottom plate for $(\beta,\lambda)=(1.6, 2.5)$ and $(1.85,2.5)$ with the baseline calculation. 
%A sketch of the streamlines obtained from surface oil flow measurements is also presented for reference.
%The predicted footprint of the flow with all three $k-\epsilon$ variants agrees reasonably well with the experiment.
Based on the description by %Gaitonde \emph{et al.}~
\cite{gaitonde1995structure}, the experimentally observed indicators of 3D separation can be identified based on the streamline structure. %as shown in Fig.~\ref{fig:DF_stream}a;
%These include: 
The principle line of separation (PLS) is the line of coalescence joining the lambda shock edges (Fig.~\ref{fig:DFFlow}).
The incoming flow separates along the line PLS as discussed earlier. 
The region beneath the separated flow is occupied by the spanwise movement of the fluid from the line of primary attachment (LPA) near the fin surface. 
%The line LPA stays very close to the fin surface up to the shoulder.
%Downstream of the shoulder the line LPA exhibits a wavy pattern.
The fluid emanating from the line LPA also forms another line of coalescence called the line of secondary separation (LSS) between the symmetry plane and the fin surface.
As expected, the flow along the lines LPA and LSS is accompanied by an adverse pressure gradient.
Interestingly, the spanwise movement of the fluid from the line LPA towards the symmetry plane and the line LSS is initiated with a favorable pressure gradient near LPA.
%a detailed investigation of the structure of this flowfield can be found in Ref.~\cite{gaitonde1995structure}.
%The key difference in the streamlines between the three $k-\epsilon$ variants is near the line of secondary separation.
For the baseline case, the line LSS asymptotically approaches the center symmetry plane with increase in downstream distance. 
%See comment
For the $(\beta,\lambda)=(1.6, 2.5)$ case which gives accurate surface pressure predictions but overpredicts heat transfer, the distance between the line LSS and the symmetry plane is larger than the baseline. 
This suggests that less influence of the line LSS on the symmetry plane might be responsible for the accurate centerline pressure predictions.
%However, due to this increased distance, the line of SS now interacts with the line of primary attachment (PA) near the fin surface. 
%This suggests that less influence of the LSS on the centerline might be responsible for the accurate centerline pressure predictions.
However, due to this increased distance, the line LSS now interacts with the line LPA near the shoulder. 
Because the value of $\lambda$ used in the length-scale limiting is calibrated based on distance from the bottom plate ($y-$coordinate), the interaction of the lines LSS with LPA near the shoulder may result in a localized dominance of spanwise length-scales; this reduces the effectiveness of $y-$coordinate based limiting. %less effective.
%This can be prevented by 
Specifying a slightly higher $\beta$ value circumvents this complication as shown. %.
The combination of $(\beta,\lambda)=(1.85, 2.5)$ predicts a footprint where LSS does not interact with the LPA while being at a sufficient distance from the symmetry plane.
This results in the best simultaneous surface pressure and heat transfer predictions.
%Therefore, further prediction improvement requires a deeper understanding of the full 3D nature of the flow to devise better correction guidelines.
%This is being pursued by the authors and will be presented in the future.

\begin{figure}
    \centering
    %\subfloat[]{\includegraphics[width=0.45\textwidth,trim=100 250 100 270,clip]{Figures/DF_Press_Final_v3.pdf}}
    %\subfloat[]{\includegraphics[width=0.45\textwidth,trim=100 250 100 270,clip]{Figures/DF_Qw_Final_v3.pdf}} \\
    \subfloat[]{\includegraphics[width=0.47\textwidth,trim=100 250 100 270,clip]{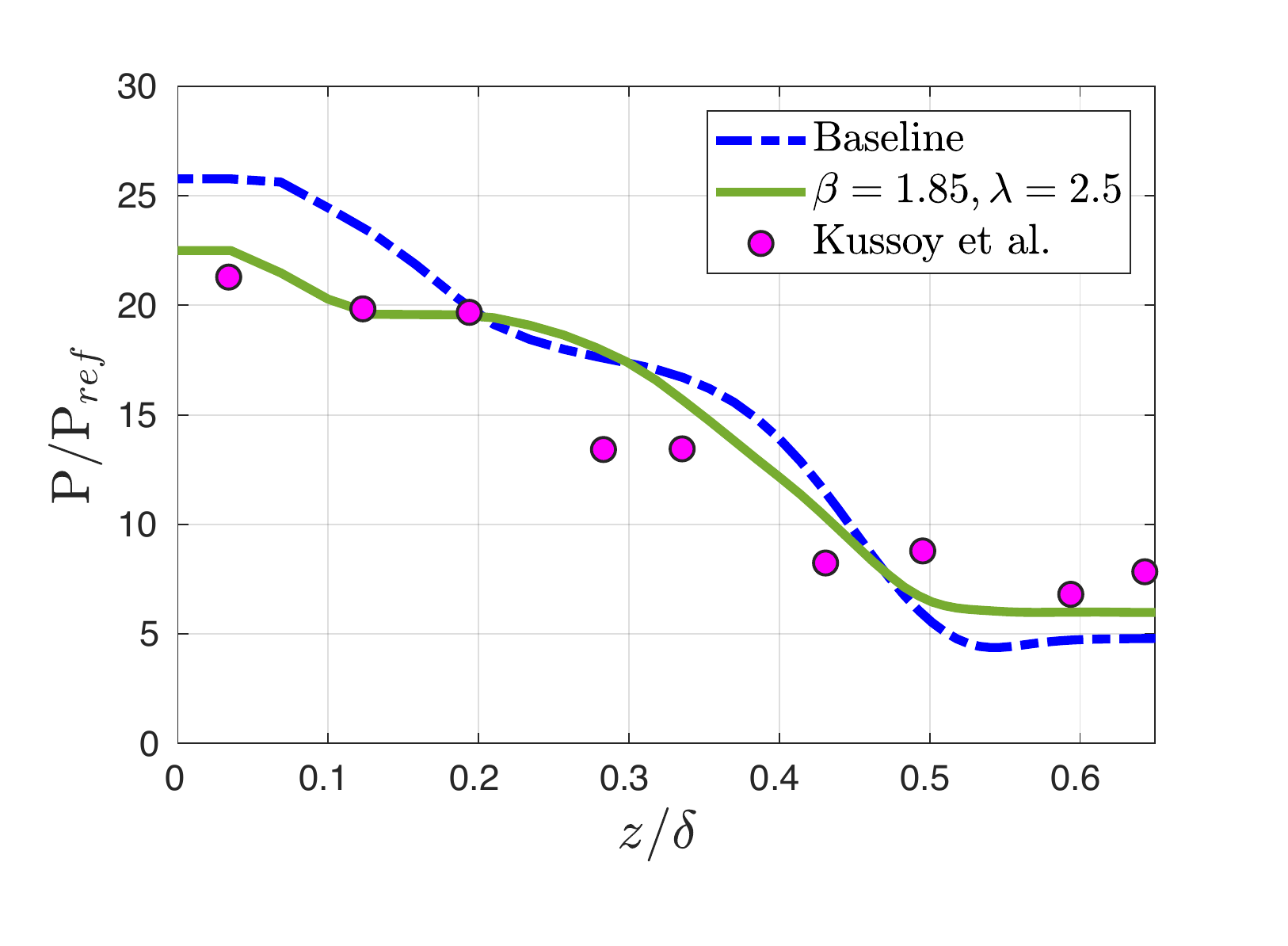}}
    \subfloat[]{\includegraphics[width=0.47\textwidth,trim=100 250 100 270,clip]{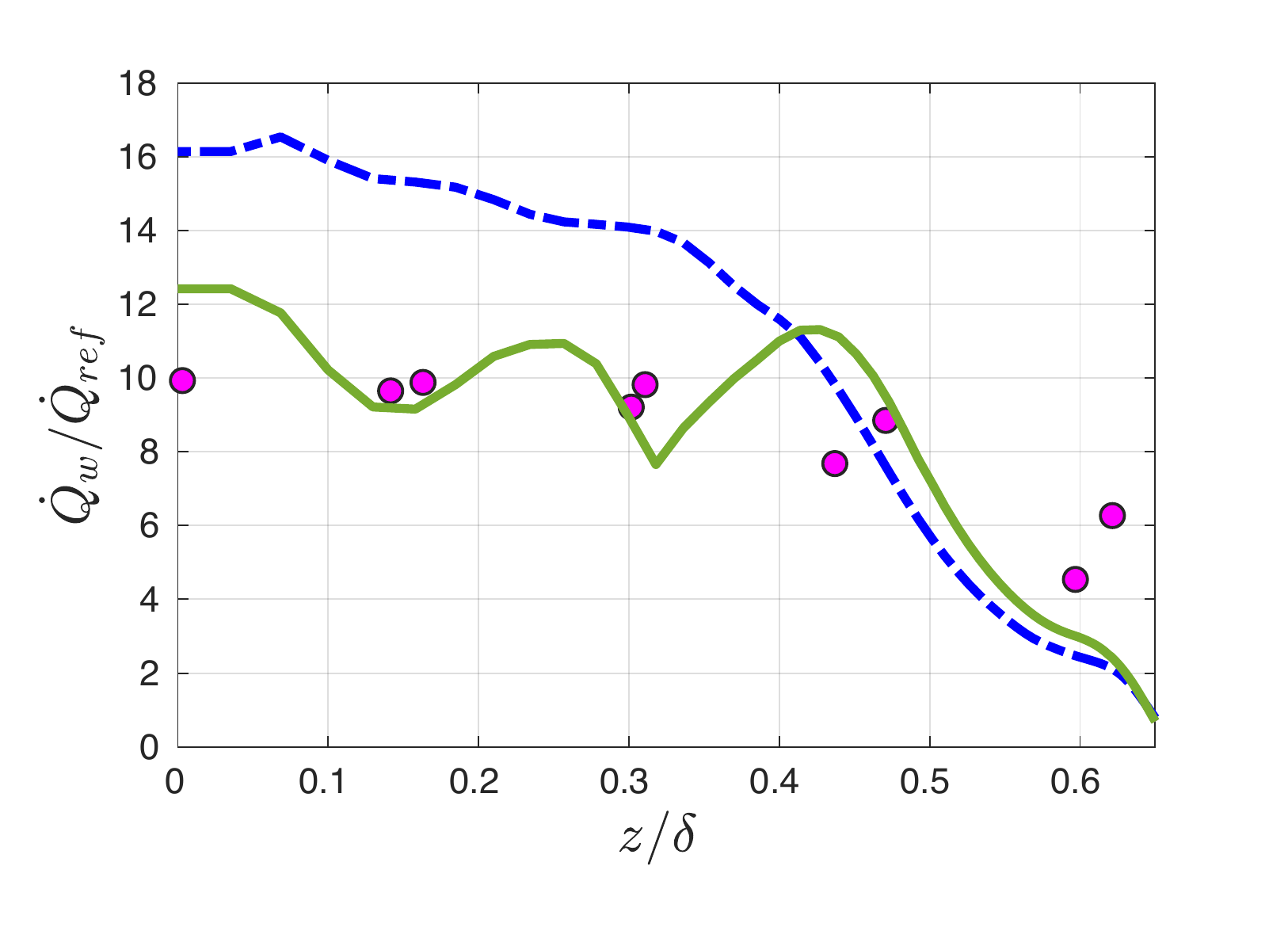}}
    \caption{DFM8: Spanwise predictions of surface pressure (left) and surface heat flux (right) with $(\beta,\lambda)=(1.85, 2.5)$  at the planes marked P and Q in Fig.~\ref{fig:DF_stream}b.}
    \label{fig:DFBestspan}
\end{figure}

%This improved behavior is also reflected in the spanwise distributions
%To further test the accuracy of the combination $(\beta,\lambda)=(1.6, 2.5)$, 
For completeness, Fig.~\ref{fig:DFBestspan} presents the spanwise surface pressure and heat flux predictions along the dashed vertical lines P and Q (see Fig.~\ref{fig:DF_stream}b) respectively.
The use of $(\beta,\lambda)=(1.85, 2.5)$ significantly improves the surface predictions along the span when compared to the baseline model.
Similar to the LCFM6 test case, these improvements are obtained at the same computational cost as the baseline calculations ($\sim 3{,}100$ CPU hours with $288$ processors).
This is highly encouraging and demonstrates that including these simple limiting coefficients can significantly improve predictions from existing RANS codes while retaining the original computational efficiency. %can 
Moreover, since the terms used to incorporate these corrections are derived from a general two-equation model formulation (~$\S~\ref{section:Turbmodel}$), %common even though the present investigation focuses on the $k-\epsilon$ model, following~$\S~\ref{section:Turbmodel}$, 
the analysis can be easily extended to other two-equation turbulence models. 

%As a final point, it is worth noting that although the upstream boundary layer dictates the threshold values of $\beta$ and $\lambda$, the optimal values of these coefficients appear to be dependant on the test case.
%Thus, a general recommendation on the optimal values of these coefficients remains an open question, especially for 3D flows. 
As a final point, we note that %it is worth noting that 
although the upstream boundary layer dictates the threshold values of $\beta$ and $\lambda$, coefficients derived for a given test case may apply for small variations in flow parameters, and are thus suitable for trend studies.
%appear to be dependant on the test case.
However, a general recommendation for values of RANS coefficients remains an open question, especially for the rich variety of features encountered in 3D flows. 

\section{Conclusion}\label{section:conclude}
This investigation presents a methodology to improve the performance of the standard $k-\epsilon$ turbulence model for a series of increasingly complicated test cases that exhibit flow phenomenon expected to occur in the current and future hypersonic vehicles. %can be significantly improved by placing simple limiters on turbulence production and length-scale without any increase in cost.
For a Mach~$11.1$ zero pressure gradient flow over a highly cooled flat plate, the baseline $k-\epsilon$ model yields very accurate estimates of surface shear stress and heat transfer rate.
However, the results show different transition locations depending on the parameters chosen.
A high inlet turbulent viscosity ratio overcomes the problem of the boundary layer avoiding transition.
%For a Mach~11 flow over a highly cooled flat plate, the baseline model accurately predicts the wall shear stress and heat transfer observed in experiments.
When applied to more complicated shock boundary layer interaction (SBLI) cases, the baseline model %fails to predict the separation bubble size and grossly overpredicts the wall heat flux.
under-predicts the size of the separation zone leading to incorrect predictions of heat transfer and surface pressure. 
The use of a compressibility correction significantly improves surface pressure predictions by reducing eddy viscosity in the upstream boundary layer. 
However, the correction degrades performance for the flat plate boundary layer flow. 
To circuvent this limitation, two coefficients are introduced that 
%This is achieved by defining two coefficients that 
independently control the amplification of turbulence production and the turbulent length-scale downstream of a shock wave.
The values of these limiting coefficients are carefully chosen based on the information from the upstream boundary layer;  this allows the modified $k-\epsilon$ model to retain its accuracy in zero pressure gradient boundary layers while significantly improving predictions in the presence of shock waves.
%In general, a decrease in the prediction limiting coefficient increase 
For a Mach~$6.17$ flow over a $7^\circ$ cone with a $40^\circ$ flare, the modified model accurately predicts the separation bubble size and the peak pressure, with only a modest disagreement in the wall heat flux.
When applied to the more complicated, three-dimensional Mach~$8.3$ flow in a $15^\circ$ double fin geometry, improved results for surface pressure and heat transfer are obtained along the centerline as well as along the span.
%Moreover, these improvements are obtained without any increase in computational cost.
The simplicity of these modifications allows a straightforward implementation in any existing code, while retaining the same computational cost.
Currently, efforts are focused on exploring a general recommendation for the optimal values of these coefficients based on a range of Mach numbers. 

%\section*{Appendix}

\section*{Acknowledgements}
This work was performed in part under the sponsorship of the DoD HPCMP Hypersonic Vehicle Simulation Institute with Dr. R. Cummings (USAFA) serving as Project Monitor. 
DVG also acknowledges partial support from the Collaborative Center for Aeronautical Sciences.
The views and conclusions contained herein are those of the authors and do not represent the opinion of USAFA or the U.S. government. 
The authors are grateful for computational resource grants from the DoD HPCMP and the Ohio Supercomputer Center.
%Several figures were made using FieldView software with licenses obtained from the IntelligentLight University Partnership Program. 
%The authors would also like to thank Pa

%\clearpage
\bibliography{Ref}

\end{document}